\documentclass[twocolumn]{aastex701}

\input{lib/patch/aastex}

\usepackage{import}
\usepackage[svgnames]{xcolor}
\usepackage{pgfplots}
\usepackage{tikz}

\usepackage{etoolbox}
\usepackage{etextools}
\usepackage{suffix}
\usepackage{xparse}
\usepackage{environ}
\usepackage{xspace}
\usepackage{iftex}

\usepackage[tbtags, intlimits, sumlimits, namelimits]{amsmath}
\usepackage{amssymb}
\usepackage{dsfont}
\usepackage{bm}
\usepackage[normalem]{ulem}
\usepackage{fancyvrb}
\usepackage[style=american]{csquotes}
\MakeOuterQuote{"}

\usepackage[allowspaces]{mathtools}

\undef{\interval}
\usepackage{interval}
\intervalconfig{soft open fences, separator symbol=;}

\usepackage{cancel}
\usepackage{esvect}
\usepackage{physics}

\makeatletter
	\usepackage[stable, flushmargin, bottom]{footmisc}
\makeatother
\usepackage{pbox}
\usepackage{makecell,multirow}
\makeatletter
\@ifclassloaded{aastex701}{}{\usepackage{multicol}}
\makeatother
\usepackage{booktabs}
\usepackage[subrefformat=parens]{subcaption}
\usepackage{appendix}
\AtBeginEnvironment{appendices}{\crefalias{section}{appendix}\crefalias{subsection}{appendix}}

\usepackage[hyphens]{url}
\usepackage{hyperref}
\usepackage{cleveref}

\crefname{appendix}{appendix}{appendices}
\crefname{subsection}{subsection}{subsections}

\makeatletter\@ifclassloaded{aastex701}{}{
\usepackage[numberlinked=true]{footnotebackref}
}\makeatother

\ifx\usetxfonts\undefined\else
\usepackage{newtxtext,newtxmath}
\defaultfontfeatures{}
\fi

\usepackage{siunitx}
\usepackage[shortcuts]{extdash}

\usepackage{graphbox}

\usepackage{pgfplotstable}
\pgfplotsset{compat=newest, filter discard warning=false}
\usepgfplotslibrary{colormaps,colorbrewer,groupplots,fillbetween,patchplots}

\usepackage{currfile}
\usetikzlibrary{arrows.meta, backgrounds, calc, colorbrewer, decorations, decorations.markings, external, fadings, fit, intersections, math, patterns, plotmarks, positioning, shapes, spath3}
\usetikzlibrary{graphs, graphs.standard}
\ifLuaTeX
\usetikzlibrary{graphdrawing}
\usegdlibrary{layered,trees,force}
\fi

\def\tikzauxdir{build }
\tikzset{
	external/system call={%
		lualatex %
		\tikzexternalcheckshellescape %
		-halt-on-error %
		-interaction=batchmode %
		-aux-directory=\tikzauxdir %
		-jobname "\image" %
		"\texsource" %
	},
	/pgf/images/external info,
}

\pgfkeys{
	/tikz/text=black, /pgfplots/axis line style={black},
	/tikz/truth star/.style={star, star point ratio=2.618, inner sep=0pt, minimum size=0.8em}
}

\pgfplotsset{
	every axis/.append style={
		axis on top,
		scaled ticks=false,
		title style={font=\footnotesize},
		label style={font=\footnotesize},
		tick label style={font=\footnotesize},
		tick style={thin, black}, tickwidth=5pt,
		tick pos=left, tick align=outside,
	},
	every axis plot/.append style={
		ultra thick, line join=round, line cap=butt,
		domain={\pgfkeysvalueof{/pgfplots/xmin}:\pgfkeysvalueof{/pgfplots/xmax}},
	},
	every axis legend/.append style={
		font=\footnotesize, cells={anchor=west}, fill=none,
	},
	filled legend/.style={legend style={fill=white, fill opacity=0.9}},
	colorbar base/.style={
		axis equal image=false, tickwidth=3pt,
		xlabel style={at={(ticklabel cs:0.5)}, inner sep=4pt, anchor=near ticklabel, text depth=0pt},
		ticklabel pos=right, tick pos=right,
	},
	colorbar top/.style={
		colorbar horizontal,
		colorbar style={
			colorbar base, height=0.5em,
			at={(parent axis.north west)}, anchor=south west,
		},
		colorbar shift/.style={yshift=0.1cm},
		colorbar is horizontal=true,
	},
	colorbar bottom/.style={
		colorbar horizontal,
		colorbar style={
			colorbar base, height=0.5em,
			at={(parent axis.south west)}, anchor=north west,
			xtick pos=bottom, xticklabel pos=bottom,
		},
		colorbar shift/.style={yshift=-0.1cm},
		colorbar is horizontal=true,
	},
	colorbar right/.append style={
		colorbar style={colorbar base, width=0.5em, at={(parent axis.north east)}},
		colorbar shift/.style={xshift=0.1cm},
		colorbar is horizontal=false,
	},
	colorbar left/.append style={
		colorbar style={ytick pos=left},
		colorbar is horizontal=false
	}
}

\definecolor{viridis0}{rgb}{0.267004, 0.004874, 0.329415}
\definecolor{viridiss0}{rgb}{0.283072, 0.130895, 0.449241}

\colorlet{Set1-red}{Set1-A}
\colorlet{Set1-blue}{Set1-B}
\colorlet{Set1-green}{Set1-C}
\colorlet{Set1-purple}{Set1-D}
\colorlet{Set1-orange}{Set1-E}
\colorlet{Set1-yellow}{Set1-F}
\colorlet{Set1-brown}{Set1-G}
\colorlet{Set1-pink}{Set1-H}
\colorlet{Set1-grey}{Set1-I}
\colorlet{Set1-gray}{Set1-I}

\pgfplotsset{
	invert cmap/.code={\pgfplotsset{colormap={#1 reverse}{indices of colormap={\pgfplotscolormaplastindexof{#1},...,0 of #1}}}},
	colormap={traffic}{rgb=(0.8,0,0), rgb=(0.9,0.45,0.07), rgb=(1, 0.9, 0.14), rgb=(0,0.7,0)},
	invert cmap=traffic,
	colormap={midtraffic}{rgb(-1)=(0,0,1) color(0)=(ForestGreen) color(0.5)=(Goldenrod) color(1)=(FireBrick)},
	invert cmap=midtraffic,
	colormap={seismic}{rgb=(0,0,0.3) rgb=(0,0,1) rgb=(1,1,1) rgb=(1,0,0) rgb=(0.5,0,0)},
	invert cmap=seismic,
	colormap={bwr}{rgb=(0,0,1) rgb=(1,1,1) rgb=(1,0,0)},
	invert cmap=bwr,
    colormap={brg}{rgb=(0,0,1) rgb=(1,0,0) rgb=(0,1,0)},
    invert cmap=brg,
	colormap={inferno}{rgb=(0.001,0.000,0.014) rgb=(0.002,0.001,0.019) rgb=(0.003,0.002,0.024) rgb=(0.005,0.003,0.031) rgb=(0.006,0.005,0.039) rgb=(0.008,0.006,0.047) rgb=(0.010,0.008,0.055) rgb=(0.012,0.009,0.063) rgb=(0.014,0.011,0.072) rgb=(0.017,0.013,0.080) rgb=(0.019,0.015,0.089) rgb=(0.022,0.017,0.097) rgb=(0.026,0.019,0.106) rgb=(0.029,0.022,0.115) rgb=(0.033,0.024,0.123) rgb=(0.038,0.026,0.132) rgb=(0.042,0.028,0.141) rgb=(0.047,0.030,0.150) rgb=(0.052,0.032,0.159) rgb=(0.056,0.035,0.168) rgb=(0.061,0.037,0.178) rgb=(0.066,0.039,0.187) rgb=(0.071,0.040,0.196) rgb=(0.077,0.042,0.206) rgb=(0.082,0.043,0.215) rgb=(0.087,0.045,0.225) rgb=(0.093,0.046,0.234) rgb=(0.099,0.046,0.244) rgb=(0.105,0.047,0.253) rgb=(0.111,0.047,0.263) rgb=(0.117,0.048,0.272) rgb=(0.123,0.048,0.282) rgb=(0.129,0.047,0.291) rgb=(0.136,0.047,0.300) rgb=(0.142,0.046,0.309) rgb=(0.149,0.045,0.317) rgb=(0.156,0.045,0.325) rgb=(0.163,0.044,0.333) rgb=(0.170,0.042,0.341) rgb=(0.176,0.041,0.348) rgb=(0.183,0.040,0.355) rgb=(0.190,0.039,0.361) rgb=(0.197,0.038,0.368) rgb=(0.204,0.038,0.373) rgb=(0.211,0.037,0.379) rgb=(0.218,0.037,0.384) rgb=(0.225,0.036,0.388) rgb=(0.232,0.036,0.392) rgb=(0.238,0.037,0.396) rgb=(0.245,0.037,0.400) rgb=(0.252,0.038,0.403) rgb=(0.258,0.039,0.406) rgb=(0.265,0.040,0.409) rgb=(0.271,0.041,0.412) rgb=(0.278,0.042,0.414) rgb=(0.284,0.044,0.417) rgb=(0.291,0.046,0.419) rgb=(0.297,0.047,0.420) rgb=(0.304,0.049,0.422) rgb=(0.310,0.051,0.424) rgb=(0.316,0.053,0.425) rgb=(0.323,0.056,0.426) rgb=(0.329,0.058,0.428) rgb=(0.335,0.060,0.429) rgb=(0.342,0.062,0.429) rgb=(0.348,0.065,0.430) rgb=(0.354,0.067,0.431) rgb=(0.360,0.069,0.431) rgb=(0.367,0.072,0.432) rgb=(0.373,0.074,0.432) rgb=(0.379,0.076,0.433) rgb=(0.385,0.079,0.433) rgb=(0.391,0.081,0.433) rgb=(0.398,0.083,0.433) rgb=(0.404,0.086,0.433) rgb=(0.410,0.088,0.433) rgb=(0.416,0.090,0.433) rgb=(0.423,0.093,0.433) rgb=(0.429,0.095,0.432) rgb=(0.435,0.097,0.432) rgb=(0.441,0.099,0.432) rgb=(0.447,0.102,0.431) rgb=(0.454,0.104,0.430) rgb=(0.460,0.106,0.430) rgb=(0.466,0.108,0.429) rgb=(0.472,0.111,0.428) rgb=(0.479,0.113,0.427) rgb=(0.485,0.115,0.427) rgb=(0.491,0.117,0.426) rgb=(0.497,0.119,0.424) rgb=(0.503,0.122,0.423) rgb=(0.510,0.124,0.422) rgb=(0.516,0.126,0.421) rgb=(0.522,0.128,0.420) rgb=(0.528,0.130,0.418) rgb=(0.535,0.133,0.417) rgb=(0.541,0.135,0.415) rgb=(0.547,0.137,0.414) rgb=(0.553,0.139,0.412) rgb=(0.560,0.141,0.410) rgb=(0.566,0.144,0.408) rgb=(0.572,0.146,0.406) rgb=(0.578,0.148,0.404) rgb=(0.585,0.150,0.402) rgb=(0.591,0.153,0.400) rgb=(0.597,0.155,0.398) rgb=(0.603,0.157,0.396) rgb=(0.609,0.159,0.394) rgb=(0.616,0.162,0.391) rgb=(0.622,0.164,0.389) rgb=(0.628,0.167,0.386) rgb=(0.634,0.169,0.384) rgb=(0.640,0.171,0.381) rgb=(0.646,0.174,0.378) rgb=(0.652,0.176,0.376) rgb=(0.658,0.179,0.373) rgb=(0.665,0.182,0.370) rgb=(0.671,0.184,0.367) rgb=(0.677,0.187,0.364) rgb=(0.683,0.190,0.361) rgb=(0.689,0.192,0.358) rgb=(0.695,0.195,0.354) rgb=(0.701,0.198,0.351) rgb=(0.707,0.201,0.348) rgb=(0.712,0.204,0.344) rgb=(0.718,0.207,0.341) rgb=(0.724,0.210,0.337) rgb=(0.730,0.213,0.334) rgb=(0.736,0.216,0.330) rgb=(0.741,0.219,0.327) rgb=(0.747,0.222,0.323) rgb=(0.753,0.226,0.319) rgb=(0.758,0.229,0.315) rgb=(0.764,0.233,0.311) rgb=(0.770,0.236,0.307) rgb=(0.775,0.240,0.304) rgb=(0.781,0.243,0.300) rgb=(0.786,0.247,0.295) rgb=(0.791,0.251,0.291) rgb=(0.797,0.255,0.287) rgb=(0.802,0.259,0.283) rgb=(0.807,0.263,0.279) rgb=(0.812,0.267,0.275) rgb=(0.817,0.271,0.270) rgb=(0.822,0.275,0.266) rgb=(0.827,0.280,0.262) rgb=(0.832,0.284,0.257) rgb=(0.837,0.288,0.253) rgb=(0.842,0.293,0.249) rgb=(0.847,0.298,0.244) rgb=(0.851,0.302,0.240) rgb=(0.856,0.307,0.235) rgb=(0.861,0.312,0.231) rgb=(0.865,0.317,0.226) rgb=(0.869,0.322,0.221) rgb=(0.874,0.327,0.217) rgb=(0.878,0.332,0.212) rgb=(0.882,0.337,0.208) rgb=(0.886,0.343,0.203) rgb=(0.890,0.348,0.198) rgb=(0.894,0.353,0.194) rgb=(0.898,0.359,0.189) rgb=(0.902,0.364,0.184) rgb=(0.906,0.370,0.179) rgb=(0.909,0.376,0.175) rgb=(0.913,0.382,0.170) rgb=(0.916,0.387,0.165) rgb=(0.920,0.393,0.160) rgb=(0.923,0.399,0.155) rgb=(0.926,0.405,0.150) rgb=(0.930,0.411,0.145) rgb=(0.933,0.418,0.140) rgb=(0.936,0.424,0.135) rgb=(0.939,0.430,0.130) rgb=(0.942,0.436,0.125) rgb=(0.944,0.443,0.120) rgb=(0.947,0.449,0.115) rgb=(0.950,0.456,0.110) rgb=(0.952,0.462,0.105) rgb=(0.955,0.469,0.100) rgb=(0.957,0.475,0.095) rgb=(0.959,0.482,0.089) rgb=(0.961,0.489,0.084) rgb=(0.963,0.495,0.079) rgb=(0.965,0.502,0.074) rgb=(0.967,0.509,0.069) rgb=(0.969,0.516,0.063) rgb=(0.971,0.523,0.058) rgb=(0.973,0.530,0.053) rgb=(0.974,0.537,0.048) rgb=(0.976,0.544,0.044) rgb=(0.977,0.551,0.039) rgb=(0.978,0.558,0.035) rgb=(0.980,0.565,0.031) rgb=(0.981,0.572,0.029) rgb=(0.982,0.579,0.026) rgb=(0.983,0.587,0.025) rgb=(0.984,0.594,0.024) rgb=(0.985,0.601,0.024) rgb=(0.985,0.608,0.024) rgb=(0.986,0.616,0.026) rgb=(0.987,0.623,0.028) rgb=(0.987,0.630,0.031) rgb=(0.987,0.638,0.035) rgb=(0.988,0.645,0.040) rgb=(0.988,0.653,0.046) rgb=(0.988,0.660,0.052) rgb=(0.988,0.668,0.058) rgb=(0.988,0.675,0.065) rgb=(0.988,0.683,0.072) rgb=(0.987,0.690,0.080) rgb=(0.987,0.698,0.088) rgb=(0.987,0.706,0.096) rgb=(0.986,0.713,0.104) rgb=(0.986,0.721,0.112) rgb=(0.985,0.728,0.121) rgb=(0.984,0.736,0.130) rgb=(0.983,0.744,0.138) rgb=(0.982,0.751,0.148) rgb=(0.981,0.759,0.157) rgb=(0.980,0.767,0.166) rgb=(0.979,0.775,0.176) rgb=(0.977,0.782,0.186) rgb=(0.976,0.790,0.196) rgb=(0.975,0.798,0.206) rgb=(0.973,0.805,0.217) rgb=(0.971,0.813,0.228) rgb=(0.970,0.821,0.239) rgb=(0.968,0.829,0.250) rgb=(0.966,0.836,0.262) rgb=(0.964,0.844,0.273) rgb=(0.963,0.851,0.286) rgb=(0.961,0.859,0.298) rgb=(0.959,0.867,0.311) rgb=(0.957,0.874,0.324) rgb=(0.955,0.882,0.337) rgb=(0.953,0.889,0.351) rgb=(0.952,0.896,0.366) rgb=(0.950,0.903,0.380) rgb=(0.949,0.910,0.395) rgb=(0.948,0.917,0.411) rgb=(0.947,0.924,0.426) rgb=(0.946,0.931,0.442) rgb=(0.946,0.937,0.459) rgb=(0.947,0.943,0.475) rgb=(0.948,0.949,0.491) rgb=(0.950,0.955,0.508) rgb=(0.952,0.961,0.524) rgb=(0.955,0.966,0.540) rgb=(0.958,0.971,0.556) rgb=(0.962,0.976,0.572) rgb=(0.966,0.981,0.587) rgb=(0.971,0.985,0.602) rgb=(0.977,0.990,0.617) rgb=(0.982,0.994,0.631) rgb=(0.988,0.998,0.645)},
	invert cmap=inferno,
	colormap={blackwhite}{color(0cm)=(black); color(1cm)=(white)}, invert cmap=blackwhite,
}

\tikzset{
	/cntrs/do mark/.style={mark=at position #1 with {
		\node[
			transform shape,
			inner sep=0.3em, outer sep=0pt
		] (cntrlbl) {\num{\pgfkeysvalueof{/cntrs/current contour label}}};
		\stepcounter{n}
		\path[spath/save global={cntrlbl \then}]
			let \p1=($(cntrlbl.east) - (cntrlbl)$), \p2=($(cntrlbl.north) - (cntrlbl)$)
			in
			(cntrlbl) circle [x radius=\x1, y radius=\y2];
	}},
	/pgfplots/axis with contours/.style={
		execute at begin axis={
			\newcounter{n}
			\setcounter{n}{0}
			\path[spath/save=bb] (\pgfkeysvalueof{/pgfplots/xmin},\pgfkeysvalueof{/pgfplots/ymin}) rectangle (\pgfkeysvalueof{/pgfplots/xmax},\pgfkeysvalueof{/pgfplots/ymax});
		}
	},
	cntr/.style={},
	cntr'/.style={cntr=#1, /pgf/decoration/reverse path}
}

\pgfplotsset{
	corner/.style={
		group/x descriptions at=edge bottom,
		group/y descriptions at=edge left,
		group/horizontal sep=0.1cm, group/vertical sep=0.1cm,
		height={\pgfkeysvalueof{/pgfplots/width}},
		domain={\pgfkeysvalueof{/pgfplots/xmin}:\pgfkeysvalueof{/pgfplots/xmax}},
		scaled ticks=false,
		tickwidth=2pt,
		tick label style={
			rotate=45, anchor=east, font=\scriptsize,
			/pgf/number format/fixed, /pgf/number format/zerofill
		},
		label style={at={(ticklabel* cs:0.5,3em)}, anchor=base, inner sep=0pt,},
		ylabel style={at={(ticklabel* cs:0.5,2.5em)}},
	},
}

\tikzset{
    cntr/.style={ultra thin,},
	cntr 1/.style={cntr, fill=green},
	cntr 2/.style={cntr, fill=yellow},
	cntr 3/.style={cntr, fill=red},
	hist1d/.style={no marks, samples=51},
	truth line/.style={black, no marks, thin},
	truth/.style={black, mark=*, mark size=1pt},
}

\input{ellipse}
\usetikzlibrary{bending}
\colorlet{graphdrawcolor}{black}
\colorlet{graphfillcolor}{graphdrawcolor!20}

\ifLuaTeX
\tikzset{subgraph text none}
\fi

\tikzset{
	unfilled/.style={fill=none},
	filled/.style={fill=graphfillcolor},
	graph edge/.style={shorten >=1.5pt},
    simple graph node/.style={minimum size=1.8em, draw=graphdrawcolor},
	graph node/.style={simple graph node, >graph edge, <graph edge},
	input/.style={graph node, draw=none, minimum size=0pt, inner sep=2pt},
	param/.style={graph node, filled},
	calc/.style={graph node, unfilled, circle},
	dist/.style={graph node, unfilled, densely dash dot, >shorten >=0pt},
	guide edge/.style={graph edge, densely dotted},
	guide*/.style={draw=graphdrawcolor, densely dash dot},
	guide/.style={dist, <guide edge},
	output/.style={calc, filled},
	sum/.style={"+"},
 	>={Stealth},
    graph group/.style={draw, densely dotted, inner sep=3pt},
    graph group label/.style={inner sep=2pt, font=\scriptsize},
    graph plate/.style={graph group, solid, inner sep=4pt},
    graph plate label/.style={graph group label, inner sep=4pt}
}

\makeatletter
\tikzset{
	coordinate node/.style={inner sep=0pt, outer sep=0pt, minimum size=0pt}
}
\makeatother

\newcommand{\nnkeys}[1]{\pgfkeys{/nngraph/.cd, #1}}
\newcommand{\nnglobalkeys}[1]{\begingroup\ifnum\the\globaldefs>0\relax \else \globaldefs=1\fi \nnkeys{#1}\endgroup}
\newcommand{\nnkeysvalueof}[1]{\pgfkeysvalueof{/nngraph/#1}}

\colorlet{nngraphfadecolor}{white}
\colorlet{nngraphneurondrawcolor}{black}
\colorlet{nngraphneuronfillcolor}{white}
\colorlet{nngraphconnectiondrawcolor}{black}

\nnkeys{%
	blackness/.initial={\pgfkeysnovalue},
	draw color/.initial={\pgfkeysnovalue},
	fill color/.initial={\pgfkeysnovalue},
	layer/.style={
		layer name/.initial={},
		n batch/.initial={1},
		n neurons/.initial={1},
		r decay opacity/.initial={0},
		yshift/.initial={1},
		zshift/.initial={-1},
		yindex/.initial={0},
		zindex/.initial={0},
		zopacity0/.initial={1},
		zopacity/.initial={\pgfkeysnovalue},
		blackness/.initial={\pgfkeysnovalue},
		current neuron name/.initial={},
	},
	neuron/.style={
		fill={nngraphneuronfillcolor!\nnkeysvalueof{blackness}!nngraphfadecolor},
		draw={nngraphneurondrawcolor!\nnkeysvalueof{blackness}!nngraphfadecolor}
	},
	connection/.style={
		draw={nngraphconnectiondrawcolor!\nnkeysvalueof{blackness}!nngraphfadecolor}
	},
	translation/.style={/nngraph/connection, dashed},
}

\newcommand{\iterateneurons}[3][\i]{
	\pgfmathtruncatemacro{\neuronimax}{div(#2, 2)}
	\ifthenelse{#2 > 1}{%
		\foreach #1 in {\neuronimax, ..., 0.1} {#3[((mod(#2, 2) == 0) ? -0.5 : 0)]{#1} #3[((mod(#2, 2) == 0) ? 0.5 : 0)]{-#1}}%
		\ifthenelse{\isodd{#2}}{#3{0}}{}%
	}{#3{0}}
}

\newcommand{\nnasquare}[2]{%
	++({-(#1)/2}, {-(#1)/2}) coordinate (#2_south west)
	-- ++(#1, 0) coordinate (#2_south east)
	-- coordinate (#2_out) ++(0, #1) coordinate (#2_north east)
	-- ++({-(#1)}, 0) coordinate (#2_north west)
	-- coordinate (#2_in) ++(0, {-(#1)})
	node[fit={(#2_south west) (#2_south east) (#2_north east) (#2_north west)}, coordinate node] (#2) {}
}

\newcommand{\nnaneuron}{\nnasquare}

\newcommand{\aneuronpath}[2][0]{
	\nnkeys{%
		blackness/.evaluated={\nnkeysvalueof{zopacity} * 100 * (1 - \nnkeysvalueof{r decay opacity} * abs(#2))},
		current neuron name={\nnkeysvalueof{layer name} batch \nnkeysvalueof{zindex} neuron #2},
	}
	\path[/nngraph/neuron, name path={\nnkeysvalueof{current neuron name}}]
	(0, {(#2 + #1) * (\nnkeysvalueof{yshift})})
	\nnaneuron{1}{\nnkeysvalueof{current neuron name}};
	\nnglobalkeys{%
		current neurons/\nnkeysvalueof{current neuron name}/.cd,
		blackness/.initial/.expanded={\nnkeysvalueof{blackness}}
	}
	\nnglobalkeys{%
		current layers/\nnkeysvalueof{layer name}/.cd,
		all neurons/.append/.expanded={\expandafter\ifblank\expandafter{\expanded{\nnkeysvalueof{current layers/\nnkeysvalueof{layer name}/all neurons}}}{}{,} \nnkeysvalueof{current neuron name}},
		batch \nnkeysvalueof{zindex} neurons/.append/.expanded={\expandafter\ifblank\expandafter{\expanded{\nnkeysvalueof{current layers/\nnkeysvalueof{layer name}/batch \nnkeysvalueof{zindex} neurons}}}{}{,} \nnkeysvalueof{current neuron name}}
	}
}

\newcommand{\nngraphout}{_out}
\newcommand{\nngraphin}{_in}

\newcommand{\connectneurons}[3][]{
	\nnkeys{blackness/.evaluated={
		100
		* (\nnkeysvalueof{current neurons/#2/blackness}) / 100
		* (\nnkeysvalueof{current neurons/#3/blackness}) / 100
	}}
	\draw[/nngraph/connection, #1] (#2\nngraphout) to (#3\nngraphin);
}

\newcommand{\connectalayer}[1]{
	\ifthenelse{\maxj > 0}{
		\foreach \j in {\maxj, ..., 0.1} {
			\connectneurons{\nnkeysvalueof{l1n} #1}{\nnkeysvalueof{l2n} \j}
			\connectneurons{\nnkeysvalueof{l1n} #1}{\nnkeysvalueof{l2n} -\j}
		}
	}{}
	\ifthenelse{\isodd{\nnkeysvalueof{current layers/\nnkeysvalueof{layer 2 name}/n neurons}}}{\connectneurons{\nnkeysvalueof{l1n} #1}{\nnkeysvalueof{l2n} 0}}{}
}

\newcommand{\connectlayers}[1][]{
	\nnkeys{%
		layer 1 name/.initial={}, layer 2 name/.initial={},
		layer 1 batch/.initial={0}, layer 2 batch/.initial={0},
		#1,
		l1n/.initial={\nnkeysvalueof{layer 1 name} batch \nnkeysvalueof{layer 1 batch} neuron},
		l2n/.initial={\nnkeysvalueof{layer 2 name} batch \nnkeysvalueof{layer 2 batch} neuron},
	}
	\pgfmathtruncatemacro{\maxi}{\nnkeysvalueof{current layers/\nnkeysvalueof{layer 1 name}/n neurons} / 2}
	\pgfmathtruncatemacro{\maxj}{\nnkeysvalueof{current layers/\nnkeysvalueof{layer 2 name}/n neurons} / 2}

	\ifthenelse{\maxi > 0}{
		\foreach \i in {\maxi, ..., 0.1} {
			\connectalayer{\i}
			\connectalayer{-\i}
		}
	}{}
	\ifthenelse{\isodd{\nnkeysvalueof{current layers/\nnkeysvalueof{layer 1 name}/n neurons}}}{\connectalayer{0}}{}
}

\input{utils}

\ifLuaTeX
\usepackage{luacode}

\begin{luacode*}
function erf(x)
	-- constants
	a1 =  0.254829592
	a2 = -0.284496736
	a3 =  1.421413741
	a4 = -1.453152027
	a5 =  1.061405429
	p  =  0.3275911
	
	-- Save the sign of x
	sign = 1
	if x < 0 then
		sign = -1
	end
	x = math.abs(x)
	
	-- A\&S formula 7.1.26
	t = 1.0/(1.0 + p*x)
	y = 1.0 - (((((a5*t + a4)*t) + a3)*t + a2)*t + a1)*t*math.exp(-x*x)
	
	return sign*y
end
\end{luacode*}

\pgfmathdeclarefunction{erf}{1}{%
	\edef\pgfmathresult{%
		\directlua{tex.print(erf(\pgfmathfloatvalueof{#1}))}%
	}%
}

\fi

\input{foreach}
\let\oldcaption\caption
\renewcommand{\caption}[1]{\oldcaption{\protecting{#1}}}
\appto{\textellipsis}{\xspace}
\appto{\textrightarrow}{\xspace}

\newcommand{\symfootnotetext}[2][]{%
\let\oldthefootnote=\thefootnote%
\renewcommand{\thefootnote}{\fnsymbol{footnote}}%
\footnotetext[#1]{#2}%
\let\thefootnote\oldthefootnote%
}

\newcommand{\SISSA}{%
	Theoretical and Scientific Data Science group,       
	Scuola Internazionale Superiore di Studi Avanzati (SISSA), 
	via Bonomea 265, 34136 Trieste Italy}

\defcitealias{Carrick_2015}{C15}

\defcitealias{Taylor_2021}{T21}
\defcitealias{Brout_2022}{B22}

\defcitealias{Planck2018-1}{\emph{Planck}}
\defcitealias{Riess_2022}{\SHoES}

\defcitealias{Rigault_2025}{R25}

\AtBeginDocument{
    \dummylabel[figure]{ext:Jones_2019:fig:hists}{1}
    \dummylabel[table]{ext:Brout_2022a:tbl:sims}{1}
    \dummylabel[section]{ext:Riess_2022:sec:formalism}{2.1}
    \dummylabel[table]{ext:Riess_2022:tbl:dists}{6}
    \dummylabel[figure]{ext:Riess_2022:fig:covdistmod}{16}

    \dummylabel[section]{ext:Weinberg_2008:sec:Taylor-cosmo}{1.4}

    \dummylabel[figure]{ext:Amenouche_2025:fig:bias}{9}
    \dummylabel[equation]{ext:Carreres_2025a:eqn:vcov}{17}
    \dummylabel[figure]{ext:Lacroix_2025:fig:offset}{12}
    \dummylabel[figure]{ext:Rigault_2025:fig:pocket}{2}
    \dummylabel[subsection]{ext:Rigault_2025:sec:selection}{4.6}

    \dummylabel[figure]{ext:Stiskalek_2026c:fig:post}{9}
}
\usepackage{dowith}

\DeclareDocumentCommand\myfbraces{ s t\big t\Big t\bigg t\Bigg m m m m }
{ 
	#8%
	\IfBooleanTF{#1}
	{\argopen#6\smash{#9}\argclose#7\vphantom{#9}}
	{
		\IfBooleanTF{#2}{\argopen*\bigl#6#9\argclose*\bigr#7}{
			\IfBooleanTF{#3}{\argopen*\Bigl#6#9\argclose*\Bigr#7}{
				\IfBooleanTF{#4}{\argopen*\biggl#6#9\argclose*\biggr#7}{
					\IfBooleanTF{#5}
					{\argopen*\Biggl#6#9\argclose*\Biggr#7}
					{\argopen#6#9\argclose#7}
				}
			}
		}
	}
}

\DeclareDocumentCommand\myopbraces{ m g o d() }{%
	\IfNoValueTF{#2}{%
		\IfNoValueTF{#3}{%
			\IfNoValueTF{#4}%
			{#1}%
			{\myfbraces{\lparen}{\rparen}{#1}{#4}}%
		}%
		{%
			\myfbraces{\lbrack}{\rbrack}{#1}{#3}%
			\IfNoValueTF{#4}{}{(#4)}%
		}%
	}%
	{
		\myfbraces{\lbrace}{\rbrace}{#1}{#2}%
		\IfNoValueTF{#3}{}{[#3]}%
		\IfNoValueTF{#4}{}{(#4)}%
	}%
}

\begingroup\lccode`\|=`\\
\lowercase{\endgroup\def\removebs#1{\if#1|\else#1\fi}}
\newcommand{\macroname}[1]{\expandafter\removebs\string#1}

\NewDocumentCommand{\newmathcommand}{m O{0} m}{\newcommand{#1}[#2]{\ensuremath{#3}\xspace}}
\NewDocumentCommand{\renewmathcommand}{m O{0} m}{\renewcommand{#1}[#2]{\ensuremath{#3}\xspace}}

\makeatletter

\def\@upgreekone #1{\csdef{#1}{\csuse{up#1}}}
\newcommand{\upgreek}[1]{{%
	\forcsvlist\@upgreekone{alpha, beta, gamma, Gamma, delta, Delta, epsilon, varepsilon, zeta, eta, theta, vartheta, Theta, iota, kappa, lambda, Lambda, mu, nu, xi, Xi, pi, Pi, rho, varrho, sigma, Sigma, tau, upsilon, Upsilon, phi, varphi, Phi, chi, psi, Psi, omega, Omega}%
	#1
}}

\newcommand{\clearextraction}{
	\providecommand{\extractedmain}{} \renewcommand{\extractedmain}{}
	\providecommand{\extractedsub}{} \renewcommand{\extractedsub}{}
	\providecommand{\extractedsuper}{} \renewcommand{\extractedsuper}{}
}
\clearextraction

\newcommand{\splitargs}[4]{%
	\def#3{#1}%
	\def#4{#2}%
}
\NewDocumentCommand{\extractsubsuper}{> {\SplitArgument{1}{_}} u{\relax}}{%
	\splitargs#1{\gfirst}{\gsecond}%
	\expandafter\SplitArgument\expandafter1\expandafter^\expandafter{\gfirst}%
	\expandafter\splitargs\ProcessedArgument\first\second%
	\expandafter\SplitArgument\expandafter1\expandafter^\expandafter{\gsecond}%
	\expandafter\splitargs\ProcessedArgument\third\fourth%
	\edef\extractedmain{\expandonce{\first}}%
	\def\extractedsub{}%
	\def\extractedsuper{}%
	\expandafter\IfNoValueTF\expandafter{\second}{%
		\expandafter\IfNoValueF\expandafter{\third}{%
			\edef\extractedsub{\expandonce{\third}}%
			\expandafter\IfNoValueF\expandafter{\fourth}{%
				\edef\extractedsuper{\expandonce{\fourth}}%
			}%
		}%
	}{\global\edef\extractedsuper{\expandonce{\second}}}%
}

\NewDocumentCommand{\baseextensive}{o m E{_^}{{}{}}}{%
	\ifiscs{#2}{\expandafter\extractsubsuper\romannumeral-`0#2\relax}{\extractsubsuper#2\relax}%
	\def\thesub{\extractedsub#3}%
	\ifcsempty{extractedsub}{\ifstrempty{#3}{\def\thesub{}}{}}{}%
	\def\thesuper{\extractedsuper#4}%
	\ifcsempty{extractedsuper}{\ifstrempty{#4}{\def\thesuper{}}{}}{}%
	\expanded{%
	    \ifmmode\unexpanded{\myopbraces}\else\unexpanded{\ensuremath}\fi%
	    \unexpanded{{%
	        \noexpand{\IfNoValueTF{#1}{\extractedmain}{#1{\extractedmain}}}%
    		\ifcsempty{thesub}{}{_{\thesub}}%
    		\ifcsempty{thesuper}{}{^{\thesuper}}
	    }}%
	    \ifmmode\else\unexpanded{\xspace}\fi%
	}%
}

\newcommand{\variabling}[1]{\bm{\mathrm{\upgreek{#1}}}}
\newcommand{\variable}{\baseextensive[\variabling]}

\newcommand{\starredbe}[3]{%
	\expandafter\NewDocumentCommand\csname\macroname{#1}@handler\endcsname{s m}{%
		\IfBooleanTF{##1}{}{\dosmash}{#2{##2}} \vphantom{##2}
	}
	\providecommand{#1}{}%
	\renewcommand{#1}{\@ifstar{\csuse{\macroname{#1}@@}}{\csuse{\macroname{#1}@}}}%
	\csdef{\macroname{#1}@}{\baseextensive[\csuse{\macroname{#1}@handler}]}%
	\csdef{\macroname{#1}@@}{\baseextensive[\csuse{\macroname{#1}@handler}*]}%
}

\newcommand{\vectoring}[1]{\mathmbox{\vec{#1}}}
\starredbe{\vector}{\vectoring}

\newcommand{\vectmatrixing}[1]{\mathmbox{\uuline{#1}}}
\starredbe{\vectmatrix}{\vectmatrixing}

\newcommand{\varmatrix}{\variable}

\newcommand{\varcomp}[1]{{\renewcommand{\variabling}[1]{##1}#1}}

\newmathcommand{\vectdet}[1]{\abs{#1}}
\newmathcommand{\vardet}[1]{\abs{#1}}

\NewDocumentCommand{\shortcut}{s m m}{%
	\def #2{\protecting{\IfBooleanTF{#1}{#3}{\baseextensive{#3}}}}%
}

\newcommand{\starredshortcut}[2]{
	\NewDocumentCommand{#1}{m m o o}{
		\providecommand{##1}{}
		\renewcommand{##1}{\@ifstar{\csuse{\macroname{##1}@star}}{\csuse{\macroname{##1}@nostar}}}
		\csdef{\macroname{##1}@nostar}{\protecting{#2{##2}}}
		\csdef{\macroname{##1}@star}{\protecting{#2*{##2}}}
	}
}

\starredshortcut{\vectorshortcut}{\vector}
\starredshortcut{\vectmatrixshortcut}{\vectmatrix}

\makeatother

\newcommand{\newoperator}[3][\operatorname]{%
	\newcommand{#2}[1][]{\myopbraces{#1{#3}\ifblank{##1}{}{_{##1}}}}%
}

\NewEnviron{eq}{\begin{equation}\begin{split}\BODY\end{split}\end{equation}}

\makeatletter
\DeclareRobustCommand{\smartdot}{\@ifnextchar .{}{.\@\xspace}}
\makeatother

\newcommand{\cf}{cf\smartdot}
\newcommand{\eg}{e.g\smartdot}
\newcommand{\ie}{i.e\smartdot}
\newcommand{\etc}{etc\smartdot}
\newcommand{\vs}{{\textit vs.}\@\xspace}

\newcommand{\latin}[1]{\textit{#1}\xspace}
\newcommand{\adhoc}{\latin{ad hoc}}
\WithSuffix\newcommand\adhoc*{\latin{ad-hoc}}
\newcommand{\defacto}{\latin{de facto}}
\WithSuffix\newcommand\defacto*{\latin{de-facto}}
\newcommand{\apriori}{\latin{a priori}}
\WithSuffix\newcommand\apriori*{\latin{a-priori}}
\newcommand{\aposteriori}{\latin{a posteriori}}
\WithSuffix\newcommand\aposteriori*{\latin{a-posteriori}}
\newcommand{\postfactum}{\latin{post factum}}
\WithSuffix\newcommand\postfactum*{\latin{post-factum}}

\newoperator{\sinc}{sinc}

\shortcut{\solidangle}{\Omega}

\newmathcommand{\reals}{\mathbb{R}}
\newmathcommand{\units}[1]{\qty[#1]}

\NewExpandableDocumentCommand{\set}{o o m}{\ensuremath{\qty{#3}\IfNoValueF{#1}{_{#1}}\IfNoValueF{#2}{^{#2}}}}
\NewExpandableDocumentCommand{\setdef}{o o m m }{\set[#1][#2]{#3 :\ #4}}
\NewExpandableDocumentCommand{\tuple}{o o m}{\ensuremath{\qty[#3]\IfNoValueF{#1}{_{#1}}\IfNoValueF{#2}{^{#2}}}}

\DeclareDocumentCommand{\vartotuple}{O{i} o o m}{
	\tuple{\varcomp{#4}_{#1}}_{\IfValueTF{#2}{#2}{#1}}\IfValueT{#3}{^{#3}}
}
\DeclareDocumentCommand{\vartoset}{O{i} o o m}{
    \set{\varcomp{#4}_{#1}}_{\IfValueTF{#2}{#2}{#1}}\IfValueT{#3}{^{#3}}
}

\newoperator{\diag}{diag}
\newoperator{\flatten}{flatten}

\newcommand{\given}{\,\middle|\,}
\WithSuffix\newcommand\given*{\,|\,}

\newoperator{\corr}{corr}
\newoperator{\argmin}{argmin}
\newoperator{\argmax}{argmax}

\newoperator{\Prob}{P}
\newoperator{\prob}{p}
\newoperator{\cumulative}{F}
\newoperator{\empprob}{\hat{p}}

\shortcut{\DeltaDist}{\delta}
\shortcut{\Uniform}{\mathcal{U}}
\shortcut{\Poisson}{\operatorname{Pois}}
\shortcut{\Binomial}{\operatorname{Binom}}
\shortcut{\Normal}{\mathcal{N}}
\newoperator{\HalfNormal}{HalfNormal}
\newoperator{\Cauchy}{Cauchy}
\newoperator{\HalfCauchy}{HalfCauchy}
\newoperator{\HC}{HC}
\newoperator{\Exponential}{Expon}
\shortcut{\GammaDist}{\gamma}
\shortcut{\InvGammaDist}{\gamma^{-1}}
\newoperator{\LKJ}{LKJ}

\newoperator{\nnmlp}{MLP}
\newoperator{\nnlinear}{Linear}
\newoperator{\nnrelu}{ReLU}
\newoperator{\lpop}{LeakyPOP}

\shortcut{\sigmoid}{\operatorname{\sigma}}

\usepackage[abbreviations,symbols]{glossaries-extra}
\glsdisablehyper

\def\acrshort{\glsxtrshort}

\def\acrlong{\glsxtrlong}

\def\acrlongpl{\glsxtrlongpl}

\setabbreviationstyle[wordacro]{long-short-sc}

\newabbreviation{ccd}{CCD}{charge-coupled device}
\newabbreviation{cpu}{CPU}{central processing unit}
\newabbreviation{gpu}{GPU}{graphics processing unit}
\newabbreviation{ad}{AD}{automatic differentiation}
\newabbreviation{nn}{NN}{neural network}
\newabbreviation{bnn}{BNN}{Bayesian neural network}
\newabbreviation{bce}{BCE}{binary cross-entropy}
\newabbreviation{nll}{NLL}{negative log-likelihood}

\newabbreviation{nlp}{NLP}{natural-language processing}
\newabbreviation{rnn}{RNN}{recurrent neural network}
\newabbreviation{raenn}{RAENN}{recurrent autoencoder neural network}
\newabbreviation{vae}{VAE}{variational autoencoder}
\newabbreviation{llm}{LLM}{large language model}

\newabbreviation{mlp}{MLP}{multi-layer perceptron}
\newabbreviation{cnn}{CNN}{convolutional neural network}
\newabbreviation{relu}{ReLU}{rectified linear unit}
\newabbreviation[category=wordacro]{nnpop}{pop}{parity-odd power}

\newabbreviation{iid}{i.i.d.\@}{independent and identically distributed}
\newabbreviation{bhm}{BHM}{Bayesian hierarchical model}
\newabbreviation{gp}{GP}{Gaussian process}
\newabbreviation{kde}{KDE}{kernel density estimation}
\newabbreviation{snr}{SNR}{signal-to-noise ratio}
\newabbreviation{map}{MAP}{maximum \aposteriori}
\newabbreviation{mle}{MLE}{maximum likelihood estimator}
\newabbreviation{mse}{MSE}{mean squared error}
\newabbreviation{pca}{PCA}{principal component analysis}
\newabbreviation{cca}{CCA}{canonical correlation analysis}
\newabbreviation{bic}{BIC}{Bayesian information criterion}

\newabbreviation{ns}{NS}{nested sampling}
\newabbreviation{mcmc}{MCMC}{Markov chain Monte Carlo}
\newabbreviation{mh}{MH}{Metropolis--Hastings}
\newabbreviation{lmc}{LMC}{Langevin Monte Carlo}
\newabbreviation[category=wordacro]{mala}{mala}{Metropolis-adjusted Langevin algorithm}
\newabbreviation{hmc}{HMC}{Hamiltonian Monte Carlo}
\newabbreviation[category=wordacro]{nuts}{nuts}{no-u-turn sampler}
\newabbreviation{vi}{VI}{variational inference}
\newabbreviation{elbo}{ELBO}{evidence lower bound}
\newabbreviation{pp}{PP}{probabilistic programming}

\newabbreviation[category=wordacro]{moped}{moped}{{\scshape m}assively {\scshape o}ptimised {\scshape p}arameter {\scshape e}stimation and {\scshape d}ata compression}
\newabbreviation[category=wordacro]{borg}{Borg}{Bayesian {\scshape o}rigin {\scshape r}econstruction from {\scshape g}alaxies}

\newabbreviation{ml}{ML}{machine learning}
\newabbreviation{sbi}{SBI}{simulation-based inference}
\newabbreviation{lfi}{LFI}{likelihood-free inference}
\newabbreviation{lfsbi}{LFSBI}{likelihood-free simulation-based inference}
\newabbreviation{abc}{ABC}{approximate Bayesian computation}
\newabbreviation{nje}{NJE}{neural joint estimation}
\newabbreviation{nle}{NLE}{neural likelihood estimation}
\newabbreviation{npe}{NPE}{neural posterior estimation}
\newabbreviation{npse}{NPSE}{neural posterior score estimation}
\newabbreviation{nre}{NRE}{neural ratio estimation}
\newabbreviation{tmnre}{TMNRE}{truncated marginal neural ratio estimation}
\newabbreviation{arnre}{ARNRE}{autoregressive neural ratio estimation}
\newabbreviation{starnre}{STAR NRE}{set-based truncated autoregressive neural ratio estimation}

\newabbreviation{mvn}{MVN}{multivariate normal}
\newabbreviation{pmvn}{PMVN}{partial multivariate normal}
\newabbreviation{nde}{NDE}{neural density estimation}
\newabbreviation[category=wordacro]{gan}{gan}{generative adversarial network}
\newabbreviation{mdn}{MDN}{mixture density network}
\newabbreviation{nf}{NF}{normalising flow}
\newabbreviation{iaf}{IAF}{inverse autoregressive flow}
\newabbreviation{maf}{MAF}{masked autoregressive flow}
\newabbreviation{ddpm}{DDPM}{de-noising diffusion probabilistic model}
\newabbreviation{ddim}{DDIM}{de-noising diffusion implicit model}
\newabbreviation{ddrm}{DDIM}{de-noising diffusion restoration model}

\NewDocumentCommand{\defineverb}{m m}{
	\newcommand{#1}{\UseVerb{\string#1}\xspace}
	\DefineShortVerb{\|}\SaveVerb{\string#1}|#2|\UndefineShortVerb{\|}
}

\newignoredglossary{software}
\setabbreviationstyle[package]{short-sc-long}

\defineverb{\cuda}{CUDA}

\defineverb{\BUGS}{BUGS}
\newabbreviation[type=software, category=package]{bugs}{bugs}{{\scshape B}ayesian inference {\scshape u}sing {\scshape G}ibbs {\scshape s}ampling}

\defineverb{\stan}{Stan}
\defineverb{\turingjl}{Turing.jl}

\defineverb{\python}{Python}
\defineverb{\numpy}{NumPy}
\defineverb{\scipy}{SciPy}
\defineverb{\astropy}{Astropy}
\defineverb{\tensorflow}{TensorFlow}
\defineverb{\jax}{JAX}
\defineverb{\pytorch}{PyTorch}
\defineverb{\arviz}{ArviZ}
\defineverb{\pyro}{Pyro}
    
\defineverb{\numpyro}{NumPyro}
\defineverb{\funsor}{Funsor}
    
\defineverb{\pymc}{PyMC}

\defineverb{\ptlightning}{PyTorch Lightning}
\defineverb{\torchdiffeq}{torchdiffeq}

\defineverb{\emcee}{emcee}
    
\defineverb{\dynesty}{dynesty}
    
\defineverb{\torchns}{torchns}

\defineverb{\mpmath}{mpmath}
\defineverb{\jaxcosmo}{jax-cosmo}

\defineverb{\clipppy}{clipppy}
\newabbreviation[type=software, category=package]{clipppy}{Clipppy}{{\scshape c}onvenience {\scshape l}ayer for {\scshape i}nference and {\scshape p}robabilistic {\scshape p}rogramming in \python}
\defineverb{\phytorch}{phytorch}
\newglossaryentry{phytorch}{type=software, name=φυtorch, description=Spark-up your physics}
\defineverb{\ptcosmo}{phytorch.cosmology}

\defineverb{\swyft}{Swyft}

\defineverb{\FSPS}{FSPS}
\newabbreviation[type=software, category=package]{fsps}{FSPS}{flexible stellar population synthesis}

\newcommand\MTkillspecial[1]{
	\bgroup
	\catcode`\&=9
	\let\\\relax%
	\scantokens{#1}%
	\egroup
}
\newcommand{\opener}[2]{\mathopen{#1\vphantom{\MTkillspecial{#2}}\kern-\nulldelimiterspace\right.}}
\newcommand{\closer}[2]{\mathclose{\left.\kern-\nulldelimiterspace\vphantom{\MTkillspecial{#1}}#2}}

\newcommand{\ifdata}[3]{\pgfkeysifdefined{/private/#1}{#2}{#3}}
\newcommand{\setdata}[2]{\pgfkeys{/private/#1/.initial={#2}}}
\newcommand{\getdata}[1]{\pgfkeysvalueof{/private/#1}}

\newcommand{\setparamlabel}[2]{\setdata{labels/#1}{#2}}
\newcommand{\getparamlabel}[1]{\getdata{labels/#1}}
\newcommand{\setparamtruth}[2]{\setdata{truths/#1}{#2}}

\newcommand{\setparamextent}[3]{
	\setdata{extents/#1/lo}{#2}
	\setdata{extents/#1/hi}{#3}
}
\newcommand{\getparamlo}[1]{\getdata{extents/#1/lo}}
\newcommand{\getparamhi}[1]{\getdata{extents/#1/hi}}
\newcommand{\getparamzoomlo}[2]{\ifdata{zoom/#1/#2/lo}{\getdata{zoom/#1/#2/lo}}{\getparamlo{#2}}}
\newcommand{\getparamzoomhi}[2]{\ifdata{zoom/#1/#2/hi}{\getdata{zoom/#1/#2/hi}}{\getparamhi{#2}}}

\newcommand{\defineparam}[5]{
	\setparamlabel{#1}{#2}
	\setparamtruth{#1}{#3}
	\setparamextent{#1}{#4}{#5}
}

\pgfplotsset{
	param axis extents/.style={xmin/.evaluated={\getparamlo{#1}}, xmax/.evaluated={\getparamhi{#1}}},
	param axis labels/.style={xlabel={\getparamlabel{#1}}},
	param axis/.style={param axis extents=#1, param axis labels=#1},
	2 param axis extents/.style 2 args={param axis extents=#1, ymin/.evaluated={\getparamlo{#2}}, ymax/.evaluated={\getparamhi{#2}}},
	2 param axis labels/.style 2 args={xlabel={\getparamlabel{#1}}, ylabel={\getparamlabel{#2}}},
	2 param axis/.style 2 args={2 param axis extents={#1}{#2}, 2 param axis labels={#1}{#2}},
    zoom axis/.style 2 args={xmin={\getparamzoomlo{#1}{#2}}, xmax={\getparamzoomhi{#1}{#2}}},
    2 zoom axis/.style n args={3}{
        zoom axis={#1}{#2},
        ymin={\getparamzoomlo{#1}{#3}}, ymax={\getparamzoomhi{#1}{#3}},
    },
}

\newcommand{\lastsmashed}{}
\newcommand{\dosmash}[1]{\global\def\lastsmashed{#1}\smash{#1}}

\NewDocumentCommand{\smashedint}{O{\int} G{} G{}}{\dosmash{#1_{#2}^{#3}}\vphantom{#1}}
\input{utils}

\defcitealias{Fitzpatrick_1999}{F99}
\defcitealias{Cardelli_1989}{CCM89}

\newcommand{\Friedmann}{Friedman(n)\xspace}
\newcommand{\Lemaitre}{Lema\^itre\xspace}

\newcommand{\LCDM}{\ensuremath{\Lambda}CDM\xspace}
\newabbreviation{lcdm}{\LCDM}{$\Lambda$\Endash\acrlong{cdm}}

\DeclareSIUnit{\clight}{c}
\DeclareSIUnit{\planckh}{h}
\DeclareSIUnit{\Ggrav}{G}

\DeclareSIUnit{\erg}{erg}
\DeclareSIUnit{\year}{yr}
\DeclareSIUnit{\AU}{AU}
\DeclareSIUnit{\ly}{ly}
\DeclareSIUnit{\Msun}{M_{\odot}}
\DeclareSIUnit{\Zsun}{Z_{\odot}}
\DeclareSIUnit{\parsec}{pc}
\DeclareSIUnit{\mag}{mag}
\DeclareSIUnit{\dex}{dex}

\newabbreviation{gr}{GR}{general theory of relativity}
\newabbreviation{flrw}{FLRW}{\Friedmann{}\Endash \Lemaitre{}\Endash Robertson\Endash Walker}
\newabbreviation{ltb}{LTB}{\Lemaitre{}\Endash Tolman\Endash Bondi}
\newabbreviation{hf}{HF}{Hubble flow}

\newabbreviation{lss}{LSS}{large-scale structure}
\newabbreviation{bao}{BAO}{baryon acoustic oscillations}
\newabbreviation{sfr}{SFR}{star-formation rate}
\newabbreviation{sfh}{SFH}{star-formation history}
\newabbreviation{sps}{SPS}{stellar population synthesis}

\newabbreviation{de}{DE}{dark energy}
\newabbreviation{dm}{DM}{dark matter}
\newabbreviation{cdm}{CDM}{cold dark matter}
\newabbreviation[shortplural={EsOS}, longplural={equations of state}]{eos}{EOS}{equation of state}

\newabbreviation{cmb}{CMB}{cosmic microwave background}
\newabbreviation{gw}{GW}{gravitational wave}
\newabbreviation{lc}{LC}{light curve}
\newabbreviation{sed}{SED}{spectral energy distribution}
\newabbreviation{sfd}{SFD}{spectral flux density}

\newabbreviation{mw}{MW}{Milky Way}
\newabbreviation{wd}{WD}{white dwarf}
\newabbreviation{agn}{AGN}{active galactic nucleus}

\newabbreviation{ir}{IR}{infrared}
\newabbreviation{nir}{NIR}{near infrared}

\newabbreviation{wmap}{WMAP}{Wilkinson Microwave Anisotropy Probe}
\newabbreviation{sdss}{SDSS}{Sloan Digital Sky Survey}
\newabbreviation{des}{DES}{Dark Energy Survey}
\newabbreviation{desi}{DESI}{Dark Energy Spectroscopic Instrument}
\newabbreviation{panstarrs}{Pan-STARRS}{Panoramic Survey Telescope and Rapid Response System}
\newabbreviation{goto}{GOTO}{Gravitational-wave Optical Transient Observer}
\newabbreviation{ztf}{ZTF}{Zwicky Transient Facility}
\newabbreviation{lsst}{LSST}{Legacy Survey of Space and Time}
\newabbreviation{sst}{SST}{Simonyi Survey Telescope}
\newabbreviation{wfirs}{WFIRS}{Wide-Field Infrared Survey}
\newabbreviation{rst}{RST}{Roman Space Telescope}
\newabbreviation{elt}{ELT}{Extremely Large Telescope}

\newmathcommand{\cosmofluidlabel}{i}
\newcommand{\cosmofluid}[2][\cosmofluidlabel]{#2_{#1}}
\newcommand{\cosmonull}[1]{#1_{0}}

\shortcut{\cosmoparams}{\mathcal{C}}
\shortcut{\H}{H}
\shortcut\Ho{H_0}
\DeclareSIUnit{\cosmoh}{h}
\DeclareSIQualifier{\seventy}{70}
\DeclareSIQualifier{\hundred}{100}
\shortcut{\rhocrit}{\mathrm{\rho}_{\textrm{c}}}

\shortcut{\Efunc}{E}
\shortcut{\cosmodecel}{q}
\def\cosmodecelo{\cosmonull{\cosmodecel}}
\shortcut{\cosmojerk}{j}

\shortcut{\cosmoO}{\Omega}
\shortcut{\Or}{\cosmofluid[\mathrm{r}]{\cosmoO}}

\shortcut{\Om}{\cosmofluid[\mathrm{m}]{\cosmoO}}
\def\Omo{\cosmonull{\Om}}
\shortcut{\Ok}{\cosmofluid[k]{\cosmoO}}

\shortcut{\Ol}{\cosmofluid[\Lambda]{\cosmoO}}

\shortcut{\Ode}{\cosmofluid[\mathrm{de}]{\cosmoO}}

\shortcut{\cosmoeos}{\mathfrak{w}}
\shortcut{\cosmow}{w}
\shortcut{\wde}{\cosmofluid[\mathrm{de}]{\cosmow}}

\shortcut{\scalefactor}{a}
\shortcut{\z}{z}
\newcommand{\zc}{\z_{\mathrm{c}}}
\newcommand{\zsun}{\z_{\odot}}
\newcommand{\zpec}{\z_{\mathrm{pec}}}
    \shortcut{\vpec}{v_{\mathrm{pec}}}

\shortcut{\distmod}{\mu}

\shortcut{\cosmodist}{D}
\shortcut{\cosmotime}{T}

\newmathcommand{\dHubble}{\mathrm{\cosmodist_{H}}}
\newmathcommand{\tHubble}{\mathrm{\cosmotime_{H}}}

\newcommand{\dlum}{\cosmodist_{\mathrm{L}}}

\shortcut{\Vc}{V_{\mathrm{c}}}

\newabbreviation{mjd}{MJD}{modified Julian day}
\newabbreviation{fp}{FP}{forced photometry}
\newabbreviation{smp}{SMP}{scene-modelling photometry}

\DeclareSIUnit{\ADU}{ADU}
\DeclareSIUnit{\photoe}{\ensuremath{\mathrm{e}^{-}}}
\newabbreviation{adu}{ADU}{analog-to-digital unit}

\newmathcommand{\restlabel}{\text{r}}
\newmathcommand{\obslabel}{\text{o}}
\newmathcommand{\framing}[2]{#2_{#1}}
\newcommand{\restframe}{\framing{\restlabel}}
\newcommand{\obsframe}{\framing{\obslabel}}

\shortcut{\phase}{t}
\shortcut{\restphase}{\restframe{\phase}}
\shortcut{\obsphase}{\obsframe{\phase}}
\shortcut{\wave}{\lambda}
\shortcut{\restwave}{\restframe{\wave}}
\shortcut{\obswave}{\obsframe{\wave}}

\shortcut{\radiantflux}{L}
\shortcut{\spectralflux}{\Phi}
\shortcut{\fluxdensity}{E}
\shortcut{\spectralfluxdensity}{F}
\shortcut{\photonratedensity}{R}

\makeatletter
\newcommand\@passband[1]{{\itshape #1}}
\newcommand{\passband}[1]{\ifmmode\text{\@passband{#1}}\else\@passband{#1}\fi}
\makeatother
\shortcut{\band}{\passband{b}}
\shortcut{\transmission}{T}
\shortcut{\ZP}{\textrm{ZP}}
\shortcut{\gain}{\textrm{gain}}
\shortcut{\datarate}{\expval{d}}

\shortcut{\srcrate}{\datarate_{\textrm{src}}}
\shortcut{\bgrate}{\datarate_{\dosmash{\textrm{bg}}\vphantom{\textrm{src}}}}

\shortcut{\fluxcal}{\mathtt{FLUXCAL}}
\shortcut{\fluxcalerr}{\mathtt{FLUXCALERR}}

\newcommand{\ztfdrtwo}{\acrshort*{ztf} DR2\xspace}
\renewcommand{\SHoES}{SH0ES\xspace}

\renewcommand{\variabling}[1]{\bm{\mathrm{#1}}}
\renewcommand{\lg}{\opbraces{\log_{10}}}

\shortcut{\cosmoh}{h}
\newcommand{\Hounit}{\ensuremath{\frac{\si[per-mode=symbol]{\kilo\meter\per\s}}{\si{\mega\parsec}}}}
\newcommand{\Hoval}[2][]{\ensuremath{\num[round-mode=places,round-precision=1,#1]{#2}\,\Hounit}\xspace}
\WithSuffix\newcommand\Hoval*[2][]{\ensuremath{\qty(\num[round-mode=uncertainty,#1]{#2})\,\Hounit}\xspace}
\newcommand{\Hores}[3][]{\Hoval*[#1]{\getdata{#2/#3/mean}+-\getdata{#2/#3/std}}}
\newcommand{\Hooff}[3][]{\Hoval[print-implicit-plus,#1]{\getdata{#2/#3/off}}}
\newcommand{\Horoff}[3][]{\ensuremath{\num[#1,round-mode=figures,round-precision=1]{\getdata{#2/#3/roff}}\sigma}}

\renewcommand{\saltx}{\saltgenx}
\shortcut{\saltxobs}{\hat{\saltgenx}}
\shortcut{\saltxnullobssn}{\snindex{\widehat{\saltgenx_{0}}}}

\shortcut{\distmodcalib}{\variable{\distmod}_{\mathrm{host}}}
\shortcut{\distmodobscalib}{\hat{\variable{\distmod}}_{\mathrm{host}}}
\shortcut{\distmodcovcalib}{\hat{\varmatrix{C}}_{\mathrm{host}}}

\renewcommand{\Mo}[1][]{\ensuremath{M_{0\ifblank{#1}{}{,#1}}}\xspace}
\renewcommand{\saltalpha}[1][]{\ensuremath{\alpha\ifblank{#1}{}{_{#1}}}\xspace}
\renewcommand{\meansaltx}[1][]{\ensuremath{\mu_{\saltx\ifblank{#1}{}{, #1}}}\xspace}
\renewcommand{\stdsaltx}[1][]{\ensuremath{\sigma_{\saltx\ifblank{#1}{}{,#1}}}\xspace}
\shortcut{\xfrac}{f_1}

\renewcommand{\meansaltc}{\baseextensive{\mu}_{\saltc}}
\renewcommand{\stdsaltc}{\baseextensive{\sigma}_{\smash{\saltc}}}

\newmathcommand{\priorMo}{\Uniform(\num{-20}, \num{-18})}
\newmathcommand{\priorsaltalpha}{\Uniform(\num{-0.55}, \num{0.25})}
\newmathcommand{\priorstdsaltx}{\Uniform(\num{0.01}, \num{3})}

\newmathcommand{\priorMintsn}{\Normal(\Mo + \saltalpha\saltxintsn + \saltbeta\saltcintsn, \sigmares^2)}
\newmathcommand{\priorsaltcintsn}{\Normal(\meansaltc+\saltalphac\saltxintsn, \stdsaltc^2)}
\newmathcommand{\priorsaltxintsn}{\priorsaltxsn}

\shortcut{\zcvar}{\variable{\z}_{\mathrm{c}}}
\shortcut{\zcmbobsvar}{\hat{\variable{\z}}_{\mathrm{CMB}}}
\shortcut{\zcov}{\varmatrix{C}_{\zc}}
\shortcut{\vcov}{\varmatrix{C}_{\vpec}}

\pgfkeys{,
/private/.cd,
pplus-cdd_B22-CCM89+ztfdr2-vlhq_B22-CCM89-pcorr-sm/hqj-latz-distR-AV-1x/mean/.initial={71.72729969024658},
pplus-cdd_B22-CCM89+ztfdr2-vlhq_B22-CCM89-pcorr-sm/hqj-latz-distR-AV-1x/std/.initial={1.5077270567417145},
pplus-cdd_B22-CCM89+ztfdr2-vlhq_B22-CCM89-pcorr-sm/hqj-latz-distR-AV-1x/off/.initial={0.1270294189453125},
pplus-cdd_B22-CCM89+ztfdr2-vlhq_B22-CCM89-pcorr-sm/hqj-latz-distR-AV-1x/roff/.initial={0.08671451037988777},
pplus-cdd_B22-F99+ztfdr2-vlhq_B22-F99-pcorr-sm/hqj-latz-distR-AV-1x/mean/.initial={71.30364775657654},
pplus-cdd_B22-F99+ztfdr2-vlhq_B22-F99-pcorr-sm/hqj-latz-distR-AV-1x/std/.initial={1.4575895853340626},
pplus-cdd_B22-F99+ztfdr2-vlhq_B22-F99-pcorr-sm/hqj-latz-distR-AV-1x/off/.initial={-0.29662251472473145},
pplus-cdd_B22-F99+ztfdr2-vlhq_B22-F99-pcorr-sm/hqj-latz-distR-AV-1x/roff/.initial={-0.20248440357803651},
pplus-cdd_T21-CCM89+ztfdr2-vlhq_T21-CCM89-pcorr-sm/hqj-latz-distR-AV-1x/mean/.initial={71.70302867889404},
pplus-cdd_T21-CCM89+ztfdr2-vlhq_T21-CCM89-pcorr-sm/hqj-latz-distR-AV-1x/std/.initial={1.5070218592882156},
pplus-cdd_T21-CCM89+ztfdr2-vlhq_T21-CCM89-pcorr-sm/hqj-latz-distR-AV-1x/off/.initial={0.10275840759277344},
pplus-cdd_T21-CCM89+ztfdr2-vlhq_T21-CCM89-pcorr-sm/hqj-latz-distR-AV-1x/roff/.initial={0.07014631001075756},
pplus-cdd_T21-F99+ztfdr2-vlhq_T21-F99-pcorr-sm/hqj-latz-distR-AV-1x/mean/.initial={71.33561372756958},
pplus-cdd_T21-F99+ztfdr2-vlhq_T21-F99-pcorr-sm/hqj-latz-distR-AV-1x/std/.initial={1.4566351659595966},
pplus-cdd_T21-F99+ztfdr2-vlhq_T21-F99-pcorr-sm/hqj-latz-distR-AV-1x/off/.initial={-0.26465654373168945},
pplus-cdd_T21-F99+ztfdr2-vlhq_T21-F99-pcorr-sm/hqj-latz-distR-AV-1x/roff/.initial={-0.1806633675810706},
pplus-cdd_nom+ztfdr2-vlhq_nom-pborg-sm/hqj-latz-distR-AV-1x-step/mean/.initial={71.45360112190247},
pplus-cdd_nom+ztfdr2-vlhq_nom-pborg-sm/hqj-latz-distR-AV-1x-step/std/.initial={1.5027874149382114},
pplus-cdd_nom+ztfdr2-vlhq_nom-pborg-sm/hqj-latz-distR-AV-1x-step/off/.initial={-0.1466691493988037},
pplus-cdd_nom+ztfdr2-vlhq_nom-pborg-sm/hqj-latz-distR-AV-1x-step/roff/.initial={-0.10012124422475123},
pplus-cdd_nom+ztfdr2-vlhq_nom-pborg-sm/hqj-latz-distR-AV-1x/mean/.initial={71.01412415504456},
pplus-cdd_nom+ztfdr2-vlhq_nom-pborg-sm/hqj-latz-distR-AV-1x/std/.initial={1.4746978878974915},
pplus-cdd_nom+ztfdr2-vlhq_nom-pborg-sm/hqj-latz-distR-AV-1x/off/.initial={-0.5861461162567139},
pplus-cdd_nom+ztfdr2-vlhq_nom-pborg-sm/hqj-latz-distR-AV-1x/roff/.initial={-0.40012285267679154},
pplus-cdd_nom+ztfdr2-vlhq_nom-pborg-sm/hqj-latz-distR-AV-2x-step/mean/.initial={71.57275676727295},
pplus-cdd_nom+ztfdr2-vlhq_nom-pborg-sm/hqj-latz-distR-AV-2x-step/std/.initial={1.5001746825873852},
pplus-cdd_nom+ztfdr2-vlhq_nom-pborg-sm/hqj-latz-distR-AV-2x-step/off/.initial={-0.027513504028320312},
pplus-cdd_nom+ztfdr2-vlhq_nom-pborg-sm/hqj-latz-distR-AV-2x-step/roff/.initial={-0.01878163381726548},
pplus-cdd_nom+ztfdr2-vlhq_nom-pborg-sm/hqj-latz-distR-AV-2x/mean/.initial={71.17968201637268},
pplus-cdd_nom+ztfdr2-vlhq_nom-pborg-sm/hqj-latz-distR-AV-2x/std/.initial={1.4855154789984226},
pplus-cdd_nom+ztfdr2-vlhq_nom-pborg-sm/hqj-latz-distR-AV-2x/off/.initial={-0.42058825492858887},
pplus-cdd_nom+ztfdr2-vlhq_nom-pborg-sm/hqj-latz-distR-AV-2x/roff/.initial={-0.2871075448543553},
pplus-cdd_nom+ztfdr2-vlhq_nom-pborg-sm/hqj-latz-globalR-AV-1x-step/mean/.initial={70.86329460144043},
pplus-cdd_nom+ztfdr2-vlhq_nom-pborg-sm/hqj-latz-globalR-AV-1x-step/std/.initial={1.494170818477869},
pplus-cdd_nom+ztfdr2-vlhq_nom-pborg-sm/hqj-latz-globalR-AV-1x-step/off/.initial={-0.7369756698608398},
pplus-cdd_nom+ztfdr2-vlhq_nom-pborg-sm/hqj-latz-globalR-AV-1x-step/roff/.initial={-0.5030841273184516},
pplus-cdd_nom+ztfdr2-vlhq_nom-pborg-sm/hqj-latz-globalR-AV-1x/mean/.initial={70.32760977745056},
pplus-cdd_nom+ztfdr2-vlhq_nom-pborg-sm/hqj-latz-globalR-AV-1x/std/.initial={1.5547750517725945},
pplus-cdd_nom+ztfdr2-vlhq_nom-pborg-sm/hqj-latz-globalR-AV-1x/off/.initial={-1.272660493850708},
pplus-cdd_nom+ztfdr2-vlhq_nom-pborg-sm/hqj-latz-globalR-AV-1x/roff/.initial={-0.8687604219586382},
pplus-cdd_nom+ztfdr2-vlhq_nom-pborg-sm/hqj-latz-globalR-AV-2x-step/mean/.initial={71.07571959495544},
pplus-cdd_nom+ztfdr2-vlhq_nom-pborg-sm/hqj-latz-globalR-AV-2x-step/std/.initial={1.5348589047789574},
pplus-cdd_nom+ztfdr2-vlhq_nom-pborg-sm/hqj-latz-globalR-AV-2x-step/off/.initial={-0.5245506763458252},
pplus-cdd_nom+ztfdr2-vlhq_nom-pborg-sm/hqj-latz-globalR-AV-2x-step/roff/.initial={-0.3580757547851925},
pplus-cdd_nom+ztfdr2-vlhq_nom-pborg-sm/hqj-latz-globalR-AV-2x/mean/.initial={70.64189910888672},
pplus-cdd_nom+ztfdr2-vlhq_nom-pborg-sm/hqj-latz-globalR-AV-2x/std/.initial={1.5159032307565212},
pplus-cdd_nom+ztfdr2-vlhq_nom-pborg-sm/hqj-latz-globalR-AV-2x/off/.initial={-0.9583711624145508},
pplus-cdd_nom+ztfdr2-vlhq_nom-pborg-sm/hqj-latz-globalR-AV-2x/roff/.initial={-0.6542160611374528},
pplus-cdd_nom+ztfdr2-vlhq_nom-pborg-sm+highz/hqj-latz-distR-AV-1x-step/mean/.initial={71.11237645149231},
pplus-cdd_nom+ztfdr2-vlhq_nom-pborg-sm+highz/hqj-latz-distR-AV-1x-step/std/.initial={1.5019758604466915},
pplus-cdd_nom+ztfdr2-vlhq_nom-pborg-sm+highz/hqj-latz-distR-AV-1x-step/off/.initial={-0.48789381980895996},
pplus-cdd_nom+ztfdr2-vlhq_nom-pborg-sm+highz/hqj-latz-distR-AV-1x-step/roff/.initial={-0.3330525641491044},
pplus-cdd_nom+ztfdr2-vlhq_nom-pborg-sm+highz/hqj-latz-distR-AV-1x/mean/.initial={70.69451808929443},
pplus-cdd_nom+ztfdr2-vlhq_nom-pborg-sm+highz/hqj-latz-distR-AV-1x/std/.initial={1.486273854970932},
pplus-cdd_nom+ztfdr2-vlhq_nom-pborg-sm+highz/hqj-latz-distR-AV-1x/off/.initial={-0.9057521820068359},
pplus-cdd_nom+ztfdr2-vlhq_nom-pborg-sm+highz/hqj-latz-distR-AV-1x/roff/.initial={-0.6182965933430812},
pplus-cdd_nom+ztfdr2-vlhq_nom-pborg-sm+highz/hqj-latz-distR-AV-2x-step/mean/.initial={71.21537327766418},
pplus-cdd_nom+ztfdr2-vlhq_nom-pborg-sm+highz/hqj-latz-distR-AV-2x-step/std/.initial={1.4933290891349316},
pplus-cdd_nom+ztfdr2-vlhq_nom-pborg-sm+highz/hqj-latz-distR-AV-2x-step/off/.initial={-0.38489699363708496},
pplus-cdd_nom+ztfdr2-vlhq_nom-pborg-sm+highz/hqj-latz-distR-AV-2x-step/roff/.initial={-0.26274350167892513},
pplus-cdd_nom+ztfdr2-vlhq_nom-pborg-sm+highz/hqj-latz-distR-AV-2x/mean/.initial={70.85794806480408},
pplus-cdd_nom+ztfdr2-vlhq_nom-pborg-sm+highz/hqj-latz-distR-AV-2x/std/.initial={1.4259603805840015},
pplus-cdd_nom+ztfdr2-vlhq_nom-pborg-sm+highz/hqj-latz-distR-AV-2x/off/.initial={-0.7423222064971924},
pplus-cdd_nom+ztfdr2-vlhq_nom-pborg-sm+highz/hqj-latz-distR-AV-2x/roff/.initial={-0.5067338512209836},
pplus-cdd_nom+ztfdr2-vlhq_nom-pborg-sm+highz/hqj-latz-globalR-AV-1x-step/mean/.initial={70.51941752433777},
pplus-cdd_nom+ztfdr2-vlhq_nom-pborg-sm+highz/hqj-latz-globalR-AV-1x-step/std/.initial={1.4679481275379658},
pplus-cdd_nom+ztfdr2-vlhq_nom-pborg-sm+highz/hqj-latz-globalR-AV-1x-step/off/.initial={-1.080852746963501},
pplus-cdd_nom+ztfdr2-vlhq_nom-pborg-sm+highz/hqj-latz-globalR-AV-1x-step/roff/.initial={-0.7378260683538713},
pplus-cdd_nom+ztfdr2-vlhq_nom-pborg-sm+highz/hqj-latz-globalR-AV-1x/mean/.initial={70.02440094947815},
pplus-cdd_nom+ztfdr2-vlhq_nom-pborg-sm+highz/hqj-latz-globalR-AV-1x/std/.initial={1.4992350712418556},
pplus-cdd_nom+ztfdr2-vlhq_nom-pborg-sm+highz/hqj-latz-globalR-AV-1x/off/.initial={-1.5758693218231201},
pplus-cdd_nom+ztfdr2-vlhq_nom-pborg-sm+highz/hqj-latz-globalR-AV-1x/roff/.initial={-1.0757408622281994},
pplus-cdd_nom+ztfdr2-vlhq_nom-pborg-sm+highz/hqj-latz-globalR-AV-2x-step/mean/.initial={70.76921463012695},
pplus-cdd_nom+ztfdr2-vlhq_nom-pborg-sm+highz/hqj-latz-globalR-AV-2x-step/std/.initial={1.4854461885988712},
pplus-cdd_nom+ztfdr2-vlhq_nom-pborg-sm+highz/hqj-latz-globalR-AV-2x-step/off/.initial={-0.8310556411743164},
pplus-cdd_nom+ztfdr2-vlhq_nom-pborg-sm+highz/hqj-latz-globalR-AV-2x-step/roff/.initial={-0.567306247806259},
pplus-cdd_nom+ztfdr2-vlhq_nom-pborg-sm+highz/hqj-latz-globalR-AV-2x/mean/.initial={70.35169005393982},
pplus-cdd_nom+ztfdr2-vlhq_nom-pborg-sm+highz/hqj-latz-globalR-AV-2x/std/.initial={1.4723789878189564},
pplus-cdd_nom+ztfdr2-vlhq_nom-pborg-sm+highz/hqj-latz-globalR-AV-2x/off/.initial={-1.2485802173614502},
pplus-cdd_nom+ztfdr2-vlhq_nom-pborg-sm+highz/hqj-latz-globalR-AV-2x/roff/.initial={-0.8523224235570453},
pplus-cdd_nom+ztfdr2-vlhq_nom-pborg-sm+late/hqj-latz-distR-AV-1x-step/mean/.initial={72.51635789871216},
pplus-cdd_nom+ztfdr2-vlhq_nom-pborg-sm+late/hqj-latz-distR-AV-1x-step/std/.initial={1.714305765926838},
pplus-cdd_nom+ztfdr2-vlhq_nom-pborg-sm+late/hqj-latz-distR-AV-1x-step/off/.initial={0.9160876274108887},
pplus-cdd_nom+ztfdr2-vlhq_nom-pborg-sm+late/hqj-latz-distR-AV-1x-step/roff/.initial={0.6253519124590123},
pplus-cdd_nom+ztfdr2-vlhq_nom-pborg-sm+late/hqj-latz-distR-AV-1x/mean/.initial={72.11650013923645},
pplus-cdd_nom+ztfdr2-vlhq_nom-pborg-sm+late/hqj-latz-distR-AV-1x/std/.initial={1.6782527789473534},
pplus-cdd_nom+ztfdr2-vlhq_nom-pborg-sm+late/hqj-latz-distR-AV-1x/off/.initial={0.5162298679351807},
pplus-cdd_nom+ztfdr2-vlhq_nom-pborg-sm+late/hqj-latz-distR-AV-1x/roff/.initial={0.35239569395137477},
pplus-cdd_nom+ztfdr2-vlhq_nom-pborg-sm+late/hqj-latz-distR-AV-2x-step/mean/.initial={71.96715474128723},
pplus-cdd_nom+ztfdr2-vlhq_nom-pborg-sm+late/hqj-latz-distR-AV-2x-step/std/.initial={1.677253283560276},
pplus-cdd_nom+ztfdr2-vlhq_nom-pborg-sm+late/hqj-latz-distR-AV-2x-step/off/.initial={0.3668844699859619},
pplus-cdd_nom+ztfdr2-vlhq_nom-pborg-sm+late/hqj-latz-distR-AV-2x-step/roff/.initial={0.25044755336961483},
pplus-cdd_nom+ztfdr2-vlhq_nom-pborg-sm+late/hqj-latz-distR-AV-2x/mean/.initial={71.63416147232056},
pplus-cdd_nom+ztfdr2-vlhq_nom-pborg-sm+late/hqj-latz-distR-AV-2x/std/.initial={1.5739811584353447},
pplus-cdd_nom+ztfdr2-vlhq_nom-pborg-sm+late/hqj-latz-distR-AV-2x/off/.initial={0.03389120101928711},
pplus-cdd_nom+ztfdr2-vlhq_nom-pborg-sm+late/hqj-latz-distR-AV-2x/roff/.initial={0.023135262106796256},
pplus-cdd_nom+ztfdr2-vlhq_nom-pborg-sm+late/hqj-latz-globalR-AV-1x-step/mean/.initial={71.92113995552063},
pplus-cdd_nom+ztfdr2-vlhq_nom-pborg-sm+late/hqj-latz-globalR-AV-1x-step/std/.initial={1.7466269433498383},
pplus-cdd_nom+ztfdr2-vlhq_nom-pborg-sm+late/hqj-latz-globalR-AV-1x-step/off/.initial={0.32086968421936035},
pplus-cdd_nom+ztfdr2-vlhq_nom-pborg-sm+late/hqj-latz-globalR-AV-1x-step/roff/.initial={0.21903632870122458},
pplus-cdd_nom+ztfdr2-vlhq_nom-pborg-sm+late/hqj-latz-globalR-AV-1x/mean/.initial={71.63002490997314},
pplus-cdd_nom+ztfdr2-vlhq_nom-pborg-sm+late/hqj-latz-globalR-AV-1x/std/.initial={1.7426982522010803},
pplus-cdd_nom+ztfdr2-vlhq_nom-pborg-sm+late/hqj-latz-globalR-AV-1x/off/.initial={0.029754638671875},
pplus-cdd_nom+ztfdr2-vlhq_nom-pborg-sm+late/hqj-latz-globalR-AV-1x/roff/.initial={0.020311506935829567},
pplus-cdd_nom+ztfdr2-vlhq_nom-pborg-sm+late/hqj-latz-globalR-AV-2x-step/mean/.initial={72.19377160072327},
pplus-cdd_nom+ztfdr2-vlhq_nom-pborg-sm+late/hqj-latz-globalR-AV-2x-step/std/.initial={1.7243780195713043},
pplus-cdd_nom+ztfdr2-vlhq_nom-pborg-sm+late/hqj-latz-globalR-AV-2x-step/off/.initial={0.5935013294219971},
pplus-cdd_nom+ztfdr2-vlhq_nom-pborg-sm+late/hqj-latz-globalR-AV-2x-step/roff/.initial={0.4051437660499514},
pplus-cdd_nom+ztfdr2-vlhq_nom-pborg-sm+late/hqj-latz-globalR-AV-2x/mean/.initial={71.64797782897949},
pplus-cdd_nom+ztfdr2-vlhq_nom-pborg-sm+late/hqj-latz-globalR-AV-2x/std/.initial={1.6978122293949127},
pplus-cdd_nom+ztfdr2-vlhq_nom-pborg-sm+late/hqj-latz-globalR-AV-2x/off/.initial={0.047707557678222656},
pplus-cdd_nom+ztfdr2-vlhq_nom-pborg-sm+late/hqj-latz-globalR-AV-2x/roff/.initial={0.032566767130284426},
pplus-cdd_nom+ztfdr2-vlhq_nom-pborg-sm+late+highz/hqj-latz-distR-AV-1x-step/mean/.initial={72.698575258255},
pplus-cdd_nom+ztfdr2-vlhq_nom-pborg-sm+late+highz/hqj-latz-distR-AV-1x-step/std/.initial={1.7277341336011887},
pplus-cdd_nom+ztfdr2-vlhq_nom-pborg-sm+late+highz/hqj-latz-distR-AV-1x-step/off/.initial={1.0983049869537354},
pplus-cdd_nom+ztfdr2-vlhq_nom-pborg-sm+late+highz/hqj-latz-distR-AV-1x-step/roff/.initial={0.7497395483835407},
pplus-cdd_nom+ztfdr2-vlhq_nom-pborg-sm+late+highz/hqj-latz-distR-AV-1x/mean/.initial={72.40139842033386},
pplus-cdd_nom+ztfdr2-vlhq_nom-pborg-sm+late+highz/hqj-latz-distR-AV-1x/std/.initial={1.740022748708725},
pplus-cdd_nom+ztfdr2-vlhq_nom-pborg-sm+late+highz/hqj-latz-distR-AV-1x/off/.initial={0.8011281490325928},
pplus-cdd_nom+ztfdr2-vlhq_nom-pborg-sm+late+highz/hqj-latz-distR-AV-1x/roff/.initial={0.5468767453373486},
pplus-cdd_nom+ztfdr2-vlhq_nom-pborg-sm+late+highz/hqj-latz-distR-AV-2x-step/mean/.initial={72.32165336608887},
pplus-cdd_nom+ztfdr2-vlhq_nom-pborg-sm+late+highz/hqj-latz-distR-AV-2x-step/std/.initial={1.683625765144825},
pplus-cdd_nom+ztfdr2-vlhq_nom-pborg-sm+late+highz/hqj-latz-distR-AV-2x-step/off/.initial={0.7213830947875977},
pplus-cdd_nom+ztfdr2-vlhq_nom-pborg-sm+late+highz/hqj-latz-distR-AV-2x-step/roff/.initial={0.4924401164722718},
pplus-cdd_nom+ztfdr2-vlhq_nom-pborg-sm+late+highz/hqj-latz-distR-AV-2x/mean/.initial={72.0881462097168},
pplus-cdd_nom+ztfdr2-vlhq_nom-pborg-sm+late+highz/hqj-latz-distR-AV-2x/std/.initial={1.683097518980503},
pplus-cdd_nom+ztfdr2-vlhq_nom-pborg-sm+late+highz/hqj-latz-distR-AV-2x/off/.initial={0.48787593841552734},
pplus-cdd_nom+ztfdr2-vlhq_nom-pborg-sm+late+highz/hqj-latz-distR-AV-2x/roff/.initial={0.3330403577146478},
pplus-cdd_nom+ztfdr2-vlhq_nom-pborg-sm+late+highz/hqj-latz-globalR-AV-1x-step/mean/.initial={71.98248505592346},
pplus-cdd_nom+ztfdr2-vlhq_nom-pborg-sm+late+highz/hqj-latz-globalR-AV-1x-step/std/.initial={1.7975538969039917},
pplus-cdd_nom+ztfdr2-vlhq_nom-pborg-sm+late+highz/hqj-latz-globalR-AV-1x-step/off/.initial={0.3822147846221924},
pplus-cdd_nom+ztfdr2-vlhq_nom-pborg-sm+late+highz/hqj-latz-globalR-AV-1x-step/roff/.initial={0.2609125365104309},
pplus-cdd_nom+ztfdr2-vlhq_nom-pborg-sm+late+highz/hqj-latz-globalR-AV-1x/mean/.initial={71.73385620117188},
pplus-cdd_nom+ztfdr2-vlhq_nom-pborg-sm+late+highz/hqj-latz-globalR-AV-1x/std/.initial={1.783967763185501},
pplus-cdd_nom+ztfdr2-vlhq_nom-pborg-sm+late+highz/hqj-latz-globalR-AV-1x/off/.initial={0.13358592987060547},
pplus-cdd_nom+ztfdr2-vlhq_nom-pborg-sm+late+highz/hqj-latz-globalR-AV-1x/roff/.initial={0.09119020301398482},
pplus-cdd_nom+ztfdr2-vlhq_nom-pborg-sm+late+highz/hqj-latz-globalR-AV-2x-step/mean/.initial={72.3889946937561},
pplus-cdd_nom+ztfdr2-vlhq_nom-pborg-sm+late+highz/hqj-latz-globalR-AV-2x-step/std/.initial={1.7461949959397316},
pplus-cdd_nom+ztfdr2-vlhq_nom-pborg-sm+late+highz/hqj-latz-globalR-AV-2x-step/off/.initial={0.788724422454834},
pplus-cdd_nom+ztfdr2-vlhq_nom-pborg-sm+late+highz/hqj-latz-globalR-AV-2x-step/roff/.initial={0.5384095486359342},
pplus-cdd_nom+ztfdr2-vlhq_nom-pborg-sm+late+highz/hqj-latz-globalR-AV-2x/mean/.initial={71.9156801700592},
pplus-cdd_nom+ztfdr2-vlhq_nom-pborg-sm+late+highz/hqj-latz-globalR-AV-2x/std/.initial={1.7292773351073265},
pplus-cdd_nom+ztfdr2-vlhq_nom-pborg-sm+late+highz/hqj-latz-globalR-AV-2x/off/.initial={0.31540989875793457},
pplus-cdd_nom+ztfdr2-vlhq_nom-pborg-sm+late+highz/hqj-latz-globalR-AV-2x/roff/.initial={0.21530929738046736},
pplus-cdd_nom+ztfdr2-vlhq_nom-pcorr/hqj-latz-distR-AV-1x/mean/.initial={73.33788871765137},
pplus-cdd_nom+ztfdr2-vlhq_nom-pcorr/hqj-latz-distR-AV-1x/std/.initial={1.184901874512434},
pplus-cdd_nom+ztfdr2-vlhq_nom-pcorr/hqj-latz-distR-AV-1x/off/.initial={1.7376184463500977},
pplus-cdd_nom+ztfdr2-vlhq_nom-pcorr/hqj-latz-distR-AV-1x/roff/.initial={1.1861561995113739},
pplus-cdd_nom+ztfdr2-vlhq_nom-pcorr-sm/hqj-latz-distR-AV-1x-step/mean/.initial={72.16033935546875},
pplus-cdd_nom+ztfdr2-vlhq_nom-pcorr-sm/hqj-latz-distR-AV-1x-step/std/.initial={1.481202244758606},
pplus-cdd_nom+ztfdr2-vlhq_nom-pcorr-sm/hqj-latz-distR-AV-1x-step/off/.initial={0.5600690841674805},
pplus-cdd_nom+ztfdr2-vlhq_nom-pcorr-sm/hqj-latz-distR-AV-1x-step/roff/.initial={0.38232180242754193},
pplus-cdd_nom+ztfdr2-vlhq_nom-pcorr-sm/hqj-latz-distR-AV-1x/mean/.initial={71.60027027130127},
pplus-cdd_nom+ztfdr2-vlhq_nom-pcorr-sm/hqj-latz-distR-AV-1x/std/.initial={1.464915368705988},
pplus-cdd_nom+ztfdr2-vlhq_nom-pcorr-sm/hqj-latz-distR-AV-1x/off/.initial={0.0},
pplus-cdd_nom+ztfdr2-vlhq_nom-pcorr-sm/hqj-latz-distR-AV-1x/roff/.initial={0.0},
pplus-cdd_nom+ztfdr2-vlhq_nom-pcorr-sm/hqj-latz-distR-AV-2x-step/mean/.initial={72.16362357139587},
pplus-cdd_nom+ztfdr2-vlhq_nom-pcorr-sm/hqj-latz-distR-AV-2x-step/std/.initial={1.4568516053259373},
pplus-cdd_nom+ztfdr2-vlhq_nom-pcorr-sm/hqj-latz-distR-AV-2x-step/off/.initial={0.5633533000946045},
pplus-cdd_nom+ztfdr2-vlhq_nom-pcorr-sm/hqj-latz-distR-AV-2x-step/roff/.initial={0.38456371755607605},
pplus-cdd_nom+ztfdr2-vlhq_nom-pcorr-sm/hqj-latz-distR-AV-2x/mean/.initial={71.69493436813354},
pplus-cdd_nom+ztfdr2-vlhq_nom-pcorr-sm/hqj-latz-distR-AV-2x/std/.initial={1.4766314066946507},
pplus-cdd_nom+ztfdr2-vlhq_nom-pcorr-sm/hqj-latz-distR-AV-2x/off/.initial={0.09466409683227539},
pplus-cdd_nom+ztfdr2-vlhq_nom-pcorr-sm/hqj-latz-distR-AV-2x/roff/.initial={0.06462086401339046},
pplus-cdd_nom+ztfdr2-vlhq_nom-pcorr-sm/hqj-latz-globalR-AV-1x-step/mean/.initial={71.41474485397339},
pplus-cdd_nom+ztfdr2-vlhq_nom-pcorr-sm/hqj-latz-globalR-AV-1x-step/std/.initial={1.5164137817919254},
pplus-cdd_nom+ztfdr2-vlhq_nom-pcorr-sm/hqj-latz-globalR-AV-1x-step/off/.initial={-0.18552541732788086},
pplus-cdd_nom+ztfdr2-vlhq_nom-pcorr-sm/hqj-latz-globalR-AV-1x-step/roff/.initial={-0.12664582629900462},
pplus-cdd_nom+ztfdr2-vlhq_nom-pcorr-sm/hqj-latz-globalR-AV-1x/mean/.initial={70.77776193618774},
pplus-cdd_nom+ztfdr2-vlhq_nom-pcorr-sm/hqj-latz-globalR-AV-1x/std/.initial={1.4790615066885948},
pplus-cdd_nom+ztfdr2-vlhq_nom-pcorr-sm/hqj-latz-globalR-AV-1x/off/.initial={-0.8225083351135254},
pplus-cdd_nom+ztfdr2-vlhq_nom-pcorr-sm/hqj-latz-globalR-AV-1x/roff/.initial={-0.5614715721359906},
pplus-cdd_nom+ztfdr2-vlhq_nom-pcorr-sm/hqj-latz-globalR-AV-2x-step/mean/.initial={71.59481048583984},
pplus-cdd_nom+ztfdr2-vlhq_nom-pcorr-sm/hqj-latz-globalR-AV-2x-step/std/.initial={1.4906438998878002},
pplus-cdd_nom+ztfdr2-vlhq_nom-pcorr-sm/hqj-latz-globalR-AV-2x-step/off/.initial={-0.005459785461425781},
pplus-cdd_nom+ztfdr2-vlhq_nom-pcorr-sm/hqj-latz-globalR-AV-2x-step/roff/.initial={-0.0037270313207571877},
pplus-cdd_nom+ztfdr2-vlhq_nom-pcorr-sm/hqj-latz-globalR-AV-2x/mean/.initial={71.11145853996277},
pplus-cdd_nom+ztfdr2-vlhq_nom-pcorr-sm/hqj-latz-globalR-AV-2x/std/.initial={1.5011759474873543},
pplus-cdd_nom+ztfdr2-vlhq_nom-pcorr-sm/hqj-latz-globalR-AV-2x/off/.initial={-0.488811731338501},
pplus-cdd_nom+ztfdr2-vlhq_nom-pcorr-sm/hqj-latz-globalR-AV-2x/roff/.initial={-0.333679161117878},
pplus-cdd_nom+ztfdr2-vlhq_nom-pcorr-sm+highz/hqj-latz-distR-AV-1x-step/mean/.initial={72.23166823387146},
pplus-cdd_nom+ztfdr2-vlhq_nom-pcorr-sm+highz/hqj-latz-distR-AV-1x-step/std/.initial={1.5598471276462078},
pplus-cdd_nom+ztfdr2-vlhq_nom-pcorr-sm+highz/hqj-latz-distR-AV-1x-step/off/.initial={0.6313979625701904},
pplus-cdd_nom+ztfdr2-vlhq_nom-pcorr-sm+highz/hqj-latz-distR-AV-1x-step/roff/.initial={0.43101326947503243},
pplus-cdd_nom+ztfdr2-vlhq_nom-pcorr-sm+highz/hqj-latz-distR-AV-1x/mean/.initial={71.67174816131592},
pplus-cdd_nom+ztfdr2-vlhq_nom-pcorr-sm+highz/hqj-latz-distR-AV-1x/std/.initial={1.5114150010049343},
pplus-cdd_nom+ztfdr2-vlhq_nom-pcorr-sm+highz/hqj-latz-distR-AV-1x/off/.initial={0.07147789001464844},
pplus-cdd_nom+ztfdr2-vlhq_nom-pcorr-sm+highz/hqj-latz-distR-AV-1x/roff/.initial={0.04879318733462904},
pplus-cdd_nom+ztfdr2-vlhq_nom-pcorr-sm+highz/hqj-latz-distR-AV-2x-step/mean/.initial={72.37097024917603},
pplus-cdd_nom+ztfdr2-vlhq_nom-pcorr-sm+highz/hqj-latz-distR-AV-2x-step/std/.initial={1.5642471611499786},
pplus-cdd_nom+ztfdr2-vlhq_nom-pcorr-sm+highz/hqj-latz-distR-AV-2x-step/off/.initial={0.7706999778747559},
pplus-cdd_nom+ztfdr2-vlhq_nom-pcorr-sm+highz/hqj-latz-distR-AV-2x-step/roff/.initial={0.5261054627036528},
pplus-cdd_nom+ztfdr2-vlhq_nom-pcorr-sm+highz/hqj-latz-distR-AV-2x/mean/.initial={71.80054783821106},
pplus-cdd_nom+ztfdr2-vlhq_nom-pcorr-sm+highz/hqj-latz-distR-AV-2x/std/.initial={1.5007886104285717},
pplus-cdd_nom+ztfdr2-vlhq_nom-pcorr-sm+highz/hqj-latz-distR-AV-2x/off/.initial={0.20027756690979004},
pplus-cdd_nom+ztfdr2-vlhq_nom-pcorr-sm+highz/hqj-latz-distR-AV-2x/roff/.initial={0.136716134725723},
pplus-cdd_nom+ztfdr2-vlhq_nom-pcorr-sm+highz/hqj-latz-globalR-AV-1x-step/mean/.initial={71.41860723495483},
pplus-cdd_nom+ztfdr2-vlhq_nom-pcorr-sm+highz/hqj-latz-globalR-AV-1x-step/std/.initial={1.5336005948483944},
pplus-cdd_nom+ztfdr2-vlhq_nom-pcorr-sm+highz/hqj-latz-globalR-AV-1x-step/off/.initial={-0.18166303634643555},
pplus-cdd_nom+ztfdr2-vlhq_nom-pcorr-sm+highz/hqj-latz-globalR-AV-1x-step/roff/.initial={-0.12400923645637289},
pplus-cdd_nom+ztfdr2-vlhq_nom-pcorr-sm+highz/hqj-latz-globalR-AV-1x/mean/.initial={70.82070708274841},
pplus-cdd_nom+ztfdr2-vlhq_nom-pcorr-sm+highz/hqj-latz-globalR-AV-1x/std/.initial={1.5745870769023895},
pplus-cdd_nom+ztfdr2-vlhq_nom-pcorr-sm+highz/hqj-latz-globalR-AV-1x/off/.initial={-0.7795631885528564},
pplus-cdd_nom+ztfdr2-vlhq_nom-pcorr-sm+highz/hqj-latz-globalR-AV-1x/roff/.initial={-0.5321557853826548},
pplus-cdd_nom+ztfdr2-vlhq_nom-pcorr-sm+highz/hqj-latz-globalR-AV-2x-step/mean/.initial={71.61791920661926},
pplus-cdd_nom+ztfdr2-vlhq_nom-pcorr-sm+highz/hqj-latz-globalR-AV-2x-step/std/.initial={1.5332226641476154},
pplus-cdd_nom+ztfdr2-vlhq_nom-pcorr-sm+highz/hqj-latz-globalR-AV-2x-step/off/.initial={0.017648935317993164},
pplus-cdd_nom+ztfdr2-vlhq_nom-pcorr-sm+highz/hqj-latz-globalR-AV-2x-step/roff/.initial={0.012047750808692177},
pplus-cdd_nom+ztfdr2-vlhq_nom-pcorr-sm+highz/hqj-latz-globalR-AV-2x/mean/.initial={71.13418579101562},
pplus-cdd_nom+ztfdr2-vlhq_nom-pcorr-sm+highz/hqj-latz-globalR-AV-2x/std/.initial={1.5338096767663956},
pplus-cdd_nom+ztfdr2-vlhq_nom-pcorr-sm+highz/hqj-latz-globalR-AV-2x/off/.initial={-0.46608448028564453},
pplus-cdd_nom+ztfdr2-vlhq_nom-pcorr-sm+highz/hqj-latz-globalR-AV-2x/roff/.initial={-0.31816478292350336},
pplus-cdd_nom+ztfdr2-vlhq_nom-pcorr-sm+late/hqj-latz-distR-AV-1x-step/mean/.initial={72.88298606872559},
pplus-cdd_nom+ztfdr2-vlhq_nom-pcorr-sm+late/hqj-latz-distR-AV-1x-step/std/.initial={1.6976185142993927},
pplus-cdd_nom+ztfdr2-vlhq_nom-pcorr-sm+late/hqj-latz-distR-AV-1x-step/off/.initial={1.2827157974243164},
pplus-cdd_nom+ztfdr2-vlhq_nom-pcorr-sm+late/hqj-latz-distR-AV-1x-step/roff/.initial={0.8756245069347487},
pplus-cdd_nom+ztfdr2-vlhq_nom-pcorr-sm+late/hqj-latz-distR-AV-1x/mean/.initial={72.42017984390259},
pplus-cdd_nom+ztfdr2-vlhq_nom-pcorr-sm+late/hqj-latz-distR-AV-1x/std/.initial={1.6377264633774757},
pplus-cdd_nom+ztfdr2-vlhq_nom-pcorr-sm+late/hqj-latz-distR-AV-1x/off/.initial={0.8199095726013184},
pplus-cdd_nom+ztfdr2-vlhq_nom-pcorr-sm+late/hqj-latz-distR-AV-1x/roff/.initial={0.5596975703282939},
pplus-cdd_nom+ztfdr2-vlhq_nom-pcorr-sm+late/hqj-latz-distR-AV-2x-step/mean/.initial={72.40227460861206},
pplus-cdd_nom+ztfdr2-vlhq_nom-pcorr-sm+late/hqj-latz-distR-AV-2x-step/std/.initial={1.6250254586338997},
pplus-cdd_nom+ztfdr2-vlhq_nom-pcorr-sm+late/hqj-latz-distR-AV-2x-step/off/.initial={0.802004337310791},
pplus-cdd_nom+ztfdr2-vlhq_nom-pcorr-sm+late/hqj-latz-distR-AV-2x-step/roff/.initial={0.5474748606257235},
pplus-cdd_nom+ztfdr2-vlhq_nom-pcorr-sm+late/hqj-latz-distR-AV-2x/mean/.initial={72.02827334403992},
pplus-cdd_nom+ztfdr2-vlhq_nom-pcorr-sm+late/hqj-latz-distR-AV-2x/std/.initial={1.5753917396068573},
pplus-cdd_nom+ztfdr2-vlhq_nom-pcorr-sm+late/hqj-latz-distR-AV-2x/off/.initial={0.42800307273864746},
pplus-cdd_nom+ztfdr2-vlhq_nom-pcorr-sm+late/hqj-latz-distR-AV-2x/roff/.initial={0.2921691463423705},
pplus-cdd_nom+ztfdr2-vlhq_nom-pcorr-sm+late/hqj-latz-globalR-AV-1x-step/mean/.initial={72.25683927536011},
pplus-cdd_nom+ztfdr2-vlhq_nom-pcorr-sm+late/hqj-latz-globalR-AV-1x-step/std/.initial={1.6955168917775154},
pplus-cdd_nom+ztfdr2-vlhq_nom-pcorr-sm+late/hqj-latz-globalR-AV-1x-step/off/.initial={0.6565690040588379},
pplus-cdd_nom+ztfdr2-vlhq_nom-pcorr-sm+late/hqj-latz-globalR-AV-1x-step/roff/.initial={0.4481958603784796},
pplus-cdd_nom+ztfdr2-vlhq_nom-pcorr-sm+late/hqj-latz-globalR-AV-1x/mean/.initial={71.8402087688446},
pplus-cdd_nom+ztfdr2-vlhq_nom-pcorr-sm+late/hqj-latz-globalR-AV-1x/std/.initial={1.7109861597418785},
pplus-cdd_nom+ztfdr2-vlhq_nom-pcorr-sm+late/hqj-latz-globalR-AV-1x/off/.initial={0.23993849754333496},
pplus-cdd_nom+ztfdr2-vlhq_nom-pcorr-sm+late/hqj-latz-globalR-AV-1x/roff/.initial={0.16379000635052468},
pplus-cdd_nom+ztfdr2-vlhq_nom-pcorr-sm+late/hqj-latz-globalR-AV-2x-step/mean/.initial={72.53901958465576},
pplus-cdd_nom+ztfdr2-vlhq_nom-pcorr-sm+late/hqj-latz-globalR-AV-2x-step/std/.initial={1.6780231148004532},
pplus-cdd_nom+ztfdr2-vlhq_nom-pcorr-sm+late/hqj-latz-globalR-AV-2x-step/off/.initial={0.9387493133544922},
pplus-cdd_nom+ztfdr2-vlhq_nom-pcorr-sm+late/hqj-latz-globalR-AV-2x-step/roff/.initial={0.6408215337270459},
pplus-cdd_nom+ztfdr2-vlhq_nom-pcorr-sm+late/hqj-latz-globalR-AV-2x/mean/.initial={72.00392484664917},
pplus-cdd_nom+ztfdr2-vlhq_nom-pcorr-sm+late/hqj-latz-globalR-AV-2x/std/.initial={1.656840741634369},
pplus-cdd_nom+ztfdr2-vlhq_nom-pcorr-sm+late/hqj-latz-globalR-AV-2x/off/.initial={0.4036545753479004},
pplus-cdd_nom+ztfdr2-vlhq_nom-pcorr-sm+late/hqj-latz-globalR-AV-2x/roff/.initial={0.27554805142392824},
pplus-cdd_nom+ztfdr2-vlhq_nom-pcorr-sm+late+highz/hqj-latz-distR-AV-1x-step/mean/.initial={73.42817783355713},
pplus-cdd_nom+ztfdr2-vlhq_nom-pcorr-sm+late+highz/hqj-latz-distR-AV-1x-step/std/.initial={1.6819527372717857},
pplus-cdd_nom+ztfdr2-vlhq_nom-pcorr-sm+late+highz/hqj-latz-distR-AV-1x-step/off/.initial={1.8279075622558594},
pplus-cdd_nom+ztfdr2-vlhq_nom-pcorr-sm+late+highz/hqj-latz-distR-AV-1x-step/roff/.initial={1.247790555894376},
pplus-cdd_nom+ztfdr2-vlhq_nom-pcorr-sm+late+highz/hqj-latz-distR-AV-1x/mean/.initial={72.96091318130493},
pplus-cdd_nom+ztfdr2-vlhq_nom-pcorr-sm+late+highz/hqj-latz-distR-AV-1x/std/.initial={1.735490933060646},
pplus-cdd_nom+ztfdr2-vlhq_nom-pcorr-sm+late+highz/hqj-latz-distR-AV-1x/off/.initial={1.360642910003662},
pplus-cdd_nom+ztfdr2-vlhq_nom-pcorr-sm+late+highz/hqj-latz-distR-AV-1x/roff/.initial={0.9288201482967351},
pplus-cdd_nom+ztfdr2-vlhq_nom-pcorr-sm+late+highz/hqj-latz-distR-AV-2x-step/mean/.initial={72.95580506324768},
pplus-cdd_nom+ztfdr2-vlhq_nom-pcorr-sm+late+highz/hqj-latz-distR-AV-2x-step/std/.initial={1.625964418053627},
pplus-cdd_nom+ztfdr2-vlhq_nom-pcorr-sm+late+highz/hqj-latz-distR-AV-2x-step/off/.initial={1.3555347919464111},
pplus-cdd_nom+ztfdr2-vlhq_nom-pcorr-sm+late+highz/hqj-latz-distR-AV-2x-step/roff/.initial={0.925333176853625},
pplus-cdd_nom+ztfdr2-vlhq_nom-pcorr-sm+late+highz/hqj-latz-distR-AV-2x/mean/.initial={72.65442609786987},
pplus-cdd_nom+ztfdr2-vlhq_nom-pcorr-sm+late+highz/hqj-latz-distR-AV-2x/std/.initial={1.6949713230133057},
pplus-cdd_nom+ztfdr2-vlhq_nom-pcorr-sm+late+highz/hqj-latz-distR-AV-2x/off/.initial={1.0541558265686035},
pplus-cdd_nom+ztfdr2-vlhq_nom-pcorr-sm+late+highz/hqj-latz-distR-AV-2x/roff/.initial={0.7196018617101253},
pplus-cdd_nom+ztfdr2-vlhq_nom-pcorr-sm+late+highz/hqj-latz-globalR-AV-1x-step/mean/.initial={72.64618277549744},
pplus-cdd_nom+ztfdr2-vlhq_nom-pcorr-sm+late+highz/hqj-latz-globalR-AV-1x-step/std/.initial={1.7801987007260323},
pplus-cdd_nom+ztfdr2-vlhq_nom-pcorr-sm+late+highz/hqj-latz-globalR-AV-1x-step/off/.initial={1.045912504196167},
pplus-cdd_nom+ztfdr2-vlhq_nom-pcorr-sm+late+highz/hqj-latz-globalR-AV-1x-step/roff/.initial={0.7139746954256196},
pplus-cdd_nom+ztfdr2-vlhq_nom-pcorr-sm+late+highz/hqj-latz-globalR-AV-1x/mean/.initial={72.27305173873901},
pplus-cdd_nom+ztfdr2-vlhq_nom-pcorr-sm+late+highz/hqj-latz-globalR-AV-1x/std/.initial={1.8321016803383827},
pplus-cdd_nom+ztfdr2-vlhq_nom-pcorr-sm+late+highz/hqj-latz-globalR-AV-1x/off/.initial={0.6727814674377441},
pplus-cdd_nom+ztfdr2-vlhq_nom-pcorr-sm+late+highz/hqj-latz-globalR-AV-1x/roff/.initial={0.45926302761915594},
pplus-cdd_nom+ztfdr2-vlhq_nom-pcorr-sm+late+highz/hqj-latz-globalR-AV-2x-step/mean/.initial={72.89748787879944},
pplus-cdd_nom+ztfdr2-vlhq_nom-pcorr-sm+late+highz/hqj-latz-globalR-AV-2x-step/std/.initial={1.7129039391875267},
pplus-cdd_nom+ztfdr2-vlhq_nom-pcorr-sm+late+highz/hqj-latz-globalR-AV-2x-step/off/.initial={1.297217607498169},
pplus-cdd_nom+ztfdr2-vlhq_nom-pcorr-sm+late+highz/hqj-latz-globalR-AV-2x-step/roff/.initial={0.8855239252790744},
pplus-cdd_nom+ztfdr2-vlhq_nom-pcorr-sm+late+highz/hqj-latz-globalR-AV-2x/mean/.initial={72.46392965316772},
pplus-cdd_nom+ztfdr2-vlhq_nom-pcorr-sm+late+highz/hqj-latz-globalR-AV-2x/std/.initial={1.738724485039711},
pplus-cdd_nom+ztfdr2-vlhq_nom-pcorr-sm+late+highz/hqj-latz-globalR-AV-2x/off/.initial={0.8636593818664551},
pplus-cdd_nom+ztfdr2-vlhq_nom-pcorr-sm+late+highz/hqj-latz-globalR-AV-2x/roff/.initial={0.589562646632178},
pplus-cdd_nom+ztfdr2-vlhq_nom-pcov-sm/hqj-latz-distR-AV-1x-step/mean/.initial={71.47302031517029},
pplus-cdd_nom+ztfdr2-vlhq_nom-pcov-sm/hqj-latz-distR-AV-1x-step/std/.initial={1.792076975107193},
pplus-cdd_nom+ztfdr2-vlhq_nom-pcov-sm/hqj-latz-distR-AV-1x-step/off/.initial={-0.12724995613098145},
pplus-cdd_nom+ztfdr2-vlhq_nom-pcov-sm/hqj-latz-distR-AV-1x-step/roff/.initial={-0.08686505640485284},
pplus-cdd_nom+ztfdr2-vlhq_nom-pcov-sm/hqj-latz-distR-AV-1x/mean/.initial={71.16243839263916},
pplus-cdd_nom+ztfdr2-vlhq_nom-pcov-sm/hqj-latz-distR-AV-1x/std/.initial={1.8240228295326233},
pplus-cdd_nom+ztfdr2-vlhq_nom-pcov-sm/hqj-latz-distR-AV-1x/off/.initial={-0.4378318786621094},
pplus-cdd_nom+ztfdr2-vlhq_nom-pcov-sm/hqj-latz-distR-AV-1x/roff/.initial={-0.29887861648203057},
pplus-cdd_nom+ztfdr2-vlhq_nom-pcov-sm/hqj-latz-distR-AV-2x-step/mean/.initial={71.55463695526123},
pplus-cdd_nom+ztfdr2-vlhq_nom-pcov-sm/hqj-latz-distR-AV-2x-step/std/.initial={1.7694545909762383},
pplus-cdd_nom+ztfdr2-vlhq_nom-pcov-sm/hqj-latz-distR-AV-2x-step/off/.initial={-0.04563331604003906},
pplus-cdd_nom+ztfdr2-vlhq_nom-pcov-sm/hqj-latz-distR-AV-2x-step/roff/.initial={-0.031150820733315536},
pplus-cdd_nom+ztfdr2-vlhq_nom-pcov-sm/hqj-latz-distR-AV-2x/mean/.initial={71.35166525840759},
pplus-cdd_nom+ztfdr2-vlhq_nom-pcov-sm/hqj-latz-distR-AV-2x/std/.initial={1.770009845495224},
pplus-cdd_nom+ztfdr2-vlhq_nom-pcov-sm/hqj-latz-distR-AV-2x/off/.initial={-0.24860501289367676},
pplus-cdd_nom+ztfdr2-vlhq_nom-pcov-sm/hqj-latz-distR-AV-2x/roff/.initial={-0.1697060582505039},
pplus-cdd_nom+ztfdr2-vlhq_nom-pcov-sm/hqj-latz-globalR-AV-1x-step/mean/.initial={70.84344625473022},
pplus-cdd_nom+ztfdr2-vlhq_nom-pcov-sm/hqj-latz-globalR-AV-1x-step/std/.initial={1.7492184415459633},
pplus-cdd_nom+ztfdr2-vlhq_nom-pcov-sm/hqj-latz-globalR-AV-1x-step/off/.initial={-0.7568240165710449},
pplus-cdd_nom+ztfdr2-vlhq_nom-pcov-sm/hqj-latz-globalR-AV-1x-step/roff/.initial={-0.5166332695653092},
pplus-cdd_nom+ztfdr2-vlhq_nom-pcov-sm/hqj-latz-globalR-AV-1x/mean/.initial={70.50561308860779},
pplus-cdd_nom+ztfdr2-vlhq_nom-pcov-sm/hqj-latz-globalR-AV-1x/std/.initial={1.8119614571332932},
pplus-cdd_nom+ztfdr2-vlhq_nom-pcov-sm/hqj-latz-globalR-AV-1x/off/.initial={-1.0946571826934814},
pplus-cdd_nom+ztfdr2-vlhq_nom-pcov-sm/hqj-latz-globalR-AV-1x/roff/.initial={-0.7472494357543885},
pplus-cdd_nom+ztfdr2-vlhq_nom-pcov-sm/hqj-latz-globalR-AV-2x-step/mean/.initial={71.03065252304077},
pplus-cdd_nom+ztfdr2-vlhq_nom-pcov-sm/hqj-latz-globalR-AV-2x-step/std/.initial={1.773267239332199},
pplus-cdd_nom+ztfdr2-vlhq_nom-pcov-sm/hqj-latz-globalR-AV-2x-step/off/.initial={-0.569617748260498},
pplus-cdd_nom+ztfdr2-vlhq_nom-pcov-sm/hqj-latz-globalR-AV-2x-step/roff/.initial={-0.38884003842738146},
pplus-cdd_nom+ztfdr2-vlhq_nom-pcov-sm/hqj-latz-globalR-AV-2x/mean/.initial={70.7719087600708},
pplus-cdd_nom+ztfdr2-vlhq_nom-pcov-sm/hqj-latz-globalR-AV-2x/std/.initial={1.7692271620035172},
pplus-cdd_nom+ztfdr2-vlhq_nom-pcov-sm/hqj-latz-globalR-AV-2x/off/.initial={-0.8283615112304688},
pplus-cdd_nom+ztfdr2-vlhq_nom-pcov-sm/hqj-latz-globalR-AV-2x/roff/.initial={-0.5654671450147937},
pplus-cdd_nom+ztfdr2-vlhq_nom-pcov-sm+highz/hqj-latz-distR-AV-1x-step/mean/.initial={71.01990580558777},
pplus-cdd_nom+ztfdr2-vlhq_nom-pcov-sm+highz/hqj-latz-distR-AV-1x-step/std/.initial={1.7463669180870056},
pplus-cdd_nom+ztfdr2-vlhq_nom-pcov-sm+highz/hqj-latz-distR-AV-1x-step/off/.initial={-0.580364465713501},
pplus-cdd_nom+ztfdr2-vlhq_nom-pcov-sm+highz/hqj-latz-distR-AV-1x-step/roff/.initial={-0.39617610553581517},
pplus-cdd_nom+ztfdr2-vlhq_nom-pcov-sm+highz/hqj-latz-distR-AV-1x/mean/.initial={70.68282961845398},
pplus-cdd_nom+ztfdr2-vlhq_nom-pcov-sm+highz/hqj-latz-distR-AV-1x/std/.initial={1.7380783334374428},
pplus-cdd_nom+ztfdr2-vlhq_nom-pcov-sm+highz/hqj-latz-distR-AV-1x/off/.initial={-0.91744065284729},
pplus-cdd_nom+ztfdr2-vlhq_nom-pcov-sm+highz/hqj-latz-distR-AV-1x/roff/.initial={-0.6262755326662305},
pplus-cdd_nom+ztfdr2-vlhq_nom-pcov-sm+highz/hqj-latz-distR-AV-2x-step/mean/.initial={71.09988927841187},
pplus-cdd_nom+ztfdr2-vlhq_nom-pcov-sm+highz/hqj-latz-distR-AV-2x-step/std/.initial={1.704404316842556},
pplus-cdd_nom+ztfdr2-vlhq_nom-pcov-sm+highz/hqj-latz-distR-AV-2x-step/off/.initial={-0.5003809928894043},
pplus-cdd_nom+ztfdr2-vlhq_nom-pcov-sm+highz/hqj-latz-distR-AV-2x-step/roff/.initial={-0.3415767242113165},
pplus-cdd_nom+ztfdr2-vlhq_nom-pcov-sm+highz/hqj-latz-distR-AV-2x/mean/.initial={71.02177739143372},
pplus-cdd_nom+ztfdr2-vlhq_nom-pcov-sm+highz/hqj-latz-distR-AV-2x/std/.initial={1.7376864328980446},
pplus-cdd_nom+ztfdr2-vlhq_nom-pcov-sm+highz/hqj-latz-distR-AV-2x/off/.initial={-0.5784928798675537},
pplus-cdd_nom+ztfdr2-vlhq_nom-pcov-sm+highz/hqj-latz-distR-AV-2x/roff/.initial={-0.3948984987293546},
pplus-cdd_nom+ztfdr2-vlhq_nom-pcov-sm+highz/hqj-latz-globalR-AV-1x-step/mean/.initial={70.4403817653656},
pplus-cdd_nom+ztfdr2-vlhq_nom-pcov-sm+highz/hqj-latz-globalR-AV-1x-step/std/.initial={1.690148003399372},
pplus-cdd_nom+ztfdr2-vlhq_nom-pcov-sm+highz/hqj-latz-globalR-AV-1x-step/off/.initial={-1.159888505935669},
pplus-cdd_nom+ztfdr2-vlhq_nom-pcov-sm+highz/hqj-latz-globalR-AV-1x-step/roff/.initial={-0.7917785086521687},
pplus-cdd_nom+ztfdr2-vlhq_nom-pcov-sm+highz/hqj-latz-globalR-AV-1x/mean/.initial={70.05313038825989},
pplus-cdd_nom+ztfdr2-vlhq_nom-pcov-sm+highz/hqj-latz-globalR-AV-1x/std/.initial={1.7436552792787552},
pplus-cdd_nom+ztfdr2-vlhq_nom-pcov-sm+highz/hqj-latz-globalR-AV-1x/off/.initial={-1.5471398830413818},
pplus-cdd_nom+ztfdr2-vlhq_nom-pcov-sm+highz/hqj-latz-globalR-AV-1x/roff/.initial={-1.0561291908678831},
pplus-cdd_nom+ztfdr2-vlhq_nom-pcov-sm+highz/hqj-latz-globalR-AV-2x-step/mean/.initial={70.66492438316345},
pplus-cdd_nom+ztfdr2-vlhq_nom-pcov-sm+highz/hqj-latz-globalR-AV-2x-step/std/.initial={1.6815541312098503},
pplus-cdd_nom+ztfdr2-vlhq_nom-pcov-sm+highz/hqj-latz-globalR-AV-2x-step/off/.initial={-0.9353458881378174},
pplus-cdd_nom+ztfdr2-vlhq_nom-pcov-sm+highz/hqj-latz-globalR-AV-2x-step/roff/.initial={-0.638498242368801},
pplus-cdd_nom+ztfdr2-vlhq_nom-pcov-sm+highz/hqj-latz-globalR-AV-2x/mean/.initial={70.3725814819336},
pplus-cdd_nom+ztfdr2-vlhq_nom-pcov-sm+highz/hqj-latz-globalR-AV-2x/std/.initial={1.7148517072200775},
pplus-cdd_nom+ztfdr2-vlhq_nom-pcov-sm+highz/hqj-latz-globalR-AV-2x/off/.initial={-1.2276887893676758},
pplus-cdd_nom+ztfdr2-vlhq_nom-pcov-sm+highz/hqj-latz-globalR-AV-2x/roff/.initial={-0.8380612393002178},
pplus-cdd_nom+ztfdr2-vlhq_nom-pcov-sm+late/hqj-latz-distR-AV-1x-step/mean/.initial={72.15942144393921},
pplus-cdd_nom+ztfdr2-vlhq_nom-pcov-sm+late/hqj-latz-distR-AV-1x-step/std/.initial={1.9152931869029999},
pplus-cdd_nom+ztfdr2-vlhq_nom-pcov-sm+late/hqj-latz-distR-AV-1x-step/off/.initial={0.5591511726379395},
pplus-cdd_nom+ztfdr2-vlhq_nom-pcov-sm+late/hqj-latz-distR-AV-1x-step/roff/.initial={0.38169520545876834},
pplus-cdd_nom+ztfdr2-vlhq_nom-pcov-sm+late/hqj-latz-distR-AV-1x/mean/.initial={71.93204760551453},
pplus-cdd_nom+ztfdr2-vlhq_nom-pcov-sm+late/hqj-latz-distR-AV-1x/std/.initial={1.8962344154715538},
pplus-cdd_nom+ztfdr2-vlhq_nom-pcov-sm+late/hqj-latz-distR-AV-1x/off/.initial={0.33177733421325684},
pplus-cdd_nom+ztfdr2-vlhq_nom-pcov-sm+late/hqj-latz-distR-AV-1x/roff/.initial={0.22648225371976782},
pplus-cdd_nom+ztfdr2-vlhq_nom-pcov-sm+late/hqj-latz-distR-AV-2x-step/mean/.initial={71.80428504943848},
pplus-cdd_nom+ztfdr2-vlhq_nom-pcov-sm+late/hqj-latz-distR-AV-2x-step/std/.initial={1.94761510938406},
pplus-cdd_nom+ztfdr2-vlhq_nom-pcov-sm+late/hqj-latz-distR-AV-2x-step/off/.initial={0.20401477813720703},
pplus-cdd_nom+ztfdr2-vlhq_nom-pcov-sm+late/hqj-latz-distR-AV-2x-step/roff/.initial={0.13926727952715834},
pplus-cdd_nom+ztfdr2-vlhq_nom-pcov-sm+late/hqj-latz-distR-AV-2x/mean/.initial={71.89502120018005},
pplus-cdd_nom+ztfdr2-vlhq_nom-pcov-sm+late/hqj-latz-distR-AV-2x/std/.initial={1.9391868263483047},
pplus-cdd_nom+ztfdr2-vlhq_nom-pcov-sm+late/hqj-latz-distR-AV-2x/off/.initial={0.2947509288787842},
pplus-cdd_nom+ztfdr2-vlhq_nom-pcov-sm+late/hqj-latz-distR-AV-2x/roff/.initial={0.2012067967715761},
pplus-cdd_nom+ztfdr2-vlhq_nom-pcov-sm+late/hqj-latz-globalR-AV-1x-step/mean/.initial={71.56006097793579},
pplus-cdd_nom+ztfdr2-vlhq_nom-pcov-sm+late/hqj-latz-globalR-AV-1x-step/std/.initial={1.913721114397049},
pplus-cdd_nom+ztfdr2-vlhq_nom-pcov-sm+late/hqj-latz-globalR-AV-1x-step/off/.initial={-0.040209293365478516},
pplus-cdd_nom+ztfdr2-vlhq_nom-pcov-sm+late/hqj-latz-globalR-AV-1x-step/roff/.initial={-0.027448202281471604},
pplus-cdd_nom+ztfdr2-vlhq_nom-pcov-sm+late/hqj-latz-globalR-AV-1x/mean/.initial={71.38308882713318},
pplus-cdd_nom+ztfdr2-vlhq_nom-pcov-sm+late/hqj-latz-globalR-AV-1x/std/.initial={1.9561687484383583},
pplus-cdd_nom+ztfdr2-vlhq_nom-pcov-sm+late/hqj-latz-globalR-AV-1x/off/.initial={-0.21718144416809082},
pplus-cdd_nom+ztfdr2-vlhq_nom-pcov-sm+late/hqj-latz-globalR-AV-1x/roff/.initial={-0.14825528409872232},
pplus-cdd_nom+ztfdr2-vlhq_nom-pcov-sm+late/hqj-latz-globalR-AV-2x-step/mean/.initial={71.61303758621216},
pplus-cdd_nom+ztfdr2-vlhq_nom-pcov-sm+late/hqj-latz-globalR-AV-2x-step/std/.initial={1.8974566832184792},
pplus-cdd_nom+ztfdr2-vlhq_nom-pcov-sm+late/hqj-latz-globalR-AV-2x-step/off/.initial={0.012767314910888672},
pplus-cdd_nom+ztfdr2-vlhq_nom-pcov-sm+late/hqj-latz-globalR-AV-2x-step/roff/.initial={0.008715394202032638},
pplus-cdd_nom+ztfdr2-vlhq_nom-pcov-sm+late/hqj-latz-globalR-AV-2x/mean/.initial={71.26286625862122},
pplus-cdd_nom+ztfdr2-vlhq_nom-pcov-sm+late/hqj-latz-globalR-AV-2x/std/.initial={1.889750175178051},
pplus-cdd_nom+ztfdr2-vlhq_nom-pcov-sm+late/hqj-latz-globalR-AV-2x/off/.initial={-0.3374040126800537},
pplus-cdd_nom+ztfdr2-vlhq_nom-pcov-sm+late/hqj-latz-globalR-AV-2x/roff/.initial={-0.23032321176212023},
pplus-cdd_nom+ztfdr2-vlhq_nom-pcov-sm+late+highz/hqj-latz-distR-AV-1x-step/mean/.initial={71.84908986091614},
pplus-cdd_nom+ztfdr2-vlhq_nom-pcov-sm+late+highz/hqj-latz-distR-AV-1x-step/std/.initial={1.9229374825954437},
pplus-cdd_nom+ztfdr2-vlhq_nom-pcov-sm+late+highz/hqj-latz-distR-AV-1x-step/off/.initial={0.24881958961486816},
pplus-cdd_nom+ztfdr2-vlhq_nom-pcov-sm+late+highz/hqj-latz-distR-AV-1x-step/roff/.initial={0.16985253546398343},
pplus-cdd_nom+ztfdr2-vlhq_nom-pcov-sm+late+highz/hqj-latz-distR-AV-1x/mean/.initial={71.65277600288391},
pplus-cdd_nom+ztfdr2-vlhq_nom-pcov-sm+late+highz/hqj-latz-distR-AV-1x/std/.initial={1.9051773473620415},
pplus-cdd_nom+ztfdr2-vlhq_nom-pcov-sm+late+highz/hqj-latz-distR-AV-1x/off/.initial={0.0525057315826416},
pplus-cdd_nom+ztfdr2-vlhq_nom-pcov-sm+late+highz/hqj-latz-distR-AV-1x/roff/.initial={0.03584216037614636},
pplus-cdd_nom+ztfdr2-vlhq_nom-pcov-sm+late+highz/hqj-latz-distR-AV-2x-step/mean/.initial={71.63483500480652},
pplus-cdd_nom+ztfdr2-vlhq_nom-pcov-sm+late+highz/hqj-latz-distR-AV-2x-step/std/.initial={1.8799595534801483},
pplus-cdd_nom+ztfdr2-vlhq_nom-pcov-sm+late+highz/hqj-latz-distR-AV-2x-step/off/.initial={0.03456473350524902},
pplus-cdd_nom+ztfdr2-vlhq_nom-pcov-sm+late+highz/hqj-latz-distR-AV-2x-step/roff/.initial={0.023595037804662593},
pplus-cdd_nom+ztfdr2-vlhq_nom-pcov-sm+late+highz/hqj-latz-distR-AV-2x/mean/.initial={71.51296734809875},
pplus-cdd_nom+ztfdr2-vlhq_nom-pcov-sm+late+highz/hqj-latz-distR-AV-2x/std/.initial={1.8620777875185013},
pplus-cdd_nom+ztfdr2-vlhq_nom-pcov-sm+late+highz/hqj-latz-distR-AV-2x/off/.initial={-0.08730292320251465},
pplus-cdd_nom+ztfdr2-vlhq_nom-pcov-sm+late+highz/hqj-latz-distR-AV-2x/roff/.initial={-0.05959588182874512},
pplus-cdd_nom+ztfdr2-vlhq_nom-pcov-sm+late+highz/hqj-latz-globalR-AV-1x-step/mean/.initial={71.21017575263977},
pplus-cdd_nom+ztfdr2-vlhq_nom-pcov-sm+late+highz/hqj-latz-globalR-AV-1x-step/std/.initial={1.9539998844265938},
pplus-cdd_nom+ztfdr2-vlhq_nom-pcov-sm+late+highz/hqj-latz-globalR-AV-1x-step/off/.initial={-0.390094518661499},
pplus-cdd_nom+ztfdr2-vlhq_nom-pcov-sm+late+highz/hqj-latz-globalR-AV-1x-step/roff/.initial={-0.26629150529431844},
pplus-cdd_nom+ztfdr2-vlhq_nom-pcov-sm+late+highz/hqj-latz-globalR-AV-1x/mean/.initial={71.10273838043213},
pplus-cdd_nom+ztfdr2-vlhq_nom-pcov-sm+late+highz/hqj-latz-globalR-AV-1x/std/.initial={1.9464123994112015},
pplus-cdd_nom+ztfdr2-vlhq_nom-pcov-sm+late+highz/hqj-latz-globalR-AV-1x/off/.initial={-0.4975318908691406},
pplus-cdd_nom+ztfdr2-vlhq_nom-pcov-sm+late+highz/hqj-latz-globalR-AV-1x/roff/.initial={-0.33963183232122707},
pplus-cdd_nom+ztfdr2-vlhq_nom-pcov-sm+late+highz/hqj-latz-globalR-AV-2x-step/mean/.initial={71.37019038200378},
pplus-cdd_nom+ztfdr2-vlhq_nom-pcov-sm+late+highz/hqj-latz-globalR-AV-2x-step/std/.initial={1.9279858097434044},
pplus-cdd_nom+ztfdr2-vlhq_nom-pcov-sm+late+highz/hqj-latz-globalR-AV-2x-step/off/.initial={-0.23007988929748535},
pplus-cdd_nom+ztfdr2-vlhq_nom-pcov-sm+late+highz/hqj-latz-globalR-AV-2x-step/roff/.initial={-0.15706019215343692},
pplus-cdd_nom+ztfdr2-vlhq_nom-pcov-sm+late+highz/hqj-latz-globalR-AV-2x/mean/.initial={71.0834801197052},
pplus-cdd_nom+ztfdr2-vlhq_nom-pcov-sm+late+highz/hqj-latz-globalR-AV-2x/std/.initial={1.8743723630905151},
pplus-cdd_nom+ztfdr2-vlhq_nom-pcov-sm+late+highz/hqj-latz-globalR-AV-2x/off/.initial={-0.5167901515960693},
pplus-cdd_nom+ztfdr2-vlhq_nom-pcov-sm+late+highz/hqj-latz-globalR-AV-2x/roff/.initial={-0.3527781622310158},
}
\pgfkeys{,
/private/.cd,
pplus-cdd_nom+ztfdr2-vlhq_nom-pcorr/hqj-latz-distR-AV-1x/qprior/mean/.initial={75.68250894546509},
pplus-cdd_nom+ztfdr2-vlhq_nom-pcorr/hqj-latz-distR-AV-1x/qprior/std/.initial={1.090617198497057},
pplus-cdd_nom+ztfdr2-vlhq_nom-pcorr-sm/hqj-latz-distR-AV-1x/qprior/mean/.initial={72.28551506996155},
pplus-cdd_nom+ztfdr2-vlhq_nom-pcorr-sm/hqj-latz-distR-AV-1x/qprior/std/.initial={1.0917569510638714},
pplus-cdd_nom+ztfdr2-vlhq_nom-pcorr-sm+late/hqj-latz-distR-AV-1x/qprior/mean/.initial={72.564035654068},
pplus-cdd_nom+ztfdr2-vlhq_nom-pcorr-sm+late/hqj-latz-distR-AV-1x/qprior/std/.initial={1.119403075426817},
pplus-cdd_nom+ztfdr2-vlhq_nom-pcorr-sm+highz/hqj-latz-distR-AV-1x/qprior/mean/.initial={72.08318710327148},
pplus-cdd_nom+ztfdr2-vlhq_nom-pcorr-sm+highz/hqj-latz-distR-AV-1x/qprior/std/.initial={1.1056719347834587},
pplus-cdd_nom+ztfdr2-vlhq_nom-pcorr-sm+late+highz/hqj-latz-distR-AV-1x/qprior/mean/.initial={72.87371158599854},
pplus-cdd_nom+ztfdr2-vlhq_nom-pcorr-sm+late+highz/hqj-latz-distR-AV-1x/qprior/std/.initial={1.1721179820597172},
}
\defineparam{h}{}{}{0.6789}{0.7730}
\defineparam{q0}{}{}{-2.5138}{1.5114}
\defineparam{M0}{\Mo}{}{-19.3011}{-18.8836}
\defineparam{sigma_res}{\sigmares}{}{0.0661}{0.1222}
\defineparam{alpha}{\saltalpha}{}{-0.1956}{-0.1046}
\defineparam{alpha_c}{\saltalphac}{}{-0.0253}{0.0154}
\defineparam{beta}{\saltbeta}{}{-0.1804}{3.5090}
\defineparam{mean_x}{\meansaltx}{}{-0.4376}{-0.0460}
\defineparam{sigma_x}{\stdsaltx}{}{0.7031}{1.0530}
\defineparam{mean_c}{\meansaltc}{}{-0.1045}{-0.0554}
\defineparam{sigma_c}{\stdsaltc}{}{0.0185}{0.0653}
\defineparam{mean_R}{\meanRv}{}{1.3826}{2.8927}
\defineparam{sigma_R}{\sigmaRv}{}{0.2083}{1.4286}
\defineparam{tau}{\bsntau}{}{0.2519}{0.3927}
\setdata{qhslope}{-2.3482429173797295}
\setdata{tension/Planck}{2.7135348946203277}
\setdata{tension-qprior/Planck}{4.068532251027092}
\setdata{tension/SH0ES}{0.7891651288264089}
\setdata{tension-qprior/SH0ES}{0.4825913785938641}
\setdata{Ncalib}{42}
\setdata{Nhf/base}{772}
\setdata{Nhf/late}{228}
\setdata{Nhf/highz}{705}
\setdata{Nhf/latehz}{205}
\newcommand{\fidh}{0.7160026431083679}
\newcommand{\fidOmo}{0.33}

\defineparam{h}{$\flatfrac{\Ho}{\Hounit}$}{}{0.635}{0.775}
\defineparam{q0}{\cosmodecelo}{}{-1.2}{1}
\defineparam{Mstar}{$\opbraces{\log_{10}}(\Mstarsn/\si{\Msun})$}{}{9}{11}

\setdata{lbls/pc/pcorr}{\citetalias{Carrick_2015} corr.}
\setdata{lbls/pc/pcov}{\acrshort*{lss} prior}
\setdata{lbls/pc/pborg}{\acrshort{borg} post.}
\setdata{lbls/step/nostep}{no step}
\setdata{lbls/step/step}{\hphantom{no} step}
\setdata{lbls/R_mode/distR}{dist.\@ $\qty{\Rvsn}$}
\setdata{lbls/R_mode/globalR}{global \Rv}
\setdata{lbls/use_2x/1x}{1\saltx}
\setdata{lbls/use_2x/2x}{2\saltx}

\newcommand{\basevariant}{hqj-latz-distR-AV-1x}

\newcommand{\dnuncorr}{pplus-cdd_nom+ztfdr2-vlhq_nom-pcorr}
\newcommand{\dnbase}{pplus-cdd_nom+ztfdr2-vlhq_nom-pcorr-sm}
\newcommand{\dnlate}{pplus-cdd_nom+ztfdr2-vlhq_nom-pcorr-sm+late}
\newcommand{\dnhz}{pplus-cdd_nom+ztfdr2-vlhq_nom-pcorr-sm+highz}
\newcommand{\dnlatehz}{pplus-cdd_nom+ztfdr2-vlhq_nom-pcorr-sm+late+highz}

\newcommand{\highz}{$\zcobs\ge\num{0.025}$\xspace}

\tikzset{
    fiducial/.style={color=black, densely dashed, semithick},
    data/.style={color=Dark2-B},
    model/.style={color=Set1-purple},
    altmark/.style={every mark/.append style={fill=white, solid, thick, draw={\pgfkeysvalueof{/tikz/the color}}}},
    uncorr/.style={
        data, 
        densely dashed,
        altmark,
        credible region/.append style={fill=none, draw={\pgfkeysvalueof{/tikz/the color}}}},
    lM/.style={color=Set1-grey},
    hM/.style={color=Set1-brown},
    late/.style={color=Set1-blue},
    hz/.style={credible region/.append style={fill=none, draw={\pgfkeysvalueof{/tikz/the color}}}, altmark},
    infer/.style={color=Set3-E},
    h result/.style={
        2 sigma region/.style={fill opacity=0.4},
        1 sigma region/.style={fill opacity=0.6}
    },
    Planck/.style={color=Set1-red},
    SH0ES/.style={color=Set1-green},
}
\pgfplotsset{
    extra x ticks as minor/.style={
        extra x ticks={#1},
        extra x tick labels=\empty,
        extra x tick style={tickwidth={\pgfkeysvalueof{/pgfplots/minor tick length}}},
    },
    extra y ticks as minor/.style={
        extra y ticks={#1},
        extra y tick labels=\empty,
        extra y tick style={tickwidth={\pgfkeysvalueof{/pgfplots/minor tick length}}},
    },
    error bars/error bar style={solid},
    every axis/.append style={tickwidth=3pt, minor tick length=1.5pt},
    repeat xticklabels in group/.style={
        group/vertical sep=16pt,
        xtick pos=both, xticklabel style={font=\scriptsize},
    },
    H0 axis/.style={param axis=h, xlabel={Hubble constant, \getparamlabel{h}}, xticklabel={$\pgfmathparse{100*\tick}\pgfmathprintnumber{\pgfmathresult}$}},
    mini x axis/.style={
        xticklabel style={font=\scriptsize, inner ysep=2pt},
        xlabel style={font=\scriptsize, anchor=near ticklabel, at={(ticklabel* cs:0.5)}, inner ysep=0pt, yshift=1.5em}
    },
    H0 as magoff/.style 2 args={
        axis x line*=top,
        x coord trafo/.code={\pgfmathparse{\fidh*10^(-(##1)/5)}},
        x coord inv trafo/.code={\pgfmathparse{-5*log10((##1)/\fidh)}},
        /pgf/number format/fixed,
        xtick={-0.3,-0.2,...,0.3},
        extra x ticks as minor={-0.3,-0.29,...,0.3},
        xmin={-5*log10((#1)/\fidh)},
        xmax={-5*log10((#2)/\fidh)},
        mini x axis,
        xlabel={equiv.\@ magnitude offset},
    },
    variants axis/.style={
        error bars/x explicit, error bars/x dir=both,
        error bars/error bar style={thick},
        every axis plot post/.append style={error bars/error mark options={solid, thick, rotate=90}},
        every axis plot/.append style={only marks, mark options={draw=black, ultra thin}},
        execute at begin axis={
            \vstrip*[Planck, h result]{0.674}{0.005}
            \vstrip*[SH0ES, h result]{0.73}{0.01}
            \axvline[fiducial, model]{\fidh};
        },
        legend image code/.code={
            \draw[/pgfplots/every error bar, mark=|, every mark/.append style={/pgfplots/every error bar, solid, color/.expanded={\pgfkeysvalueof{/tikz/the color}}}, sharp plot] plot coordinates {(-0.3cm,0cm) (0.3cm,0cm)};
            \draw plot coordinates {(0cm, 0cm)};
        }, forget plot,
    }
}

\newcommand{\vstrips}[4][]{
    \pgfplotsinvokeforeach{#4}{
        \fill[#1, ##1 sigma region, draw=none]
        ({{axis description cs:0,0}} -| {{{#2-##1*#3}, 0}}) -- ({{axis description cs:0,1}} -| {{{#2-##1*#3}, 0}})
        --
        ({{axis description cs:0,1}} -| {{{#2+##1*#3}, 0}}) -- ({{axis description cs:0,0}} -| {{{#2+##1*#3}, 0}})
        ;
        \draw[#1, ##1 sigma region] ({{axis description cs:0,0}} -| {{{#2-##1*#3}, 0}}) -- ({{axis description cs:0,1}} -| {{{#2-##1*#3}, 0}});
        \draw[#1, ##1 sigma region] ({{axis description cs:0,0}} -| {{{#2+##1*#3}, 0}}) -- ({{axis description cs:0,1}} -| {{{#2+##1*#3}, 0}});
    }
    \axvline[#1]{#2};
}
\newcommand{\vstrip}[3][]{\vstrips[#1]{#2}{#3}{2,1}}
\WithSuffix\newcommand\vstrip*[3][]{\vstrips[#1]{#2}{#3}{1}}

\pgfplotsset{
    vstrip legend 1/.style={legend image code/.code={
        \fill[##1, 1 sigma region, draw=none] (-0.3cm,-0.1cm) rectangle (0.3cm,0.1cm);
        \draw[##1, 1 sigma region, fill=none, semithick]
        (-0.3cm,-0.1cm) -- (-0.3cm,0.1cm)
        (0.3cm,-0.1cm) -- (0.3cm,0.1cm);
        \draw[/pgfplots/every axis plot, ##1] (0cm,-0.1cm) -- (0cm,0.1cm);
    }},
    vstrip legend/.style={legend image code/.code={
        \fill[##1, 2 sigma region, draw=none] (-0.3cm,-0.1cm) rectangle (0.3cm,0.1cm);
        \fill[##1, 1 sigma region, draw=none] (-0.15cm,-0.1cm) rectangle (0.15cm,0.1cm);
        \draw[##1, 2 sigma region, fill=none, semithick]
        (-0.3cm,-0.1cm) -- (-0.3cm,0.1cm)
        (0.3cm,-0.1cm) -- (0.3cm,0.1cm);
        \draw[##1, 1 sigma region, fill=none, semithick]
        (-0.15cm,-0.1cm) -- (-0.15cm,0.1cm)
        (0.15cm,-0.1cm) -- (0.15cm,0.1cm);
        \draw[/pgfplots/every axis plot, ##1] (0cm,-0.1cm) -- (0cm,0.1cm);
    }}
}
\newcommand{\acontour}[2][]{
    \pgfplotsinvokeforeach{2,1}{
        \addplot[#1, ##1 sigma region] coordinates {\getdata{corner/#2/##1}};
    }
}
\NewDocumentCommand{\acornerpanel}{smm}{
    \acontour[late]{\dnlate/\basevariant/#2-#3}
    \IfBooleanT{#1}{
        \acontour[late, hz]{\dnlatehz/\basevariant/#2-#3}
    }
    \acontour[uncorr]{\dnuncorr/\basevariant/#2-#3}
    \acontour[model]{\dnbase/\basevariant/#2-#3}
    \IfBooleanT{#1}{
        \acontour[model, hz]{\dnhz/\basevariant/#2-#3}
    }
}
\newcommand{\labeluncorr}{\acrshort{fp}, all\xspace}
\newcommand{\labelbase}{\acrshort{smp}, all\xspace}
\newcommand{\labellate}{\acrshort{smp}, late host\xspace}
\newcommand{\labelhz}{\acrshort{smp}, \highz\xspace}
\newcommand{\ztflegend}[2][]{
    \matrix[/pgfplots/every axis legend, name=ztf legend, column sep=0.3em, ampersand replacement=\&, #1] at #2 {
        \pgfplotsset{ellipses legend, legend image code={uncorr}}
        \& \node {\labeluncorr};
        \\
        \pgfplotsset{ellipses legend, legend image code={model}}
        \& \node {\labelbase (\num{\getdata{Nhf/base}})};
        \\
        \pgfplotsset{ellipses legend, legend image code={late}}
        \& \node {\labellate (\num{\getdata{Nhf/late}})};
        \\
        \pgfplotsset{ellipses legend, legend image code={hz, color=black}}
        \& \node {\labelhz};
        \\
    };
}

\definecolor{tsds-blue}{RGB}{18, 30, 51}
\definecolor{tsds-yellow}{RGB}{253, 219, 64}
\definecolor{tsds-grey}{RGB}{240, 240, 240}
\colorlet{tsds-gray}{tsds-grey}

\colorlet{graph-global}{MediumVioletRed}
\colorlet{graph-local}{DodgerBlue!60!white}
\colorlet{graph-physics}{Goldenrod!60!white}
\colorlet{graph-instrument}{Brown!60!white}
\colorlet{graph-redshift}{Chocolate}
\colorlet{graph-cosmo}{ForestGreen!60!white}
\colorlet{graph-data}{BlueViolet!60!white}

\colorlet{nn-nre}{tsds-yellow}

\defcitealias{Carrick_2015}{C15}

\defcitealias{Taylor_2021}{T21}
\defcitealias{Brout_2022}{B22}

\defcitealias{Planck2018-1}{\emph{Planck}}
\defcitealias{Riess_2022}{\SHoES}

\defcitealias{Rigault_2025}{R25}

\AtBeginDocument{
    \dummylabel[figure]{ext:Jones_2019:fig:hists}{1}
    \dummylabel[table]{ext:Brout_2022a:tbl:sims}{1}
    \dummylabel[section]{ext:Riess_2022:sec:formalism}{2.1}
    \dummylabel[table]{ext:Riess_2022:tbl:dists}{6}
    \dummylabel[figure]{ext:Riess_2022:fig:covdistmod}{16}

    \dummylabel[section]{ext:Weinberg_2008:sec:Taylor-cosmo}{1.4}

    \dummylabel[figure]{ext:Amenouche_2025:fig:bias}{9}
    \dummylabel[equation]{ext:Carreres_2025a:eqn:vcov}{17}
    \dummylabel[figure]{ext:Lacroix_2025:fig:offset}{12}
    \dummylabel[figure]{ext:Rigault_2025:fig:pocket}{2}
    \dummylabel[subsection]{ext:Rigault_2025:sec:selection}{4.6}

    \dummylabel[figure]{ext:Stiskalek_2026c:fig:post}{9}
}

\newcommand{\ECHoEZ}{EC\Ho{}EZ\xspace}

\def\tikzauxdir{. }

\shorttitle{\ECHoEZ}
\shortauthors{Karchev et al.}

\newcommand{\ICCUB}{ICCUB: Institute of Cosmos Sciences, University of Barcelona; Martí i Franquès, 1, 08028 Barcelona, Spain}
\renewcommand{\SISSA}{SISSA: Scuola Internazionale Superiore di Studi Avanzati; via Bonomea, 265, 34136 Trieste, Italy}
\newcommand{\ICREA}{ICREA: Catalan Institution for Research and Advanced Studies; Pg.~Llu\'is Companys, 23, 08010 Barcelona, Spain}
\newcommand{\INFN}{INFN: National Institute for Nuclear Physics; via Valerio 2, 34127 Trieste, Italy}
\newcommand{\ICSC}{ICSC: Centro Nazionale di Ricerca in High Performance Computing, Big Data e Quantum Computing; via Magnanelli 2, Bologna, Italy}
\newcommand{\ICL}{Astrophysics Group, Physics Department, Blackett Lab, Imperial College London; Prince Consort Road, London SW7 2AZ, UK}

\begin{document}
    \title{\ECHoEZ: Estimating the Hubble constant with ZTF \SNIa DR2}
    \author[0000-0001-9344-736X]{Konstantin Karchev}
    \affiliation{\ICCUB}
    \affiliation{\SISSA}

    \email{kkarchev@icc.ub.edu}

    \author{Marco Giunta}
    \affiliation{\SISSA}
    \email{marco.giunta@sissa.it}

    \author{Raul Jimenez}
    \affiliation{\ICCUB}
    \affiliation{\ICREA}
    \email{raul.jimenez@icc.ub.edu}

    \author{Roberto Trotta}
    \affiliation{\SISSA}
    \affiliation{\ICL}
    \affiliation{\INFN}
    \affiliation{\ICSC}
    \email{rtrotta@sissa.it}

    \author{Licia Verde}
    \affiliation{\ICCUB}
    \affiliation{\ICREA}
    \email{liciaverde@icc.ub.edu}

    \correspondingauthor{Konstantin Karchev}

    \begin{abstract}
        Type~\Ia supernov\ae (\SNaeIa) are instrumental for a precise measurement of the Hubble constant (\Ho) from the local distance ladder, yet the data used hitherto for this purpose are inhomogeneous and subject to severe selection effects. We present the first, preliminary, cosmological analysis of \num{772} \SNaeIa from the \acrlong{ztf}'s second data release: an untargeted single-telescope sample whose volume-limited subset ($\z<\num{0.06}$) is largely free of Malmquist bias. We correct the public data for known instrumental and calibration biases and the redshifts for peculiar motions. Using the \num{42} \SHoES calibrators (their host distances computed from Cepheids and geometric anchors) and a Bayesian hierarchical model of \SNIa standardisation, dust extinction, redshift and calibration uncertainties, we infer $\Ho = \Hores{\dnbase}{\basevariant}$ after marginalising over deceleration with a flat prior (or \Hores{\dnbase}{\basevariant/qprior} if a prior on matter density from \PantheonPlus is imposed), with the uncertainty dominated by the number of calibrators. We demonstrate robustness to the choice of light-curve and Milky Way dust models and to extensions of the model to include peculiar-velocity correlations, a host-mass step, and two stretch populations. These results are in agreement with previous distance-ladder measurements and $\num{2.7}\sigma$ ($\num{4.1}\sigma$) away from the value obtained from {\NoHyper\citetalias{Planck2018-1}} data, suggesting that the resolution of the Hubble tension lies outside of \SNaeIa\Emdash or the distance ladder entirely.
    \end{abstract}

    \section{Introduction}

The Hubble tension is a significant disagreement between values of the Hubble constant, \Ho\Emdash the present-day relative expansion rate of the Universe\Emdash inferred from early-Universe probes, \eg the \gls{cmb} \citep[henceforth \citetalias{Planck2018-1}]{Planck2018-1}, and local observations through the so-called "distance ladder" \citep[henceforth \citetalias{Riess_2022}]{Riess_2022}. Substantial efforts and resources have been devoted both to verifying the local measurement(s) \citep[see \eg][]{H0DNCollaboration_2025} and to resolving the tension through a modification of the physical laws underlying the \gls{cmb} \citep[see \eg][]{Schoneberg_2026,Schoneberg_2026a}.

The key component (providing much of the constraining power) of the distance ladder are type~\Ia supernov\ae (\SNaeIa): white dwarfs that explode under consistent conditions, allowing their intrinsic brightnesses to be inferred (with a residual scatter of \SI{\approx 0.1}{\mag}) from the colour and duration of their \glspl{lc}\Emdash a process known as standardisation\Emdash and used for measuring large cosmological distances (in the so-called \gls{hf}). However, to set the absolute distance scale, represented by $1/\Ho$, one first needs to fix the absolute \emph{brightness} scale by observing so-called \emph{calibrators}: \SNaeIa whose distances are inferable from other probes like Cepheid variables, themselves calibrated from geometric measurements (\eg parallax, masers). As a consequence, any unaccounted-for systematic difference between the \gls{hf} and calibrator \SNaeIa\Emdash be it on the level of the observed/selected populations or the instrumental calibration(s)\Emdash directly translates into a bias in the inferred \Ho.

To limit this possibility, the \citetalias{Riess_2022} programme observed calibrator and \gls{hf} \SNaeIa (the latter comprising the \Foundation sample \citep{Foley_2018,Jones_2019}) with the same instrument (the \acrshort{panstarrs}), thus emphasising their photometric consistency. However, candidates were collected from a variety of sources (predominantly \acrshort{asassn}, \SI{38}{\percent}, and \acrshort{psst}, \SI{20}{\percent}), introducing a complicated selection function based on both brightness and redshift. Importantly, the magnitude limits of both \acrshort{asassn} and \acrshort{psst} fall deeply within the distribution of \SNIa brightnesses, suppressing the dust-obscured tail (as recently pointed out by \citet{Giunta_2026}) and biasing\Emdash in a redshift-dependent way\Emdash the sample towards intrinsically brighter \SNaeIa, from which a larger \Ho would be inferred. While the customary "bias-correction" procedure was applied\footnote{but see \crefext{ext:Brout_2022a:tbl:sims} of \citet{Brout_2022a}, which labels the detection efficiency of \Foundation as "N/A"} by \citetalias{Riess_2022}, it would be valuable to investigate sample (in)completeness as a possible source for the Hubble tension using newer, cleaner data.

In this work, for the first time, we infer \Ho using Hubble-flow \SNaeIa discovered and observed by a single telescope in an untargeted volume-limited survey\Emdash namely, from the second data release of the \acrlong[preunset=global]{ztf} \citep[\ztfdrtwo;][]{Rigault_2025}\Emdash substantially reducing (if not fully eradicating) selection biases. We address the outstanding complications with \ztfdrtwo's photometric calibration (which have thus far deterred the \acrshort{ztf} collaboration from a cosmological analysis) using publicly available information, propagating the reported uncertainties. Nevertheless, since \gls{ztf}-observed calibrators are not yet available, we still rely on the \citetalias{Riess_2022} sample, so our results are contingent both on the Cepheid calibration therein and on the cross-calibration between \gls{ztf} and \acrshort{panstarrs}: any unaccounted-for offset would be directly absorbed by our estimate of \Ho, as we indicate on all our plots.

This paper is structured as follows. In \cref{sec:data}, we present the data we analyse and the steps we take to calibrate it photometrically and correct for peculiar velocities. In \cref{sec:model}, we describe the \gls{bhm} we use. In \cref{sec:results}, we present and discuss our key results and conclude in \cref{sec:conclusion}.

    \section{Data}\label{sec:data}

Cosmological analyses of \SN data (including the present one) are typically performed following a preliminary fitting stage that compresses the raw \acrlongpl{lc} to a small number of \emph{summary parameters}. The \latin{de-facto} standard \SALT template model \citep{Guy_2005,Guy_2007} outputs, for each \SN \snindexlabel, three best-fit parameters: \saltxobssn, \saltcobssn, \mobssn\Emdash "stretch" (or duration), colour at peak, and overall brightness\footnote{Instead of an apparent magnitude (\appmag), \SALT uses a linear scaling parameter (\saltxcomp[0]) to model the spectral flux density (irradiance). We assume the "standard" conversion between them:\begin{equation*}
    \appmag = \num{-2.5}\lg(\saltxcomp[0]) + \SI{10.635}{\mag}
\end{equation*} and the corresponding linear uncertainty transformation, leading to the interpretation of \mobssn as the \SN's rest-frame-\passband{B}-band magnitude in the Vega system.}\Emdash and their covariance matrix \datacovarsn.

\subsection{ZTF DR2}\label{sec:ztfdr2}

The \acrshort{ztf}'s second data release\footnote{publicly accessible from \url{https://ztfcosmo.in2p3.fr/download/data}} contains in total \num{3628} transients, of which \num{2492} are spectroscopically classified as "normal" \SNaeIa, of which \num{1970} have "good" light-curve coverage and fit quality (as defined by \citet{Rigault_2025}), for \num{1956} of which the fit results are actually provided. Based on realistic simulations, \citet{Amenouche_2025} define a "volume-limited" subset that is expected to be free of non-random selection effects (\ie systematic biases in the distributions of observed brightnesses, stretches, and colours) as the objects with measured (heliocentric) redshift $\zobs \le \num{0.06}$, which leaves \num{780} (normal, high-quality) \SNaeIa.

Apart from the \SALT fit results\Emdash obtained with the version of \citet[\citetalias{Taylor_2021}]{Taylor_2021}\Emdash\ztfdrtwo provides measurements \zobssn, with uncertainties \snindex{\sigmaz}, of the \SNae's redshifts (in the heliocentric frame, \ie after correcting for the motion of the Earth around the Sun), derived either from a spectrum of the \SN itself (\SI{22}{\percent}) or from its host galaxy (\SI{78}{\percent}).

\subsubsection{Redshifts and peculiar velocity correction}\label{sec:data-vpec}
The heliocentric redshift \zsn of a \SN arises as a combination of cosmological expansion (\zc) and the relative line-of-sight motion of the emitter with respect to the observer\footnote{and the gravitational redshift due to the masses of the host galaxy and the Milky Way, which we neglect}, with the latter composed of the motions of the Solar System (\zsun), the \SN's host galaxy (\zpec), and the \SN itself\footnote{and the expanding shell of emitting material} (which we neglect, especially since the majority of redshift measurements come from the hosts):
\begin{equation}\label{eqn:redshift}
    \qty(1+\zsn) = \qty(1+\zcsn) \qty(1+\snindex{\zsun}) \qty(1+\snindex{\zpec}).
\end{equation}

While the Sun's motion has been very precisely measured from the \gls{cmb} dipole \citep{Planck2018-1}, the peculiar motions of galaxies are the subject of active research; nevertheless, they can be estimated from the bulk flow expected from the uneven distribution of mass on intergalactic scales, as done by \citet[hereafter \citetalias{Carrick_2015}]{Carrick_2015}. We use the map $\zpecobs(\mathrm{RA}, \mathrm{Dec}, \zc)$ released therewith, which gives the peculiar velocity as a function of sky coordinates ($\mathrm{RA}, \mathrm{Dec}$) and distance from Earth (parametrised by \zc for a fiducial cosmology\footnote{We note that the \citetalias{Carrick_2015} map is in terms of absolute distances, which are used for charting densities and the gravitational potential that gives rise to peculiar velocities. To transform them back to redshift, we use the flat \LCDM cosmology with $\Omo=\num{0.3}$ and $\Ho=\SI{100}{\km\per\second\per\mega\parsec}$ that was assumed when constructing it.}) and an iterative procedure to estimate, as \zcobssn, the cosmological redshift of each \SN, which satisfies:
\begin{equation}\label{eqn:zcobssn}
    1+\zcobssn = \frac{1+\zobssn}{\qty(1+\snindex{\zsun}) \qty(1+\zpecobs(\mathrm{RA}, \mathrm{Dec}, \zcobssn))}.
\end{equation}
Since the \citetalias{Carrick_2015} map only extends to $\zc \lesssim \num{0.07}$, and peculiar velocities are only important below $\zc \lesssim \num{0.03}$, we assume null correction ($\zcobs = \flatfrac{\qty(1+\zobs)}{\qty(1+\zsun)}-1$) for \SNae beyond the map. Still, \num{8} of our \gls{ztf} \SNaeIa fall within "blind spots" of the map, so we exclude them from the analysis. Finally, we augment the redshift-measurement uncertainty with the typical correction uncertainty:
\begin{equation}\label{eqn:varzc}
    \qty(\snindex{\sigmazc})^2 = \qty(\snindex{\sigmaz})^2 + \qty(\frac{\SI{300}{\km\per\second}}{c})^2,
\end{equation}
where $c$ is the speed of light. More rigorous probabilistic treatment, \eg of the correlations among peculiar velocities (see \cref{sec:variants}), does not alter our conclusions.

\subsubsection{Photometric calibration}\label{sec:data-phot}
\cite{Lacroix_2025} point out that the currently available \gls{ztf} \gls{fp} is mis-calibrated by about \SI{0.09}{\mag} due to the lack of \gls{fp} time series of standard stars, and more sophisticated scene-modelling measurements (\acrshort{smp})\glsunset{smp} are required for precision analyses. Furthermore, \gls{ztf} photometry taken after 1 November 2019 is subject to a systematic effect\Emdash dubbed the "pocket effect", of the order of \SI{0.01}{\mag}\Emdash that was not taken into account during data reduction. Both in isolation and in combination the two effects mimic cosmological evolution and can bias inference of \Ho.

The influence of imperfect photometry on inferred \SALT parameters has been well documented by \citet[\crefext{ext:Rigault_2025:fig:pocket}]{Rigault_2025} and \citet[\crefext{ext:Lacroix_2025:fig:offset}]{Lacroix_2025} using realistic simulations. The former gives the average offset between measured and true peak brightnesses and stretches (the pocket effect does not affect colour significantly), while the latter presents averaged differences in \emph{distance} estimates (equivalent to peak-magnitude shifts) $\text{\acrshort{smp}}-\text{\acrshort{fp}}$ as a function of redshift. This allows us to correct\Emdash \emph{on average}\Emdash the (inaccurate) public values:
\begin{align}
    \label{eqn:pocket-x}
    \saltxobssn & \to \saltxobssn - \qty(\Delta_{\saltx}^{\text{pocket}})(\saltxobssn),
    \\
    \label{eqn:pocket-m}
    \mobssn & \to \mobssn - \qty(\Delta_{\appmag}^{\text{pocket}})(\mobssn) + (\text{\acrshort{smp}}-\text{\acrshort{fp}})(\zobssn),
\end{align}
where \citeauthor{Lacroix_2025}'s binned correction is interpolated linearly. Given that our modelling (\cref{sec:model}) places particular emphasis on disentangling noise, systematics, and population scatter, we also consider the reported uncertainties in the $\text{\acrshort{smp}}-\text{\acrshort{fp}}$ correction, which are \SI{<0.01}{\mag} (corresponding to \Hoval{\approx0.3} in \Ho), except for the lowest redshift bin (\SI{0.05}{\mag}); specifically, we marginalise them out as described in \cref{sec:model-photo}. The scatter in the pocket correction is not shown for \appmag by \citet{Rigault_2025}, but it must be smaller than the total size of the effect (\SI{0.01}{\mag}), so we ignore it.

Short of re-implementing the \gls{ztf} data-reduction pipeline, these two corrections are the only practical publicly accessible route towards credible cosmological inference from \ztfdrtwo \SNIa data, and we embark on it, cautioning that newly released data by \gls{ztf} may change the results in detail.

\subsection{Calibrator \SNae (\citetalias{Riess_2022})}\label{sec:calib}
Inferring the Hubble constant from standard(isable) candles requires knowledge of their absolute brightness (\ie \Mo, which is entirely degenerate with \Ho: see \cref{sec:model}) from observations at \emph{known} distances (or from theoretical considerations). To this end, we use the "calibrator" \SNaeIa from the \citetalias{Riess_2022} program\footnote{\label{foot:github-sh0es}\url{https://github.com/PantheonPlusSH0ES/DataRelease}}, for which host distances can be inferred from Cepheid stars, themselves calibrated using parallaxes and other geometric anchors. The formalism for doing this is described in \cref{ext:Riess_2022:sec:formalism} of \citet{Riess_2022}. We follow it (using the available resources in the data release, after removing all entries related to \SNae, \ie using only geometric and Cepheid parameters and observables) to calculate the mean \distmodobscalib and covariance matrix \distmodcovcalib of the distance moduli of the \num{37} distinct calibrator-\SNIa hosts.

For the \num{42} distinct\footnote{The \citetalias{Riess_2022} data release contains \num{73} \glspl{lc} (and fits) since some \SNae have been observed with multiple instruments. In these cases, we take the results with the smallest reported distance-estimate uncertainty as a proxy of the data+fit quality, to avoid the need to model correlations between "different" \SNae. We do not expect this to reduce our constraining power substantially since including multiple \glspl{lc} per object cannot reduce the intrinsic magnitude \emph{scatter}, which ultimately dictates the \Ho uncertainty.} calibrator \SNaeIa themselves, we take the same fit outputs (\mobssn, \saltxobs, \saltcobssn, \datacovarsn) as for \gls{ztf}. Importantly, the data release has an arbitrary $\mathtt{MAG\_OFFSET}$\footnote{stemming from the \SNANA fitting software, which re-scales the \SALT \emph{templates} at runtime, rather than\Emdash as would be more natural\Emdash modifying the zero-point offset in the $\saltxcomp[0]\to\appmag$ conversion}, which we need to apply to the released $\mobssn\to\mobssn+\SI{0.27}{\mag}$ before comparing with the Hubble-flow sample. Finally, we note that a different \SALT model \citep[\citetalias{Brout_2022}]{Brout_2022} was used in this case\Emdash a possible systematic that we address in \cref{sec:datavar}.

    \section{Bayesian hierarchical modelling}\label{sec:model}

To analyse \SNIa \gls{lc}-fit summary data, we use the Simple-\BayeSN hierarchical model \citep{Mandel_2017}, which was recently applied to \ztfdrtwo by \citet{Giunta_2026}. It can be summarised in the following sequence of (forward) sampling operations.

First, sample all global parameters (\num{14} in total, defined in \cref{tbl:model}) from their priors. Then sample "intrinsic" and dust-related parameters for each \SN (labelled \snindexlabel, there being \nsn in total):
\begin{alignat}{2}
    & \text{stretch:}\quad &
        \saltxintsn &\sim \priorsaltxintsn, \label{eqn:saltxintsn}
    \\ & \text{colour:}\quad &
        \saltcintsn &\sim \priorsaltcintsn, \label{eqn:saltcintsn}
    \\ & \text{int.\@ scatter:}\quad &
        \deltaMsn &\sim \Normal(0, \sigmares^2), \label{eqn:deltaMsn}
    \\ &\text{abs.\@ mag.:}\quad &
        \Mintsn &= \Mo + \saltalpha\saltxintsn + \saltbeta\saltcintsn + \deltaMsn, \label{eqn:Mintsn}
    \\ &\text{dust ext.:}\quad&
        \Avsn &\sim \priorAv, \label{eqn:Avsn}
    \\ &\text{dust law:}\quad&
        \Rvsn &\sim \priorRv, \label{eqn:Rvsn}
\intertext{convert those to \enquote{extrinsic} through the action of dust:}
    && \saltxextsn &= \saltxintsn,
    \\ && \saltcextsn &= \saltcintsn + \flatfrac{\Avsn}{\Rvsn},
    \\ && \Mextsn &= \Mintsn + \flatfrac{\Avsn \qty(\Rvsn+1)}{\Rvsn},
\intertext{apply cosmological distance and photometric calibration (see \cref{sec:model-cosmo,sec:model-photo}):}
    & \text{app.\@ mag.:}\quad &
        \msn &= \Mextsn + \distmodsn + \snindex{\delta m}, \label{eqn:msn}
\intertext{and add fitting noise to arrive at the recorded data:}
    && \mathllap{\tuple{\saltxobssn, \saltcobssn, \mobssn}} &\sim \Normal(\tuple{\saltxextsn, \saltcextsn, \msn}, \datacovarsn).
\end{alignat}

\subsection{Distances and cosmology}\label{sec:model-cosmo}

For calibrator \SNae, direct estimates of the distance moduli are available through those of their hosts:
\begin{align}
    \label{eqn:distmod-calib}
    \distmodcalib &\sim \Normal(\distmodobscalib, \covdistmod),
    \\ \snindexlabel\in\text{calibrators:}\quad
    \distmodsn & = \distmodcalib[\text{host of \snindexlabel}].
\intertext{For \SNae in the \acrlong{hf}, a latent cosmological redshift is sampled (see \cref{eqn:zcobssn,eqn:varzc}) and used to calculate the distance modulus using a cosmological model with the parameters \cosmoparams sampled before:}
    \snindexlabel\in\text{\acrlong{hf}:}\quad
    \label{eqn:zcsn}
    \zcsn &\sim \Normal(\zcobssn, \qty(\snindex{\sigmazc})^2),
    \\ \label{eqn:distmod-hf}
    \distmodsn &= \num{5} \lg[\frac{\dlum(\zcsn, \cosmoparams)}{\SI{10}{\parsec}}],
\end{align}
where \dlum is the luminosity distance.

Given that we are analysing low-redshift data, which is largely insensitive to cosmic expansion (see \cref{sec:results}), we will adopt a simple phenomenological model \citep[see][\crefext{ext:Weinberg_2008:sec:Taylor-cosmo}]{Weinberg_2008} parametrised by the Hubble constant \Ho and the present-day deceleration\footnote{Assuming flatness and \LCDM, it can be related to the cosmic matter density via $\Omo = \tfrac{2}{3} \qty(1+\cosmodecelo)$.} \cosmodecelo:
\begin{equation}
    \cosmoparams \equiv \tuple{\Ho, \cosmodecelo},
\end{equation}
according to which
\begin{equation}\label{eqn:dlum}
    \dlum(\zc, \cosmoparams) = \frac{c}{\Ho} \qty[\zc + \frac{1-\cosmodecelo}{2}\times\zc^2 + \order{\zc^3}],
\end{equation}
with $c$ again the speed of light.

It is obvious from \cref{eqn:msn,eqn:distmod-calib,eqn:distmod-hf,eqn:dlum}\Emdash and well known in the literature\Emdash that the Hubble constant is entirely degenerate with a (constant/global) magnitude offset ($\Delta\appmag$) between \gls{hf} and calibrator \SNae. Such may arise within the observed population if objects on the two rungs are differently selected or from differing instrumental or pipeline calibrations (\eg \SNANA's \texttt{MAG\_OFFSET}). While we assume $\Delta\appmag=0$, we show in plots the $\Delta\appmag$ required to shift the inferred \Ho to a given value.

\begin{table}\centering
    \caption{Global model parameters and their priors. $\Uniform$ denotes a uniform distribution within the given range, and $\InvGammaDist$ an inverse-gamma with specified shape and scale.\label{tbl:model}}%
    \begin{tabular}{@{\hspace{1em}}r@{\hspace{0.5em}}rlL@{\hspace{1em}}}%
        \toprule
        & \multicolumn{2}{c}{parameter} & \text{prior} \\
        \toprule
        \multirow{2}{*}{stretch pop.}
            & mean & \meansaltx & \Uniform(\num{-3}, \num{3}) \\
            & st.\@ dev. & \stdsaltx & \priorstdsaltx \\
        \tableline
        \multirow{2}{*}{colour pop.}
            & mean & \meansaltc & \Uniform(\num{-0.3}, \num{0.3}) \\
            & st.\@ dev. & \stdsaltc & \Uniform(\num{0.001}, \num{0.3}) \\
        \tableline
        \multicolumn{2}{r}{standard magnitude}
            & \Mo & \priorMo \\
        \multicolumn{2}{r}{residual mag.\@ scatter}
            & \varres & \InvGammaDist(\num{0.003}, \num{0.003}) \\
        \tableline
        \multicolumn{2}{r}{\multirow{3}{*}{\shortstack[r]{standardisation\\coefficients}}}
            & \saltalpha & \priorsaltalpha \\
            && \saltalphac & \Uniform(\num{-0.03}, \num{0.03}) \\
            && \saltbeta & \Uniform(\num{-4}, \num{4}) \\
        \tableline
        dust-ext.\@ pop.
            & mean & \bsntau & \Uniform(\num{0.001}, \num{0.5}) \\
        \multirow{2}{*}{dust-law pop.}
            & mean & \meanRv & \priormeanRv \\
            & st.\@ dev. & \sigmaRv & \Uniform(\num{0.01}, \num{2}) \\
        \tableline
        \multicolumn{2}{r}{\multirow{2}{*}{cosmology (\cosmoparams)}}
            & \cosmoh & \Uniform(\num{0.5}, \num{0.9}) \\
            & & \cosmodecelo & \Uniform(\num{-2}, \num{1}) \\
        \bottomrule
    \end{tabular}
\end{table}

\subsection{Photometric calibration}\label{sec:model-photo}
\pgfplotstableread{plotdata/ztf-offset.txt}{\ztfofftbl}

The corrections in \cref{eqn:pocket-x,eqn:pocket-m} undo the (average) bias due to inaccurate photometry but leave residual \emph{correlated} errors in the individual objects, the largest part coming from the uncertainty in $\text{\acrshort{fp}}\to\text{\acrshort{smp}}$ \citep[\crefext{ext:Lacroix_2025:fig:offset}]{Lacroix_2025}. We model it using three nuisance variables (marginalised out during inference) to represent the offset of the "true" correction from the mean estimate within the first three redshift bins (located at $\z\approx\numlist[list-final-separator={\text{, and }}]{0.013;0.034;0.054}$):
\begin{equation}
    \variable{\delta m} \sim \Normal(\variable{0}, \qty[\sisetup{round-precision=2,round-mode=places}
        \pgfplotstablegetelem{0}{err}\of{\ztfofftbl}
        \num{\pgfplotsretval}^2,
        \pgfplotstablegetelem{1}{err}\of{\ztfofftbl}
        \num{\pgfplotsretval}^2,
        \pgfplotstablegetelem{2}{err}\of{\ztfofftbl}
        \num{\pgfplotsretval}^2
    ])\,\si{\mag}.
\end{equation}
We then interpolate between the $\variable{\delta m}$ at the redshift \zobssn of each \SN to calculate its individual $\snindex{\delta m}$ (which is a deterministic rather than random latent variable).

\subsection{Posterior sampling with \acrlong{hmc}}
We implement \SimpleBayeSN in the \numpyro \citep{numpyro} probabilistic programming "language", which is based on \jax \citep{jax} and can automatically\Emdash and efficiently\Emdash calculate the gradients required for \acrshort[postunset=global]{nuts} \citep{NUTS}, a variant of \gls[prereset=global]{hmc} \citep{HMC}. This way, we can sample the full $\approx 6\nsn + \num{17} \gtrsim \num{4600}$-dimensional posterior using arbitrary analytical prior, likelihood, and model structures. We run \num{4} chains in parallel for \num{1000} draws each (after \num{1000} steps of burn-in), then check the quality of the posterior approximation using the $\hat{R}$ statistic \citep{Gelman_1992,Vehtari_2021} and effective sample size using the \arviz library \citep{arviz}.

    \section{Results and discussion}\label{sec:results}

The marginal posterior for the cosmological parameters is shown in \cref{fig:corner} in purple. We obtain $\Ho\approx\Hores{\dnbase}{\basevariant}$ and only weakly constrain the deceleration parameter to $\cosmodecelo\approx\num{0.2+-0.5}$ (implying $\Omo\approx\num{0.77+-0.34}$ for a spatially flat \LCDM model). A mild correlation is present whereby a smaller Hubble constant can be compensated by faster deceleration: a unit increase in \Omo shifts \Ho by \Hoval{\approx\getdata{qhslope}}. Breaking this degeneracy requires external higher-redshift data: adopting the constraint from \PantheonPlus \citep{Brout_2022a} of $\Omo\approx\num{0.33+-0.02} \implies \cosmodecelo\approx\num{-0.5+-0.03}$ reduces the uncertainty and results in \Hores{\dnbase}{\basevariant/qprior}.

A corner plot with all global parameters is shown in \cref{fig:corner-all} and does not reveal any further significant correlations involving the cosmological parameters, beyond the degeneracy between \Ho and \Mo, which is broken by the calibrator \SNaeIa.

\begin{figure}
    \centering%
    \tikzsetnextfilename{\currfilebase}%
\begin{tikzpicture}[
    trim axis group left, trim axis group right,
    truth line/.style={fiducial, thick, dashed}
]
    \input{plotdata/corner}

    \pgfplotsset{
        scale only axis, width={0.7\linewidth}, height={0.5\linewidth}, param axis={h},
        marginal ax/.style={height={0.18\linewidth}, axis y line=none}
    }

    \newcommand{\amarg}[3][]{
        \addplot[#1] coordinates {
            (0.01*\getdata{#2/mean}, #3) +- (0.01*\getdata{#2/std}, 0)
        };
        \addplot[#1, error bars/error bar style={densely dashed}] coordinates {
            (0.01*\getdata{#2/qprior/mean}, #3+0.4) +- (0.01*\getdata{#2/qprior/std}, 0)
        };
    }

    \begin{groupplot}[
        group style={rows=2, xlabels at=edge bottom, vertical sep=1.75em},
        param axis extents={h}, H0 axis,
        minor x tick num=4, xlabel near ticks,
    ]
        \nextgroupplot[
            2 param axis extents={h}{q0},
            minor y tick num=4,
            ylabel={deceleration, \getparamlabel{q0}},
            axis x line*=bottom, xticklabel style={font=\scriptsize},
        ]
            \vstrip[Planck, h result]{0.674}{0.005}
            \vstrip[SH0ES, h result]{0.73}{0.01}
            \axvline[fiducial, model]{\fidh};

            \acornerpanel*{h}{q0}
            \axhline[densely dashed, thick]{-0.5};

        \nextgroupplot[
            marginal ax, ymin=0.5, ymax=5, y dir=reverse,
            variants axis, xtick pos=both,
            xlabel style={inner ysep=0pt, anchor=north, at={(ticklabel cs:0.5)}},
            legend style={at={(rel axis cs:0, 0.5)}, anchor=center, xshift=2em},
            filled legend
        ]

            \amarg[model]{\dnbase/\basevariant}{1}
            \amarg[late]{\dnlate/\basevariant}{2}
            \amarg[model, hz]{\dnhz/\basevariant}{3}
            \amarg[late, hz]{\dnlatehz/\basevariant}{4}

            \addlegendimage{color=black}
                \addlegendentry{$\cosmodecelo\sim\Uniform(\num{-2}, \num{1})$}
            \addlegendimage{color=black, /pgfplots/every error bar/.append style={densely dashed}}
                \addlegendentry{$\cosmodecelo\sim\Normal(\num{-0.50}, \num{0.03}^2)$}
%
%
%
    \end{groupplot}

    \begin{axis}[
        at={(group c1r1.south west)},
        axis x line*=top, xlabel style={name={top xlabel}},
        H0 as magoff={\getparamlo{h}}{\getparamhi{h}},
        axis y line*=right,
        y coord trafo/.code={\pgfmathparse{1.5*(#1)-1}},
        y coord inv trafo/.code={\pgfmathparse{2*(1+(#1))/3}},
        ymin={2*(1+(\getparamlo{q0}))/3}, ymax={2*(1+(1))/3},
        ytick pos=right, ytick={0,0.2,...,1.4},
        extra y ticks as minor={0.1,0.3,...,1.3},
        ylabel={(assuming flatness)\\\acrshort*{dm} density, \Omo},
        ylabel style={rotate=180, align=center, inner ysep=0pt},
    ]
        \fill[black!60, fill opacity=0.5] (rel axis cs:0, 0) rectangle ({{rel axis cs:1, 0}} |- {{0, 0}});
    \end{axis}



    \ztflegend[matrix anchor=south west, yshift=0.5em, draw=none]{({{group c1r1.west}} |- {{top xlabel.north}})}
    \matrix[/pgfplots/every axis legend, matrix anchor=south east, draw=none, name=prev legend] at ({{ztf legend.south}} -| {{group c1r1.east}}) {
        \pgfplotsset{vstrip legend, legend image code={Planck, h result}}
        &
        \node {\citetalias{Planck2018-1}};
        \\
        \pgfplotsset{vstrip legend, legend image code={SH0ES, h result}}
        &
        \node {\citetalias{Riess_2022}};
        \\
    };
    \node[draw, inner sep=0pt, fit={(ztf legend) (prev legend)}] {};
\end{tikzpicture}%
    \caption{Posteriors for the cosmological parameters from: the full \acrshort{ztf} volume-limited sample before (\acrshort{fp}, orange) and after photometric corrections (\acrshort{smp}, purple); the subset with late-type hosts (blue); and respective subsets with \highz (unfilled purple and blue). \emph{Top}: $1\sigma$ and $2\sigma$ \acrlong{hpd} contours, with \SI{39}{\percent} and \SI{86}{\percent} credibility, respectively. \emph{Bottom}: marginal posterior $\text{mean}\pm1\,\text{st.\@ dev.}$ for \Ho, where dashed error bars use the external constraint on \cosmodecelo. Red and green vertical bars indicate the \Ho constraints from \citetalias{Planck2018-1} and \citetalias{Riess_2022}, respectively. Labels along the top show the magnitude shift (of \acrshort{ztf} \SNae) that would cause \emph{our} results to move to a given \Ho.\label{fig:corner}}
\end{figure}

\begin{figure}
    \centering%
    \hspace{-0.6\columnsep}\input{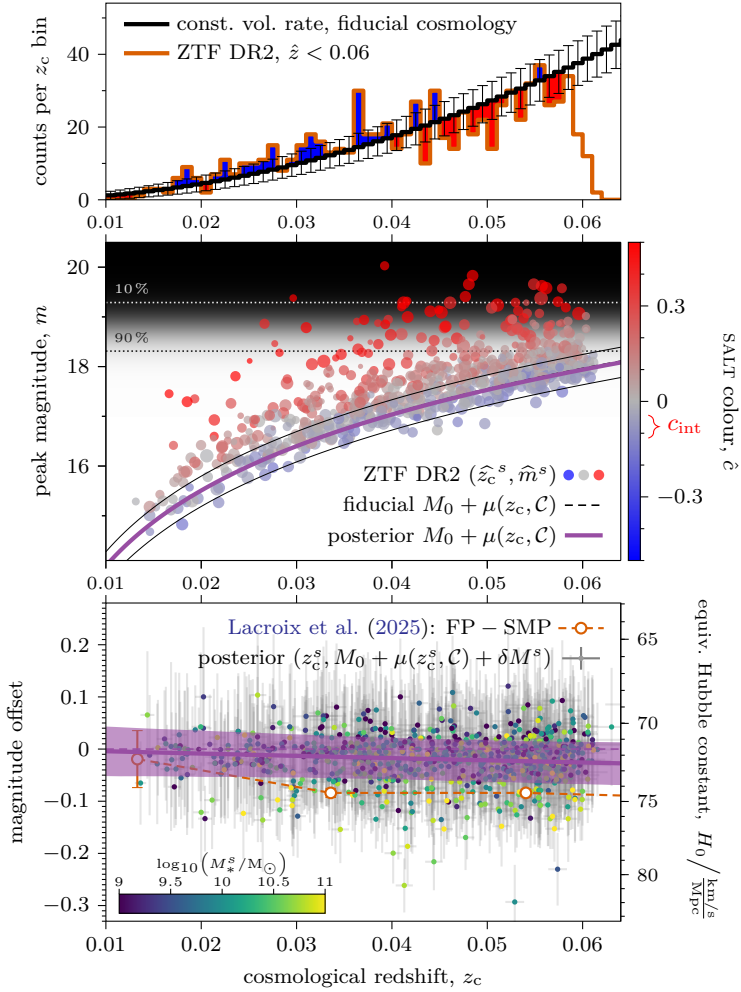}%
    \caption{The \ztfdrtwo data and our cosmological fit to it. \emph{Top:} Histogram of the cosmological redshifts (after the peculiar-velocity correction from \cref{sec:data-vpec}) in orange and the prediction (Poisson rate and corresponding standard deviation) from a constant-volumetric-rate model (under the fiducial cosmology) in black. The difference between the two is filled in blue (observed excess) and red (observed deficit). Note that \zcsn extend slightly beyond \num{0.06} since the redshift cut is applied to the measured values \zobssn. \emph{Middle:} Raw Hubble diagram (\SALT peak magnitude \vs redshift), with the colour and size of the points indicating measured \SALT colours and stretches (\saltcobssn and \saltxobssn), respectively. We also plot the inferred distance modulus relation ($1\sigma$ uncertainties of the latter are within the size of the line), overlapping the fiducial one. The background shading depicts the detection probability given by \citet[\crefext{ext:Rigault_2025:sec:selection}]{Rigault_2025}. \emph{Bottom:} A zoomed-in Hubble diagram offset by the fiducial distance modulus (extent indicated by thin lines in the middle panel) showing standardised magnitudes for all \SNae (marginal posterior means $\pm$ st.\@ dev.\@) as error bars and the inferred distance modulus relation (mean $\pm$ st.\@ dev.) as a purple band. Finally, the offset between \glsxtrfmt{fp}{forced} and \acrlong{smp} from \citet[\crefext{ext:Lacroix_2025:fig:offset}]{Lacroix_2025}\Emdash which we have corrected (\cref{sec:data-phot})\Emdash is shown as orange points with error bars (where substantial).\label{fig:hd}}
\end{figure}

In \cref{fig:hd}, we examine the inference results in terms of the individual \gls{ztf} \SNaeIa\Emdash as observed and after standardisation\Emdash and compare with a fiducial flat \LCDM cosmology with $\Omo=\num{\fidOmo}$ and the posterior mean $\Ho=\Hoval{\getdata{\dnbase/\basevariant/mean}}$. We observe mild signs of a trend towards brighter objects at higher redshifts, which can be explained by a Malmquist bias caused by missing highly reddened \SNae: this is corroborated by \citet{Amenouche_2025}, whose \crefext{ext:Amenouche_2025:fig:bias} shows a $2\sigma$ discrepancy at $\zobs = \num{0.06}$ between the input colour distribution and that recovered after selection. Further support for this interpretation comes from the increasing deficit\Emdash with respect to a constant-volumetric-rate prediction\Emdash in the observed counts at increasing redshift, as illustrated in the top panel of \cref{fig:hd}.

Nevertheless, at our current precision, the missing red \SNae do not affect the result for \Ho, owing to the small size of the \emph{intrinsic} scatter\Emdash the colour-correlated scatter correctly undone by standardisation even in the presence of non-random selection\Emdash in comparison with the transition between detected and missed objects (\SI{\approx1}{\mag} \citep{Rigault_2025}, as shown in \cref{fig:hd}). Indeed, restricting the \gls{ztf} sample to $\zobs < \num{0.05}$ only slightly \emph{increases} \Ho by \Hoval{\approx0.3}, fully within the uncertainties.

\subsection{Data variants}\label{sec:datavar}

\paragraph{Photometry corrections}
We confirm that failing to correct for non-\acrlong{smp} significantly biases the cosmological results: the corresponding contours are shown in \cref{fig:corner} (labelled "\labeluncorr{}") and indicate a \SI{+2}{\kilo\meter\per\second\per\mega\parsec} shift in \Ho and a best-fit \Omo in excess of \num{1}. This can be easily understood in terms of the redshift dependence of the correction (shown as orange points in \cref{fig:hd} (bottom)), which resembles a recently decelerated Universe and is thus primarily absorbed by a bias in \cosmodecelo, whereas \Ho is affected (to a much lesser extent) only by the \emph{average} offset. Consequently, keeping the mean correction without propagating its uncertainty (as in \cref{sec:model-photo}) shrinks the final error bars by \Hoval{0.2} but does not modify the mean inferred \Ho.

\paragraph{Low-redshift uncertainties}
At low redshifts, both corrections we have applied to the \gls{ztf} \SNae (to obtain their cosmological redshifts and calibrated brightnesses) bear substantial uncertainties. As a robustness check, we analyse the subset of \num{\getdata{Nhf/highz}} \gls{hf} objects with \highz, resulting in a mere \Hooff[round-precision=2]{\dnhz}{\basevariant} (\Horoff{\dnhz}{\basevariant}) shift. Notably, it comes with a corresponding shift towards lower \cosmodecelo (\ie it is along the correlated direction in the \Ho{\Endash}\cosmodecelo plane), so it can be interpreted as arising from the reduced statistics rather than systematic mis-correction.

\paragraph{Calibrator selection}

\begin{figure}\centering%
    \tikzsetnextfilename{\currfilebase}%
\begin{tikzpicture}[trim axis left, trim axis right]
    \input{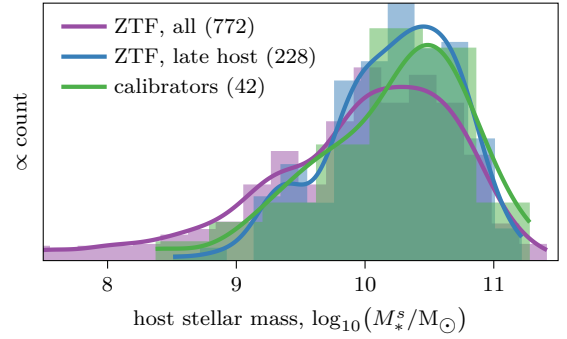}
    \newcommand{\aplot}[2][]{
        \path[name path=hist #2] plot coordinates {\getdata{masses/#2/hist}};
        \addplot[#1, draw=none, fill, fill opacity=0.5, forget plot] fill between[of=axis and hist #2, soft clip/.expanded={domain=\getdata{masses/#2/range}}];
        \addplot[#1] coordinates {\getdata{masses/#2/kde}};
    }
    \begin{axis}[
        scale only axis, width={0.8\linewidth}, height={0.4\linewidth},
        xmin=7.5, xmax=11.5, xlabel={host stellar mass, \getparamlabel{Mstar}},
        ymin=0, ytick=\empty, ylabel={$\propto \text{count}$},
        legend pos=north west, legend style={inner sep=0pt, draw=none, fill=none}
    ]
        \axhline[name path=axis, draw=none]{0};
        \aplot[model]{full}
            \addlegendentry{\acrshort{ztf}, all (\num{\getdata{Nhf/base}})}
        \aplot[late]{late}
            \addlegendentry{\acrshort{ztf}, late host (\num{\getdata{Nhf/late}})}
        \aplot[color=Set1-green]{calib}
            \addlegendentry{calibrators (\num{\getdata{Ncalib}})}
    \end{axis}
\end{tikzpicture}%
    \caption{Distributions of host masses for the full and "late-type" \ztfdrtwo samples, compared to that of the calibrators.\label{fig:masses}}
\end{figure}

Measuring \Ho relies on the consistency between the calibrator and Hubble-flow \SNae\Emdash formally, the distributions of their intrinsic brightnesses, conditional on values of the latent parameters that are informed by data. It has been repeatedly pointed out \citep[see][in the context of \gls{ztf}]{Ginolin_2025a,Giunta_2026} that host properties (total stellar mass, metallicity, \acrlong{sfr}, \etc) are correlated with \SNIa brightness\Emdash generally, high-mass galaxies tend to host faster (lower stretch, hence higher $\saltalpha\saltx$, \ie dimmer) and intrinsically brighter (\ie with lower \deltaM) \SNaeIa. Host properties can therefore act as confounders if the host population changes with redshift, or if the hosts of calibrators\Emdash which have been selected for the presence of Cepheids and are thus star-forming late-type galaxies\Emdash are different from those of the Hubble-flow \SNaeIa.

To investigate this effect, we attempt to recreate the Cepheid-host selection\Emdash "a [non-inclined] spiral type in the full range of Sa{\Endash}Sd" \citep{Riess_2022}\Emdash using morphological indicators from Galaxy Zoo \acrshort{desi} \citep{Walmsley_2023a}, which was used for an environmental study within the \ztfdrtwo \citep{Senzel_2025}. To select non-inclined late-type galaxies, \citet{Garland_2024} recommend the conditions $\mathtt{SFFDF}\ \text{("has features or disk")} \ge \num{0.27}$ and $\mathtt{DEONo}\ \text{("not edge-on")} \ge \num{0.68}$, fulfilled by \num{\getdata{Nhf/late}} hosts out of \num{507} successful (position-based) cross-matches. As shown in \cref{fig:masses}, the resulting stellar-mass distribution is in better agreement with that of the calibrator hosts than the full \gls{ztf} sample (Kolmogorov{\Endash}Smirnov $p$-value of \num{0.76} \vs \num{0.1}).

Analysing this "late-type host" sample, we obtain $\Ho \approx \Hores{\dnlate}{\basevariant}$, a \Horoff{\dnlate}{\basevariant} shift with respect to the un-host-matched data set. In the \Ho{\Endash}\cosmodecelo plane, the shift lies away from the primary degeneracy direction (see \cref{fig:corner}), indicating a systematic rather than statistical origin. If we further restrict the data to high-redshift objects (\highz, leaving \num{\getdata{Nhf/latehz}} \SNaeIa), \Ho increases further to \Hores{\dnlatehz}{\basevariant}, but as before the effect is consistent with that expected from data-set statistics.

\paragraph{Light-curve modelling}
Lastly, we check whether a systematic magnitude offset has been introduced by the difference in the \gls{lc}-fitting methodologies: while the \citetalias{Riess_2022} calibrators were fitted using the \SNANA \citep{SNANA} software, the bespoke \citetalias{Brout_2022} \SALT model, and the \citet[\citetalias{Fitzpatrick_1999}]{Fitzpatrick_1999} Milky Way dust law, \ztfdrtwo used \sncosmo \citep{sncosmo}, the earlier \citetalias{Taylor_2021} light-curve templates and the \citet[\citetalias{Cardelli_1989}]{Cardelli_1989} Milky Way dust law. We re-fit consistently \emph{all} \SNae (calibrators and the \gls{ztf} \acrlong{hf}) using \sncosmo and each of the four \SALT{}+extinction choices and infer \Ho as before. We obtain minimally\Emdash up to $\num{0.2}\sigma$\Emdash different values, listed in \cref{tbl:datavar}: concretely, the choice of a Milky Way extinction model can lead to $\Delta\Ho \approx \Hoval{0.4}$, whereas the employed \SALT version has a much smaller effect. We conclude that using the publicly released values does not represent a significant systematic, especially given the current photometric calibration uncertainties of \gls{ztf}.

\subsection{Analysis variants}

\Cref{sec:model} presented our "baseline" \gls{bhm}, which was used for all results discussed thus far. To test their sensitivity to \emph{modelling} choices, we now consider three extensions (described in detail in \cref{sec:variants}): including peculiar-velocity correlations from a \gls{lss} prior or a full Bayesian analysis of the local neighbourhood; allowing for a "mass step" (\ie an offset in intrinsic brightnesses of \SNae hosted in high- or low-mass galaxies, split at fixed stellar mass of \SI[unit-mode=math]{e10}{\Msun}); and allowing for two stretch populations (unrelated to the host) with separate \Mo and \saltalpha. These are meant to improve robustness at low redshifts and when the calibrator and \gls{hf} hosts are differently selected, two scenarios we countered above by restricting the data. We show the effect of individual modifications on the inferred value of \Ho in \cref{tbl:variants}, with the results from all \num{12} model configurations presented in \cref{fig:variants}.

\begin{table}\centering
    \caption{Offsets in inferred \Ho (absolute and in terms of the baseline posterior uncertainty of \Hoval{\getdata{\dnbase/\basevariant/std}}) for different choices in fitting the \acrlongpl{lc}.\label{tbl:datavar}}%
    \newcommand{\ares}[2]{\getdata{pplus-cdd_#1+ztfdr2-vlhq_#1-pcorr-sm/\basevariant/#2}}%
    \sisetup{table-format=+1.2, round-mode=places, round-precision=2, print-implicit-plus}%
    \begin{tabular}{llSS}
        \toprule
        \multicolumn{2}{c}{model variant} & \multicolumn{2}{c}{$\Delta\Ho$} \\
        \SALT & \acrshort{mw} dust & {$\Hounit$} & {$\sigma_{\Ho}$} \\
        \midrule
        \citetalias{Brout_2022} & \citetalias{Cardelli_1989}
        & \ares{B22-CCM89}{off} & \ares{B22-CCM89}{roff} \\
        \citetalias{Brout_2022} & \citetalias{Fitzpatrick_1999}
        & \ares{B22-F99}{off} & \ares{B22-F99}{roff} \\
        \citetalias{Taylor_2021} & \citetalias{Cardelli_1989}
        & \ares{T21-CCM89}{off} & \ares{T21-CCM89}{roff} \\
        \citetalias{Taylor_2021} & \citetalias{Fitzpatrick_1999}
        & \ares{T21-F99}{off} & \ares{T21-F99}{roff} \\
        \bottomrule
    \end{tabular}
\end{table}

We find that modelling correlated peculiar velocities tends to decrease the Hubble constant by \Hoval{-0.5} (or even \Hoval{-1} when only considering \SNae with \highz), consistently with the findings of \citet{Carreres_2025a} in realistic \gls{ztf} simulations. The shift is again along the degeneracy direction with \cosmodecelo and is accompanied by an increased \sigmares, indicating that a peculiar-velocity uncertainty of \SI{300}{\km\per\second} may have been overestimated and incorrectly absorbed some of the intrinsic scatter. Using only \apriori correlations from \gls{lss}, without correcting for the bulk flow, increases the \Ho uncertainty to \Hoval{\approx 1.8}.

On the contrary, modelling a mass step tends to increase the inferred \Ho by \Hoval{+0.5}, bringing the result close to the no-step analysis of late-type-hosted \SNae. However, the shift is also present for the host-matched subset, which would not be the case if a mass step could fully account for the host dependence of \SNIa brightnesses. Finally, including a second stretch population does limit the effect (on \Ho) of matching host properties, indicating that the primary host-related effect is on the distribution of stretches, which affects \Ho through the different correlation coefficients in the two sub-populations \citep[see][]{Ginolin_2025a}.

Consequently, the most extended model with the least assumptions (\gls{lss} \vpec prior, step, and two stretch populations) appears to be the most robust to both redshift cuts and host matching, so we designate it "fiducial". Interestingly, it infers \Hores{pplus-cdd_nom+ztfdr2-vlhq_nom-pcov-sm}{hqj-latz-distR-AV-2x-step}, in agreement with our baseline. Similarly, the most advanced\Emdash"frontier"\Emdash model (probabilistically constraining the bulk flow) infers \Hores{pplus-cdd_nom+ztfdr2-vlhq_nom-pborg-sm}{hqj-latz-distR-AV-2x-step} from the full volume-limited \ztfdrtwo \acrlong{hf} and \Hores{pplus-cdd_nom+ztfdr2-vlhq_nom-pborg-sm+late+highz}{hqj-latz-distR-AV-2x-step} from only late-type-hosted \SNae with \highz.

\begin{table*}\centering
    \caption{Marginal constraints on \Ho (in \Hounit) from the four Hubble-flow sample variants using the baseline model (first row) and after modifying a single modelling aspect (subsequent rows) as described in \cref{sec:variants}. The last columns give the change to \Ho due to each change. The results using all \num{12} distinct configurations and from all \num{4} data variants are shown in \cref{fig:variants}.\label{tbl:variants}}
    \newcommand{\ares}[2]{\getdata{pplus-cdd_nom+ztfdr2-vlhq_nom-#1/#2/mean}+-\getdata{pplus-cdd_nom+ztfdr2-vlhq_nom-#1/#2/std}}
    \renewcommand{\repeat}[1]{{\color{black!50}\NoHyper#1}}
    \newcommand{\change}[1]{#1}
    \sisetup{round-mode=uncertainty,table-alignment-mode=format}
    \begin{tabular}{cccc *{4}{S[table-format=2.1(2)]} l}
        \toprule
        & \multicolumn{3}{c|}{modelling aspect} & \multicolumn{4}{c}{data variant}
        & \multicolumn{1}{|c}{\multirow{2}{*}{\makecell{
            overall\\effect%
            \\($\Delta\Ho/\Hounit$)
        }}}
        \\
        & \makecell{peculiar\\velocities} & \makecell{host-mass\\step} &
        \multicolumn{1}{c|}{\makecell{stretch\\pop.}} &
        {all} & {late host} & {\highz} & {\makecell{late host,\\\highz}}
        & \multicolumn{1}{|c}{}
        \\
        \midrule
        baseline &
        \getdata{lbls/pc/pcorr} & \getdata{lbls/step/nostep} & 1\saltx &
        \ares{pcorr-sm}{hqj-latz-distR-AV-1x} &
        \ares{pcorr-sm+late}{hqj-latz-distR-AV-1x} &
        \ares{pcorr-sm+highz}{hqj-latz-distR-AV-1x} &
        \ares{pcorr-sm+late+highz}{hqj-latz-distR-AV-1x} &
        \\
        \cline{1-8}
        \multirow[c]{4}{*}{\makecell{single-\\change\\variants}}
        &
        \change{\getdata{lbls/pc/pcov}} & \repeat{\getdata{lbls/step/nostep}} & \repeat{1\saltx} &
        \ares{pcov-sm}{hqj-latz-distR-AV-1x} &
        \ares{pcov-sm+late}{hqj-latz-distR-AV-1x} &
        \ares{pcov-sm+highz}{hqj-latz-distR-AV-1x} &
        \ares{pcov-sm+late+highz}{hqj-latz-distR-AV-1x} &
        \num{-0.5} to \num{-1} \rlap{(high-\zcobs)}
        \\
        &
        \change{\getdata{lbls/pc/pborg}} & \repeat{\getdata{lbls/step/nostep}} & \repeat{1\saltx} &
        \ares{pborg-sm}{hqj-latz-distR-AV-1x} &
        \ares{pborg-sm+late}{hqj-latz-distR-AV-1x} &
        \ares{pborg-sm+highz}{hqj-latz-distR-AV-1x} &
        \ares{pborg-sm+late+highz}{hqj-latz-distR-AV-1x} &
        \num{-0.5} to \num{-1} \rlap{(high-\zcobs)}
        \\
        &
        \repeat{\citetalias{Carrick_2015} corr.} & \change{step} & \repeat{1\saltx} &
        \ares{pcorr-sm}{hqj-latz-distR-AV-1x-step} &
        \ares{pcorr-sm+late}{hqj-latz-distR-AV-1x-step} &
        \ares{pcorr-sm+highz}{hqj-latz-distR-AV-1x-step} &
        \ares{pcorr-sm+late+highz}{hqj-latz-distR-AV-1x-step} &
        \num{+0.5}
        \\
        &
        \repeat{\citetalias{Carrick_2015} corr.} & \repeat{\getdata{lbls/step/nostep}} & \change{2\saltx} &
        \ares{pcorr-sm}{hqj-latz-distR-AV-2x} &
        \ares{pcorr-sm+late}{hqj-latz-distR-AV-2x} &
        \ares{pcorr-sm+highz}{hqj-latz-distR-AV-2x} &
        \ares{pcorr-sm+late+highz}{hqj-latz-distR-AV-2x} &
        \num{\approx 0} to \num{-0.4} \rlap{(late)}
        \\
        \bottomrule
    \end{tabular}
\end{table*}

    \newpage
\section{Conclusion}\label{sec:conclusion}

We have analysed the volume-limited subset of the \ztfdrtwo \SNaeIa, in combination with calibrators from \citetalias{Riess_2022}, to obtain a new measurement of the Hubble constant: \Hores{\dnbase}{\basevariant} marginalising over cosmological expansion, or \Hores{\dnbase}{\basevariant/qprior} with a prior on \Omo from \PantheonPlus. Since the data we used have not yet been validated for cosmological inference by the \gls{ztf} team, we apply to it the best corrections we can construct from publicly available information: predominantly, a \SI{\approx 0.09}{\mag} photometric re-calibration (equivalent to a \Hoval[round-mode=none]{3} shift in \Ho), whose uncertainty (amounting to \Hoval{0.2}) we have propagated. We have also tested the robustness of our result by restricting the data to trusted objects (in terms of photometry and peculiar velocities), matching the distribution of host properties to those of the calibrators, and extending the hierarchical modelling, leading to a slight increase in uncertainty (to \Hoval{\approx1.8}) and no significant shift of the mean. Lastly, we have investigated the impact of the choice of a \SALT model and Milky Way extinction law and concluded that their effect is sub-dominant at present.

These findings are fully consistent ($\num{<1}\sigma$) with previous distance-ladder estimates and away from \citetalias{Planck2018-1} values by $\num[round-mode=figures,round-precision=2]{\getdata{tension/Planck}}\sigma$ ($\num[round-mode=figures,round-precision=2]{\getdata{tension-qprior/Planck}}\sigma$ with the \Omo prior). More recent early-Universe analyses \citep[\eg][]{DESICollaboration_2025} are only slightly more consistent ($\numrange[range-phrase=\Endash]{\approx 2}{3}\sigma$). The quoted statistical uncertainties are dominated by the number of calibrators (with a secondary contribution from the \gls{ztf} photometric calibration), so they are similar to those of \citetalias{Riess_2022}.

The present analysis has two major advantages over previous ones. Firstly, we use a truly homogeneous sample of Hubble-flow \SNaeIa, detected and observed with the same telescope and shown to be nearly free of Malmquist bias. Secondly, instead of propagating uncertainties and correlations only linearly in a covariance matrix, our results are derived from rigorous full posterior sampling (using \gls{hmc}) of a \acrlong{bhm} that simultaneously accounts for standardisation, dust extinction, redshift uncertainties (from peculiar-velocity corrections), and cosmological distance. We have investigated extensions to account for documented environmental dependences of the \SNIa stretches and brightnesses and shown that they do not impact the qualitative agreement with \citetalias{Riess_2022}. Our analysis thus all but rules out peculiarities of the \citetalias{Riess_2022} Hubble-flow sample (\eg the complicated selection and compilation procedure) and simplified modelling of \SNaeIa as possible sources of the Hubble tension. Its resolution, therefore, lies either in the Cepheid calibration or in the geometric anchors\Emdash or outside the distance ladder entirely. Changing this conclusion would require a further, even larger photometric offset (\SI{\sim 0.13}{\mag}) to be uncovered in \gls{ztf} data. This work therefore also offers a first assessment of their potential for precision cosmography.

Future releases, with improved data reduction and calibration, are expected to increase the confidence in \acrshort{ztf}-derived results, especially when the legacy calibrators are replaced with \acrshort{ztf}-observed ones. Larger homogeneous samples will lead to a decrease in the statistical uncertainty (and elimination of some systematics), necessitating even more careful modelling. For instance, to account for the selection effects of compiling a calibrator sample (related to the presence and detectability of Cepheid variables), we plan to extend our model to include the host galaxies: more specifically, the occurrence of \SNaeIa and Cepheids from their hosts' stellar populations and the influence of physical properties like age or metallicity on their brightnesses. Lastly, as the model grows in complexity and the data grows in size, we will transition to a more modern Bayesian method: \acrlong{sbi}, which enables rigorous validation, calibration, and model misspecification checks.

    \clearpage
    \begin{acknowledgments}
    This work was in part supported by the European Research Council (ERC Synergy Grant RedH0T, Grant Agreement No.\@ 101225120). Funded by the European Union. Views and opinions expressed are however those of the author(s) only and do not necessarily reflect those of the European Union or the European Research Council Executive Agency. Neither the European Union nor the granting authority can be held responsible for them.
    The work of RJ is supported by the Simons Foundation. Funding for the work of LV and RJ was partially provided by project PID2022-141125NB-I00, and grant CEX2024-001451-M funded by MICIU/AEI/10.13039/501100011033.
    RT acknowledges co-funding from Next Generation EU, in the context of the National Recovery and Resilience Plan, Investment PE1 Project FAIR ``Future Artificial Intelligence Research''. This resource was co-financed by the Next Generation EU [DM 1555 del 11.10.22]. RT is partially supported by the Fondazione ICSC, Spoke 3 ``Astrophysics and Cosmos Observations'', Piano Nazionale di Ripresa e Resilienza Project ID CN00000013 ``Italian Research Center on High-Performance Computing, Big Data and Quantum Computing'' funded by MUR Missione 4 Componente 2 Investimento 1.4: Potenziamento strutture di ricerca e creazione di ``campioni nazionali di R\&S (M4C2-19)'' - Next Generation EU (NGEU).
    \end{acknowledgments}
    \begin{contribution}
        This paper was\Emdash lovingly and proudly\Emdash produced by humans. KK performed the analysis, designed the figures, and wrote the paper. MG assisted in verifying the pipeline. RJ, RT, and LV interrogated the results and assisted with their interpretation and polishing the manuscript.
    \end{contribution}

    \clearpage
    \appendix
    \crefalias{section}{appendix}
    \counterwithin{figure}{section}
    \renewcommand\thefigure{\thesection\arabic{figure}}

    \section{Posterior of all global parameters}\label{sec:corner-all}

    \begin{figure*}[h!]\centering
        \tikzsetnextfilename{\currfilebase}%
\begin{tikzpicture}[trim axis group left]
    \input{plotdata/corner}

    \pgfkeys{
        /prec/h/.initial={0},
        /prec/q0/.initial={1},
        /prec/M0/.initial={1},
        /prec/sigma_res/.initial={2},
        /prec/alpha/.initial={2},
        /prec/alpha_c/.initial={2},
        /prec/beta/.initial={0},
        /prec/mean_x/.initial={1},
        /prec/sigma_x/.initial={1},
        /prec/mean_c/.initial={2},
        /prec/sigma_c/.initial={2},
        /prec/mean_R/.initial={1},
        /prec/sigma_R/.initial={1},
        /prec/tau/.initial={2},
    }

    \begin{groupplot}[
        group/rows=12, group/columns=13,
        group/horizontal sep=0.1cm, tight layout=13, corner,
        tick label style={font=\tiny},
        plot/.style 2 args={
            2 param axis extents={#1}{#2},
            execute at begin axis={\acornerpanel*{#1}{#2}},
        },
        /pgf/number format/fixed,
        left edge/.style={plot/.append style={
            ylabel={\getparamlabel{##2}},
            /pgf/number format/precision/.expanded={\pgfkeysvalueof{/prec/##2}}
        }},
        bottom edge/.style={plot/.append style={
            param axis labels=##1,
            /pgf/number format/precision/.expanded={\pgfkeysvalueof{/prec/##1}}
        }},
    ]

        \nextgroupplot[left edge, plot={h}{M0}]
        \nextgroupplot[plot={q0}{M0}]
        \nextgroupplot[group/empty plot]
        \nextgroupplot[group/empty plot]
        \nextgroupplot[group/empty plot]
        \nextgroupplot[group/empty plot]
        \nextgroupplot[group/empty plot]
        \nextgroupplot[group/empty plot]
        \nextgroupplot[group/empty plot]
        \nextgroupplot[group/empty plot]
        \nextgroupplot[group/empty plot]
        \nextgroupplot[group/empty plot]
        \nextgroupplot[group/empty plot]

        \nextgroupplot[left edge, plot={h}{sigma_res}]
        \nextgroupplot[plot={q0}{sigma_res}]
        \nextgroupplot[plot={M0}{sigma_res}]
        \nextgroupplot[group/empty plot]
        \nextgroupplot[group/empty plot]
        \nextgroupplot[group/empty plot]
        \nextgroupplot[group/empty plot]
        \nextgroupplot[group/empty plot]
        \nextgroupplot[group/empty plot]
        \nextgroupplot[group/empty plot]
        \nextgroupplot[group/empty plot]
        \nextgroupplot[group/empty plot]
        \nextgroupplot[group/empty plot]

        \nextgroupplot[left edge, plot={h}{alpha}]
        \nextgroupplot[plot={q0}{alpha}]
        \nextgroupplot[plot={M0}{alpha}]
        \nextgroupplot[plot={sigma_res}{alpha}]
        \nextgroupplot[group/empty plot]
        \nextgroupplot[group/empty plot]
        \nextgroupplot[group/empty plot]
        \nextgroupplot[group/empty plot]
        \nextgroupplot[group/empty plot]
        \nextgroupplot[group/empty plot]
        \nextgroupplot[group/empty plot]
        \nextgroupplot[group/empty plot]
        \nextgroupplot[group/empty plot]

        \nextgroupplot[left edge, plot={h}{alpha_c}]
        \nextgroupplot[plot={q0}{alpha_c}]
        \nextgroupplot[plot={M0}{alpha_c}]
        \nextgroupplot[plot={sigma_res}{alpha_c}]
        \nextgroupplot[plot={alpha}{alpha_c}]
        \nextgroupplot[group/empty plot]
        \nextgroupplot[group/empty plot]
        \nextgroupplot[group/empty plot]
        \nextgroupplot[group/empty plot]
        \nextgroupplot[group/empty plot]
        \nextgroupplot[group/empty plot]
        \nextgroupplot[group/empty plot]
        \nextgroupplot[group/empty plot]

        \nextgroupplot[left edge, plot={h}{beta}]
        \nextgroupplot[plot={q0}{beta}]
        \nextgroupplot[plot={M0}{beta}]
        \nextgroupplot[plot={sigma_res}{beta}]
        \nextgroupplot[plot={alpha}{beta}]
        \nextgroupplot[plot={alpha_c}{beta}]
        \nextgroupplot[group/empty plot]
        \nextgroupplot[group/empty plot]
        \nextgroupplot[group/empty plot]
        \nextgroupplot[group/empty plot]
        \nextgroupplot[group/empty plot]
        \nextgroupplot[group/empty plot]
        \nextgroupplot[group/empty plot]

        \nextgroupplot[left edge, plot={h}{mean_x}]
        \nextgroupplot[plot={q0}{mean_x}]
        \nextgroupplot[plot={M0}{mean_x}]
        \nextgroupplot[plot={sigma_res}{mean_x}]
        \nextgroupplot[plot={alpha}{mean_x}]
        \nextgroupplot[plot={alpha_c}{mean_x}]
        \nextgroupplot[plot={beta}{mean_x}]
        \nextgroupplot[group/empty plot]
        \nextgroupplot[group/empty plot]
        \nextgroupplot[group/empty plot]
        \nextgroupplot[group/empty plot]
        \nextgroupplot[group/empty plot]
        \nextgroupplot[group/empty plot]

        \nextgroupplot[left edge, plot={h}{sigma_x}]
        \nextgroupplot[plot={q0}{sigma_x}]
        \nextgroupplot[plot={M0}{sigma_x}]
        \nextgroupplot[plot={sigma_res}{sigma_x}]
        \nextgroupplot[plot={alpha}{sigma_x}]
        \nextgroupplot[plot={alpha_c}{sigma_x}]
        \nextgroupplot[plot={beta}{sigma_x}]
        \nextgroupplot[plot={mean_x}{sigma_x}]
        \nextgroupplot[group/empty plot]
        \nextgroupplot[group/empty plot]
        \nextgroupplot[group/empty plot]
        \nextgroupplot[group/empty plot]
        \nextgroupplot[group/empty plot]

        \nextgroupplot[left edge, plot={h}{mean_c}]
        \nextgroupplot[plot={q0}{mean_c}]
        \nextgroupplot[plot={M0}{mean_c}]
        \nextgroupplot[plot={sigma_res}{mean_c}]
        \nextgroupplot[plot={alpha}{mean_c}]
        \nextgroupplot[plot={alpha_c}{mean_c}]
        \nextgroupplot[plot={beta}{mean_c}]
        \nextgroupplot[plot={mean_x}{mean_c}]
        \nextgroupplot[plot={sigma_x}{mean_c}]
        \nextgroupplot[group/empty plot]
        \nextgroupplot[group/empty plot]
        \nextgroupplot[group/empty plot]
        \nextgroupplot[group/empty plot]

        \nextgroupplot[left edge, plot={h}{sigma_c}]
        \nextgroupplot[plot={q0}{sigma_c}]
        \nextgroupplot[plot={M0}{sigma_c}]
        \nextgroupplot[plot={sigma_res}{sigma_c}]
        \nextgroupplot[plot={alpha}{sigma_c}]
        \nextgroupplot[plot={alpha_c}{sigma_c}]
        \nextgroupplot[plot={beta}{sigma_c}]
        \nextgroupplot[plot={mean_x}{sigma_c}]
        \nextgroupplot[plot={sigma_x}{sigma_c}]
        \nextgroupplot[plot={mean_c}{sigma_c}]
        \nextgroupplot[group/empty plot]
        \nextgroupplot[group/empty plot]
        \nextgroupplot[group/empty plot]

        \nextgroupplot[left edge, plot={h}{mean_R}]
        \nextgroupplot[plot={q0}{mean_R}]
        \nextgroupplot[plot={M0}{mean_R}]
        \nextgroupplot[plot={sigma_res}{mean_R}]
        \nextgroupplot[plot={alpha}{mean_R}]
        \nextgroupplot[plot={alpha_c}{mean_R}]
        \nextgroupplot[plot={beta}{mean_R}]
        \nextgroupplot[plot={mean_x}{mean_R}]
        \nextgroupplot[plot={sigma_x}{mean_R}]
        \nextgroupplot[plot={mean_c}{mean_R}]
        \nextgroupplot[plot={sigma_c}{mean_R}]
        \nextgroupplot[group/empty plot]
        \nextgroupplot[group/empty plot]

        \nextgroupplot[left edge, plot={h}{sigma_R}]
        \nextgroupplot[plot={q0}{sigma_R}]
        \nextgroupplot[plot={M0}{sigma_R}]
        \nextgroupplot[plot={sigma_res}{sigma_R}]
        \nextgroupplot[plot={alpha}{sigma_R}]
        \nextgroupplot[plot={alpha_c}{sigma_R}]
        \nextgroupplot[plot={beta}{sigma_R}]
        \nextgroupplot[plot={mean_x}{sigma_R}]
        \nextgroupplot[plot={sigma_x}{sigma_R}]
        \nextgroupplot[plot={mean_c}{sigma_R}]
        \nextgroupplot[plot={sigma_c}{sigma_R}]
        \nextgroupplot[plot={mean_R}{sigma_R}]
        \nextgroupplot[group/empty plot]

        \nextgroupplot[bottom edge, left edge, plot={h}{tau}, xticklabel={$\pgfmathparse{100*\tick}\pgfmathprintnumber[/pgf/number format/precision=0]{\pgfmathresult}$}]
        \nextgroupplot[bottom edge, plot={q0}{tau}]
        \nextgroupplot[bottom edge, plot={M0}{tau}, xtick={-19.2,-19.1,-19,-18.9}]
        \nextgroupplot[bottom edge, plot={sigma_res}{tau}]
        \nextgroupplot[bottom edge, plot={alpha}{tau}]
        \nextgroupplot[bottom edge, plot={alpha_c}{tau}]
        \nextgroupplot[bottom edge, plot={beta}{tau}]
        \nextgroupplot[bottom edge, plot={mean_x}{tau}]
        \nextgroupplot[bottom edge, plot={sigma_x}{tau}]
        \nextgroupplot[bottom edge, plot={mean_c}{tau}]
        \nextgroupplot[bottom edge, plot={sigma_c}{tau}]
        \nextgroupplot[bottom edge, plot={mean_R}{tau}]
        \nextgroupplot[bottom edge, plot={sigma_R}{tau}]
    \end{groupplot}

    \ztflegend[matrix anchor=north west, draw=none]{({{group c8r7.north west}} |- {{group c6r5.north east}})}
\end{tikzpicture}%
        \caption{All two-dimensional marginal posteriors for pairs of global parameters of interest (except \Ho--\cosmodecelo, shown in \cref{fig:corner}). Once again, contours delimit the \SI{39}{\percent} and \SI{86}{\percent} credibility regions (\acrlong{hpd}). Results from the \enquote{uncorrected} \acrshort{ztf} data match precisely the corrected version, save for the stretch population (\meansaltx, \stdsaltx) and cosmology (\cosmoh, \cosmodecelo). Restricting the sample to \SNaeIa in late-type hosts affects primarily the distribution of stretches (\meansaltx, \stdsaltx) and the related coefficient (\saltalpha). Note also the \Mo--\saltbeta correlation, which underscores the importance of proper colour and dust modelling.\label{fig:corner-all}}
    \end{figure*}

    \clearpage
    \section{Model modifications}\label{sec:variants}

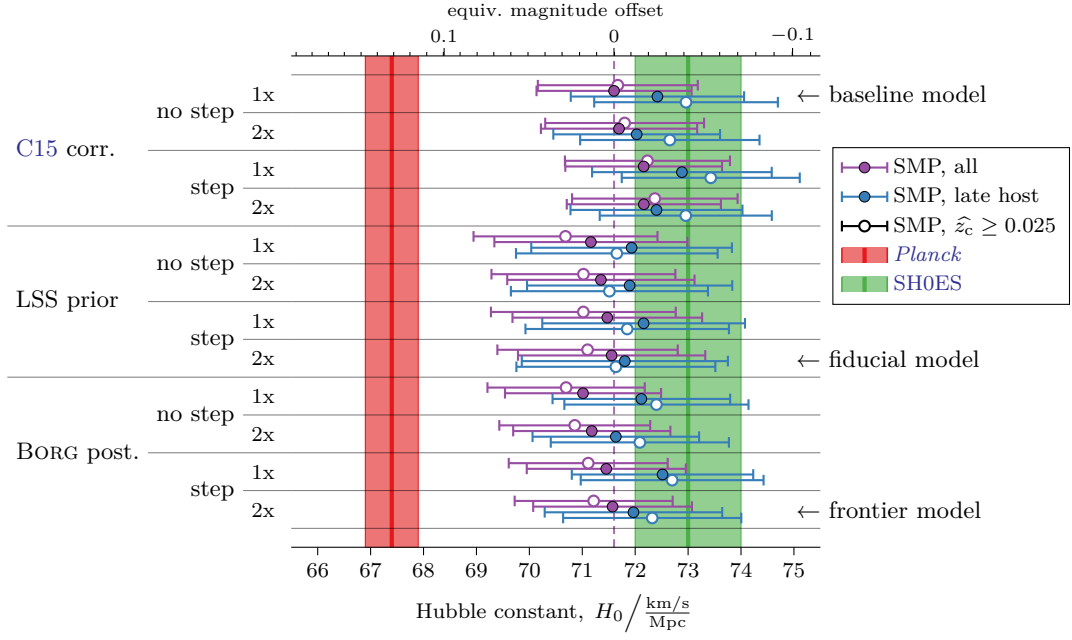
\begin{figure*}
    \hfil\DisableQuotes\tikzsetnextfilename{\currfilebase}%
\begin{tikzpicture}[trim axis left, trim axis right]
    \ExplSyntaxOn
    \cs_new:Npn \__get_max_length:nN #1#2 {
        \dim_set:Nn \l_tmpa_dim { 0pt }
        \clist_map_inline:nn { #1 } {
            \hbox_set:Nn \l_tmpa_box {##1}
            \dim_compare:nT {\l_tmpa_dim < \box_wd:N \l_tmpa_box}
            {
                \dim_set:Nn \l_tmpa_dim { \box_wd:N \l_tmpa_box }
            }
        }
        \dim_set_eq:NN #2 \l_tmpa_dim
    }
    \newcommand{\getwidest}{\__get_max_length:nN}
    \ExplSyntaxOff

    \newlength{\maxwidth}
    \directlua{
        labels = {
            ["pc"] = {
                ["pborg"] = "\\getdata{lbls/pc/pborg}",
                ["pcorr"] = "\\getdata{lbls/pc/pcorr}",
                ["pcov"] = "\\getdata{lbls/pc/pcov}"
            },
            ["dostep"] = {
                ["nostep"] = "\\getdata{lbls/step/nostep}",
                ["step"] = "\\getdata{lbls/step/step}",
            },
            ["z_mode"] = {
                ["latz"] = "\\zcsn latent",
                ["linz"] = "$\\snindex{\\sigmazc}\\to\\snindex{\\sigmadistmod}$",
            },
            ["R_mode"] = {
                ["distR"] = "\\getdata{lbls/R_mode/distR}",
                ["globalR"] = "\\getdata{lbls/R_mode/globalR}",
            },
            ["use_AV"] = {
                ["AV"] = "\\Avsn{}",
                ["E"] = "$\\snindex{E}$",
            },
            ["use_2x"] = {
                ["1x"] = "\\getdata{lbls/use_2x/1x}",
                ["2x"] = "\\getdata{lbls/use_2x/2x}",
            }
        }
        function reserve_label(cat)
            local s = {}
            for l in pairs(labels[cat]) do
                s[\string#s+1] = "{"..labels[cat][l].."}"
            end
            tex.sprint("\\getwidest{"..table.concat(s, ",").."}{\\maxwidth}")
        end
        function print_one_label(cat, lbl)
            tex.sprint(labels[cat][lbl])
        end
        function one_label(cat, lbl)
            reserve_label(cat, lbl)
            tex.sprint("\\makebox[\\maxwidth][l]{"..labels[cat][lbl].."}")
        end
        function one_phantom_label(cat, lbl)
            reserve_label(cat, lbl)
            tex.sprint("\\makebox[\\maxwidth][l]{}")
        end
    }

    \pgfplotstableread{plotdata/stats.txt}\thestatstable
    \pgfplotstablegetrowsof{\thestatstable}
    \pgfmathsetmacro{\nvariants}{\pgfplotsretval-1}

    \tikzset{lbl/.style={inner ysep=0pt}, hline/.style={thin, opacity=0.5}}

    \pgfplotsset{scale only axis, width=7cm, height=6.5cm}

    \newcommand{\Homin}{0.655}
    \newcommand{\Homax}{0.755}

    \begin{axis}[
        name=ax, H0 axis, param axis extents={h},
        xmin={\Homin}, xmax={\Homax},
        xtick={0.64,0.65,...,0.76},
        axis x line*=bottom,
        axis y line*=left, y axis line style={draw=none},
        y dir=reverse, ymin={-1}, ymax={\nvariants+1},
        ytick={0,1,...,\nvariants},
        ytick style={draw=none},
        yticklabels from table={\thestatstable}{use_2x},
        yticklabel style={name={lbl \ticknum}, lbl},
        variants axis,
        legend style={anchor=north west, at={({{rel axis cs:1,0.5}} |- {{axis cs:0.7,3.5}})}, right=0.5em},
        after end axis/.code={
            \newcommand{\lbl}[2]{\smash{\directlua{one_label("##1","##2")}}\vphantom{\strut}}
            \newcommand{\mklbl}[6][]{
                \draw[decorate, decoration={brace, mirror, amplitude=2pt}, draw=none]
                    (##2.west) -- (##3.west)
                    node[pos=##6, left=0pt, lbl, ##1] {\lbl{##4}{##5}};
            }
            %
            %
            %
            \mklbl[name=step 0]{lbl 0}{lbl 1}{dostep}{nostep}{0.5}
            \mklbl[name=step 1]{lbl 2}{lbl 3}{dostep}{step}{0.5}
            \mklbl[name=step 2]{lbl 4}{lbl 5}{dostep}{nostep}{0.5}
            \mklbl[name=step 3]{lbl 6}{lbl 7}{dostep}{step}{0.5}
            \mklbl[name=step 4]{lbl 8}{lbl 9}{dostep}{nostep}{0.5}
            \mklbl[name=step 5]{lbl 10}{lbl 11}{dostep}{step}{0.5}
            \mklbl[name=pc 0]{step 0}{step 1}{pc}{pcorr}{0.5}
            \mklbl[name=pc 1]{step 2}{step 3}{pc}{pcov}{0.5}
            \mklbl[name=pc 2]{step 4}{step 5}{pc}{pborg}{0.5}
            \newcommand{\mkline}[2]{
                \foreach \i in {##2}{
                    \draw[hline] let
                        \p1 = ({{rel axis cs:0,0}} |- {{axis cs:0.7,\i+0.5}}),
                        \p2 = (##1 0.west)
                    in (\x1, \y1) -- (\x1,\y1 -| \x2,\y2);
                }
            }
            \mkline{lbl}{0,2,4,6,8,10}
            \mkline{step}{1,5,9}
            \mkline{pc}{3,7}
            \foreach \y/\lbl in {0/baseline,7/fiducial,11/frontier}{
                \node[right] at ({{axis cs:0.7, \y}} -| {{rel axis cs:1, 0}}) {\llap{\textleftarrow\ }\lbl{} model};
            }
        },
    ]
        \pgfplotsinvokeforeach{-0.5,...,\nvariants+0.5}{
            \axhline[hline]{#1};
        }

        \addplot[model, altmark] table [y expr={\coordindex-0.225}, x=hz-mean, x error=hz-std] {\thestatstable};
        \addplot[late, altmark] table [y expr={\coordindex+0.225}, x=latehz-mean, x error=latehz-std] {\thestatstable};
        \addplot[model] table [y expr={\coordindex-0.075}, x=full-mean, x error=full-std] {\thestatstable};
        \addplot[late] table [y expr={\coordindex+0.075}, x=late-mean, x error=late-std] {\thestatstable};

        \addlegendimage{model}
            \addlegendentry{\labelbase}
        \addlegendimage{late}
            \addlegendentry{\labellate}
        \addlegendimage{color=black, altmark}
            \addlegendentry{\labelhz}
        \addlegendimage{Planck, h result, vstrip legend 1}
            \addlegendentry{\citetalias{Planck2018-1}}
        \addlegendimage{SH0ES, h result, vstrip legend 1}
            \addlegendentry{\citetalias{Riess_2022}}
    \end{axis}
    \begin{axis}[
        at={(ax.south west)},
        axis y line=none, ymin=0, ymax=1,
        H0 as magoff={\Homin}{\Homax},
    ]\end{axis}
\end{tikzpicture}\EnableQuotes\hfil%
    \caption{Inference results for the Hubble constant (posterior mean $\pm$ st.\@ dev.\@) from the data variants as in \cref{fig:corner} using the \num{12} distinct model configurations.\label{fig:variants}}
\end{figure*}

This \namecref{sec:variants} presents three extensions of the hierarchical model, whose effects on our main result (the inferred value of \Ho) we explore in \cref{tbl:variants,fig:variants}, namely:
\begin{itemize}
    %
    \item
    Modifying the treatment of peculiar velocities to include correlations among them. Recall that our baseline model applies a fixed correction based on external density-field data from \citet{Carrick_2015} and assumes a (diagonal, constant) uncertainty of $\SI{300}{\kilo\metre\per\second\per c}$. We consider two alternatives.

    The first is to \emph{not} correct for peculiar velocities but treat them as latent variables with a correlated prior from \gls[prereset=global]{lss} considerations. We assume a multivariate normal distribution with zero mean and covariance \vcov, following \citet[\crefext{ext:Carreres_2025a:eqn:vcov}]{Carreres_2025a}, who furthermore show that the required cosmological assumptions (\eg on \Ho and \Omo, within reasonable ranges) do not impact the final result for \Ho. Propagating this covariance (and measurement noise, \sigmaz) to cosmological redshifts (\cf \cref{eqn:redshift})\Emdash\ie analytically marginalising peculiar velocities at leading order\Emdash results in
    \begin{equation}\label{eqn:zcov-lss}
        \varcomp{\zcov^{ij}} = \qty(1+\zcobs^i)\qty(1+\zcobs^j) \qty[\frac{\varcomp{\vcov^{ij}}/c^2}{\qty(1+\zpecobs^i)\qty(1+\zpecobs^{j})} + \frac{\sigmaz^2}{\qty(1+\zobs^i)\qty(1+\zobs^{j})}],
    \end{equation}
    where $i, j$ run over the Hubble-flow \SNae, and \cref{eqn:zcsn} being replaced by
    \begin{equation}
        \mathllap{\text{\getdata{lbls/pc/pcov}:}\quad}
        \zcvar \sim \Normal(\zcmbobsvar, \zcov),
    \end{equation}
    where $\snindex{\hat{\z}_{\mathrm{CMB}}} = \flatfrac{\qty(1+\zobssn)}{\qty(1+\snindex{\zsun})}-1$ are the measured redshifts transformed to the \gls{cmb} frame. Note the slight inconsistency in using corrected \zcobs and \zpecobs (as in the baseline) to evaluate \zcov once, rather than re-calculating it while marginalising \zpec; its effect, however, is tiny.

    The second alternative is to adopt a correlated \emph{posterior} for the peculiar velocity field $\vector{v}_{\mathrm{pec}}(\vector{r})$, conditioned on density measurements. The \texttt{Manticore-local} project \citep{McAlpine_2025,McAlpine_2026} provides \num{80} suitable posterior samples, derived with the \acrshort{borg} algorithm \citep{BORG} from the same data as \citetalias{Carrick_2015}. Following \citet{Stiskalek_2026c}, we assume an additional ("external") uniform velocity field $\vector{v}_{\mathrm{ext}}$ and dispersion $\sigma_v^2$ at each point, sampled from the posteriors\footnote{uncorrelated; in any case, $\vector{v}_{\mathrm{ext}}$ and $\sigma_v$ are not \aposteriori* correlated with \Ho, so the choices here have little effect on our final result} from \crefext{ext:Stiskalek_2026c:fig:post} therein, approximated as Gaussians: $\abs{V_{\mathrm{ext}}}\approx\SI{200+-50}{\kilo\metre\per\second}$ and $\sigma_v \approx \SI{150+-30}{\kilo\metre\per\second}$. Given a sample $\qty(\vector{v}_{\mathrm{pec}}(\vector{r}), \vector{v}_{\mathrm{ext}}, \sigma_v)$ and latent heliocentric redshifts $\zsn \sim \Normal(\zobssn, \sigmaz^2)$, we then obtain the corresponding cosmological redshifts of the \gls{ztf} \SNae as in \cref{eqn:zcobssn}. We repeat this \num{1000} times and calculate the mean ($\variable{\mu}_{\zc}$) and covariance ($\zcov^{\text{\acrshort{borg}}}$) of the resulting samples, thus propagating (to leading order) all redshift-related uncertainties. Finally, we replace \cref{eqn:zcsn} with
    \begin{equation}
        \mathllap{\text{\getdata{lbls/pc/pborg}:}\quad}
        \zcvar \sim \Normal(\variable{\mu}_{\zc}, \zcov^{\text{\acrshort{borg}}}).
    \end{equation}
    \item
    Modelling a "mass step", \ie a (intrinsic absolute) magnitude offset between \SNaeIa hosted in "high"- and "low"-mass galaxies by adding an additional coefficient{\Endash}covariate pair to the standardisation relation \cref{eqn:Mintsn}:
    \begin{equation}
        \Mintsn = \Mo + \saltalpha\saltxintsn + \saltbeta\saltcintsn \mspace{\medmuskip} \underline{+ \mspace{\medmuskip} \gamma\saltxcompsn[\Mstar]} + \deltaMsn,
    \end{equation}
    where $\gamma$ is a global parameter, and $\saltxcompsn[\Mstar]=1$ if the stellar mass $\Mstarsn$ of the \SN's host exceeds a given threshold\Emdash here set to \SI[unit-mode=math]{e10}{\Msun} for simplicity\Emdash and $0$ otherwise. Ideally, this requires appropriate measurements, to be modelled \eg as $\Mstarobssn \sim \Normal(\Mstarsn, \qty(\snindex{\hat{\sigma}_{\Mstar}})^2)$, but uncertainties are only available (in the \citetalias{Riess_2022} data release\footref{foot:github-sh0es}) for \num{25} of the \num{42} calibrators, and furthermore, the \SHoES host-analysis methodology does not match that employed in \ztfdrtwo, which may anyway lead to systematic differences in the mass estimates. Because of this, we a) assume perfect measurements ($\Mstarobssn=\Mstarsn$), and b) do not include a mass step in our baseline (main-text) model. In any case, proper modelling of host-related/environmental effects requires inferring the distribution of stellar metallicities and ages (which is much harder to sample in a \gls{bhm} and must include the host/environment observables, \eg photometry or spectra), so that \SN brightnesses can be\Emdash in accordance with physical reasoning rather than \emph{ad hoc}\Emdash correlated with their progenitors rather than their host galaxies as a whole.
    %
    %
    \item Allowing for two \SNIa populations with different stretches, standard magnitudes, and stretch{\Endash}magnitude correlations, as have been observed in the literature \citep[\eg][]{Ginolin_2025a,Wojtak_2026}. This means replacing \cref{eqn:saltxintsn,eqn:Mintsn}\Emdash we leave the colour-related \cref{eqn:saltcintsn} as is\Emdash with a mixture distribution\footnote{During \gls{hmc} sampling, the component assignment can be efficiently marginalised using the \funsor package \citep{funsor}.}:
    \begin{equation}
        \mqty[\saltxintsn\\\Mintsn] \sim
        \xfrac \times \mqty[\begin{aligned}
            &\Normal(\meansaltx[1], \stdsaltx[1]^2)
            \\& \Normal(\Mo[1] + \saltalpha[1] \saltxintsn + \mathellipsis, \sigmares^2)
        \end{aligned}]
        + \qty(1-\xfrac) \times \mqty[\begin{aligned}
            &\Normal(\meansaltx[2], \stdsaltx[2]^2)
            \\& \Normal(\Mo[2] + \saltalpha[2] \saltxintsn + \mathellipsis, \sigmares^2)
        \end{aligned}],
    \end{equation}
    with \xfrac (representing the relative prominence\Emdash mixture amount\Emdash of the first population), \meansaltx[1], \stdsaltx[1], \meansaltx[2], \stdsaltx[2], \Mo[1], \saltalpha[1], \Mo[2], \saltalpha[2] replacing \meansaltx, \stdsaltx, \Mo, \saltalpha. To avoid the degeneracy arising from the symmetry between the two components, we set a prior that enforces $\meansaltx[1] \le \meansaltx[2]$ but still has uniform density on the respective subspace of $\interval{-3}{3} \otimes \interval{-3}{3}$. The rest of the parameters inherit the priors from their single-component equivalents:
    \begin{alignat*}{5}
        \stdsaltx[1] & \sim \priorstdsaltx &&\qq{and}& \stdsaltx[2]  & \sim \priorstdsaltx,
        \\
        \Mo[1] & \sim \priorMo &&\qq{and}& \Mo[2] &\sim \priorMo,
        \\
        \saltalpha[1] & \sim \priorsaltalpha &&\qq{and}& \saltalpha[2] &\sim \priorsaltalpha,
    \end{alignat*}
    while $\xfrac\sim\Uniform(\num{0}, \num{1})$.
    %
\end{itemize}

 \paragraph{Host dust} Finally, the distribution of host dust laws has been the topic of recent discussion in the literature \citep[see][and references therein]{Giunta_2026}, prompting us to quantify its influence on \Ho inference. Our baseline model flexibly fits a Gaussian population, allowing each \SN to be extinguished with a bespoke \Rvsn. If we instead assume a shared dust-law parameter (modifying \cref{eqn:Rvsn} to simply $\Rvsn=\Rv$ and assigning to \Rv the same prior as \meanRv, \ie \priormeanRv), the inferred \Ho shifts by \Hoval{-0.8}, the largest single-intervention impact we have encountered. Nevertheless, this can be regarded as a mis-specification since the (more extended) baseline model prefers a spread $\sigmaRv\approx\num{0.5}$ (see \cref{fig:corner-all}), so we highlight this as an example of the serious adverse effects of simplified dust modelling.

    \clearpage
    \bibliographystyle{mnras}
    \bibliography{ech0ez}

\end{document}